\documentclass[a4paper]{article}
\usepackage{a4, ams, amsmath, amssymb, graphics, subfigure, fullpage, color}%
\usepackage[pdftex]{graphicx}
\usepackage{fancyhdr}
\usepackage{multirow}
\usepackage{amsthm}
\usepackage[colorlinks]{hyperref}
\hypersetup{colorlinks,citecolor=green,filecolor=black,linkcolor=blue,urlcolor=blue}

\newtheorem{remark}{Remark}
\usepackage{upgreek} 

\newcommand{\MyBib}{./MyBib}

\usepackage{lscape}

\newcommand{\cN}{\mathcal{N}}
\newcommand{\cF}{\mathcal{F}}
\newcommand{\cG}{\mathcal{G}}

\newcommand{\FF}{\mathbb{F}}
\newcommand{\GG}{\mathbb{G}}
\newcommand{\HH}{\mathbb{H}}

\DeclareMathOperator{\E}{\mathbb{E}}
\renewcommand{\Q}{\mathbb{Q}}
\renewcommand{\Pr}{\mathbb{P}}
\DeclareMathOperator{\cov}{\mathbb{C}\textrm{ov}}
\DeclareMathOperator{\filF}{\mathcal{F}}
\DeclareMathOperator{\filG}{\mathcal{G}}

\DeclareMathOperator{\e}{\textrm{e}}

\DeclareMathOperator{\ind}{1{\hskip -2.5 pt}\textrm{I}}

\newcommand{\beq}{\begin{equation}}
\newcommand{\eeq}{\end{equation}}
\newcommand{\beqn}{\begin{eqnarray}}
\newcommand{\eeqn}{\end{eqnarray}}
\newcommand{\bfig}{\begin{figure}}
\newcommand{\efig}{\end{figure}}
\newcommand{\btab}{\begin{table}}
\newcommand{\etab}{\end{table}}

\title{{\bf Disentangling wrong-way risk: pricing CVA via change of measures and drift adjustment}\vspace{0.5cm}}

\author{ 
Damiano Brigo\thanks{Dept. of Mathematics, Imperial College London, {\small \tt{damiano.brigo@imperial.ac.uk}}}
\and
Fr\'ed\'eric Vrins\thanks{
 Louvain Finance Center \& Center for Operations Research and Econometrics (CORE), Universit\'e catholique de Louvain, {\small \tt{frederic.vrins@uclouvain.be}}}
}
\date{First version: February 29, 2016. This version: \today .}

\begin{document}
\maketitle
\begin{abstract} A key driver of Credit Value Adjustment (CVA) is the possible dependency between exposure and counterparty credit risk, known as Wrong-Way Risk (WWR). At this time, addressing WWR in a both sound and tractable way remains challenging: arbitrage-free setups have been proposed by academic research through dynamic models but are computationally intensive and hard to use in practice. Tractable alternatives based on resampling techniques have been proposed by the industry, but they lack mathematical foundations. This probably explains why WWR is not explicitly handled in the Basel III regulatory framework in spite of its acknowledged importance. The purpose of this paper is to propose a new method consisting of an appealing compromise: we start from a stochastic intensity approach and end up with a pricing problem where WWR does not enter the picture explicitly. This result is achieved thanks to a set of changes of measure: the WWR effect is now embedded in the drift of the exposure, and this adjustment can be approximated by a deterministic function without affecting the level of accuracy typically required for CVA figures. The performances of our approach are illustrated through an extensive comparison of Expected Positive Exposure (EPE) profiles and CVA figures produced either by (i) the standard method relying on a full bivariate Monte Carlo framework and (ii) our drift-adjustment approximation. Given the uncertainty inherent to CVA, the proposed method is believed to provide a promising way to handle WWR in a sound and tractable way.
\end{abstract}
\section{Introduction}
\label{sec:intro}
%
The 2008 Financial crisis stressed the importance of accounting for counterparty risk in the valuation of OTC transactions, even when the later are secured via (clearly unperfect) collateral agreements. Counterparty default risk calls for a price adjustment when valuing OTC derivatives, called Credit Value Adjustment (CVA). This adjustment depends on the traded portfolio $\Pi$ and the counterparty $C$. It represents the market value of the expected losses on the portfolio in case $C$ defaults prior to the portfolio maturity $T$. Alternatively, this can be seen as today's price of replacing the counterparty in the financial transactions constituting the portfolio, see for example~\cite{Brigo06}, \cite{Brigo13}, \cite{Grego10}. The mathematical expression of this adjustment can be derived in a rather easy way within a risk-neutral pricing framework. Yet, the computation of the resulting conditional expectation poses some problems when addressing Wrong-Way Risk (WWR) that is, accounting for the possible statistical dependence between exposure and counterparty credit risk. Several techniques have been proposed to tackle this point. At this time, there are two main approaches to tackle WWR: the dynamic approach (either structural or reduced-form) and the static (resampling) approach. The first one provides an arbitrage-free setup and is popular among academic researchers. Unfortunately, it has the major disadvantage of being computationally intensive and cumbersome, which makes its practical use difficult. On the other hand, the second approach does not have a rigorous justification but has the nice feature of providing the industry with a tractable alternative to evaluate WWR in a rather simple way. In spite of its significance, WWR is currently not explicitly accounted for in the Basel III regulatory framework; the lack of a reasonable alternative to handle CVA is probably one of the reasons.

In this paper, we revisit the CVA problem under WWR and propose an appealing way to handle it in a sound but yet tractable way. We show how CVA \textit{with WWR} can be written as CVA \textit{without WWR} provided that the exposure dynamics is modified accordingly. This will be achieved via a set of measures called ``wrong way measures''. 

The paper is organized as follows. Section~\ref{sec:CVA} recalls the fundamental CVA pricing formulae with and without WWR. Next, in Section~\ref{sec:WWR}, we briefly review the most popular techniques to address WWR in CVA computations. We then focus on the case where default risk is managed in a stochastic intensity framework and consider a Cox process setting more specifically. Section~\ref{sec:WWM} introduces a set of new num\'eraires that will generate equivalent martingale measures called \textit{wrong way measures} (WWM). Equipped with these new measures, the CVA problem with WWR takes a similar form as the CVA problem without WWR, provided that we change the measure under which one computes the expectation of the positive exposure. Section~\ref{sec:EDA} is dedicated to the computation of the exposure dynamics under the WWM. Particular attention is paid to the stochastic drift adjustment under affine intensity models. In order to reduce the complexity of the pricing problem, the stochastic drift adjustment is approximated by a deterministic function; the WWR effect is thus fully encapsulated in the exposure's drift via a deterministic adjustment. Finally, Section~\ref{sec:numexp} proposes an extensive analysis of the performances of the proposed approach in comparison with the standard stochastic intensity method featuring Euler discretizations of the bivariate stochastic differential equation (SDE) governing the joint dynamics of default intensity (credit spread risk) and portfolio value (market risk).
%
\section{Counterparty risk adjustment}
\label{sec:CVA}
%
Define the short (risk-free) rate process $r=(r_t)_{t\geq 0}$ and the corresponding bank account num\'eraire $B_t:=\e^{\int_0^t r_s ds}$ so that the deflator $B:=(B_t)_{t\geq 0}$ has dynamics :
\beq
dB_t=r_tB_tdt\;.\nonumber
\eeq
Under the no-arbitrage assumption, there exists a risk-neutral probability measure $\Q$ associated to this num\'eraire, in the sense that it makes all $B$-discounted  non-dividend paying tradeable assets $\Q$-martingales. In this setup, CVA can be computed as the $\Q$- expectation of the non-recovered losses resulting from counterparty's default, discounted according to $B$.\bigskip

More explicitly, if $R$ stands for the recovery rate of C and $V_t$ is the close out price of $\Pi$ at time $t$\footnote{Here, we assume that this corresponds to the risk-free price of the portfolio which is the most common assumption, named ``risk free closeout'', even though other choices can be made, such as replacement closeout, see for example \cite{Brigocloseout},\cite{Brigo13}}, the general formula for the CVA on portfolio $\Pi$ traded with counterparty C which default time is modeled via the random variable $\tau>0$ is given by (see for example \cite{brigomasetti}):
\beq
\hbox{CVA}=\E^B\left[(1-R)\ind_{\{\tau\leq T\}}\frac{V^+_\tau}{B_\tau}\right]=(1-R)\E^B\left[\E^B\left[H_T\frac{V^+_\tau}{B_\tau}\Big|\sigma(H_u, 0 \leq u \leq t)\right]\right]\nonumber
\eeq
%
where $\E^B$ denotes the expectation operator under $\Q$, $H:=(H_t)_{t\geq 0}$ is the default indicator process defined as $H_t:=\ind_{\{\tau\leq t\}}$, and the second equality results from the assumption that $R$ is a constant and from the tower property. The outer expectation can be written as an integral with respect to the risk-neutral \textit{survival probability }
\beq
G(t):=\Q[\tau>t]=\E^B\left[\ind_{\{\tau>t\}}\right]\nonumber\; .
\eeq
The survival probability is a deterministic positive and decreasing function satisfying $G(0)=1$ and typically expressed as $G(t)=e^{-\int_0^t h(s)ds}$ where $h$ is a non-negative function called \textit{hazard rate}. In practice, this curve is bootstrapped from market quotes of securities driven by the creditworthiness of $C$, i.e. defaultable bonds or credit default swaps (CDS). If $\tau$ admits a density, the expression for CVA then becomes
\beq
\hbox{CVA}=-(1-R)\int_{0}^T\E^B\left[\frac{V^+_t}{B_t}\Big|\tau=t\right]dG(t)\; .\label{eq:CVAgen}
\eeq
In the case where the portfolio $\Pi$ is independent of $\tau$, one can drop  conditioning in the above expectation to obtain the so-called \textit{standard} (or \textit{independent}) CVA formula:
\beq
\boxed{\hbox{CVA}^\perp=-(1-R)\int_{0}^T\E^B\left[\frac{V^+_t}{B_t}\right]dG(t)\; .\label{eq:CVAind}}
\eeq
where the superscript $\perp$ in general denotes that the related quantity is computed under the  independence assumption.
The deterministic function being integrated with respect to the survival probability is called the (\textit{discounted}) \textit{expected positive exposure}, also known under the acronym EPE:
$$\hbox{EPE}^\perp(t):=\E^B\left[\frac{V^+_t}{B_t}\right]\;.$$

Under this independence assumption, CVA takes the form of the weighted (continuous) sum of European call option prices with strike $0$ where the underlying of the option is the residual value of the portfolio $\Pi$.
%
\section{Wrong way risk}
\label{sec:WWR}
%
In the more general case where the market value of $\Pi$ depends on the default time $\tau$, we cannot drop conditioning in the expectation~(\ref{eq:CVAgen}), and one has to account for the dependency between credit and exposure. Depending on the sign of this relationship and, more generally, on the joint distribution of the portfolio value and the default time, this can increase or decrease the CVA; when CVA is increased, this effect is known as wrong way risk (WWR). When CVA decreases, this is called right way risk. In this paper we will use the term ``wrong way risk'' to loosely denote both situations. In order to capture this effect, we need to model \textit{jointly} exposure and credit. The first named author and co-authors pioneered the literature on WWR in a series of papers using a variety of modeling approaches across asset classes. In interest rate markets, the analysis of WWR on uncollateralized interest rate portfolios is studied in \cite{brigoccdsrisk} via intensity models for credit risk, while WWR on collateralized interest  rate portfolios is studied in \cite{bcpp13}. For credit markets, and uncollateralized CDS in particular, WWR is considered in \cite{brigochourdakis}, where intensity models and copula functions are used; WWR on collateralized CDS with collateral and gap risk is studied with the same technical tools in \cite{brigopallaagocollateral}. WWR on commodities, and oil swaps in particular, has been studied in \cite{brigobakkar} via intensity models, and WWR on equity is studied in \cite{brigomorinitarenghi} resorting to analytically tractable first passage (AT1P) firm value models. Most of these studies are summarized in
the monograph \cite{Brigo13}.
%
\subsection{Two approaches for one problem}
%
Two main approaches have been proposed in the literature to tackle WWR. They all aim at coupling portfolio value and default likelihood in a tractable way. The first approach (called \textit{dynamic}) consists in modeling credit worthiness using stochastic processes. Among this first class of models one distinguish two setups. The first dynamic setup (\textit{structural model}) relies on Merton's approach to model the firm value. Default is reached as soon as the firm value goes below a barrier representing the level of the firm's assets. This method is very popular in credit risk in general, except for pricing purposes as it is know to underestimate short-term default~(see e.g. \cite{Ballo15,Brigo13,brigomorinitarenghi} and references therein for CVA pricing methods using structural credit models). The second dynamic setup (\textit{reduced-form model}) consists in modeling the default likelihood via a stochastic intensity process. In this setup, default is unpredictable; only the default likelihood is modeled. In the sequel we restrict ourselves to stochastic intensity models, which is the most popular dynamic setup for CVA purposes (see for instance \cite{Brigo13},\cite{brigoccdsrisk},\cite{Hull12},\cite{Pykth11}). \footnote{Note that other methodologies have been recently proposed for credit risk modeling and CVA pricing (see e.g.~\cite{Vrins16a} and ~\cite{Vrins16}) but they will not be considered here.}

This first class of models is mathematically sound in the sense that it can be arbitrage-free if handled properly. However, as pointed in \cite{Pykth11}, dealing with this additional stochastic process may be computationally intensive. Hence, practitioners developed a second class of models called \textit{static}, to get rid of these difficulties. They consist in coupling exposure and credit using a copula, a specific function that creates a valid multivariate distribution from univariate ones (this is not to be confused with the copula connecting two default times that was used for example in \cite{brigopallaagocollateral}, resulting in a more rigorous formulation in that context). This method is also known as a \textit{resampling technique} and is very popular among practitioners as it drastically simplifies the way CVA can be evaluated under WWR. In particular, it is numerically interesting in the sense that in a first phase one can consider exposure and credit separately, and then, in a second phase, introduce the dependence effect \textit{a posteriori} by joining the corresponding distributions via a copula (see e.g. \cite{Pykth11} and \cite{Sokol11b} for further reading on this technique). Clearly, this way to couple exposure and credit risk is somewhat artificial. In particular, it is known to suffer from potential arbitrage opportunities, contrary to the WWR approaches listed earlier. 

In summary there are two classes of models: dynamic arbitrage-free models that are computationally demanding and hard to use in practice, and static resampling models that have no sound mathematical justification but providing a tractable alternative for the industry. Later in the paper we explain how one can develop a framework encompassing the best of both the static and dynamic approaches without their inconveniences. In particular, we circumvent the technical difficulty inherent to stochastic intensity models with the help of changes of measure. Before doing so we provide the reader with additional details regarding the stochastic intensity model setup. 
%
\subsection{CVA under a stochastic intensity model}
%
The reduced-form approach relies on a change of filtrations. Filtration $\GG:=(\cG_t)_{t\geq 0}$ represents the total information available to the investors on the market. In our context, this can be viewed as all relevant asset prices and/or risk factors. All stochastic processes considered here are thus defined on a complete filtered probability space $(\Omega,\cG,\GG=(\cG_t)_{0\leq t\leq T},\Q)$ where $\Q$ is the risk-neutral measure and $\cG:=\cG_T$ with $T$ the investment horizon (which can be considered here as the portfolio maturity). 
We can define $\FF:=(\cF_t)_{0\leq t\leq T}$ as the largest subfiltration of $\GG$ preventing the default time $\tau$ to be a $\FF$-stopping time. In other words, $\FF$ contains the same information as $\GG$ except that the default indicator process $H$ is not observable (i.e. $H$ is adapted to $\GG$ but not to $\FF$). 

In other terms, we are assuming the total market filtration $\GG$ to be separable in $\FF$ and the pure default monitoring filtration $\mathbb{H}$ where $\mathbb{H} = ({\mathcal H}_t)_{0\leq t\leq T}$, 
\beq
 {\mathcal H}_t = \sigma(H_u, 0 \leq u \leq t), \ \ {\mathcal G}_t = {\mathcal H}_t \vee {\mathcal F}_t\;.\nonumber
\eeq

A key quantity for tackling default is the Az\'ema $(\Q,\FF)$-supermartingale~(see \cite{Dell75}), defined as the projection of the survival indicator $H$ to the subfiltration $\FF$: 
\beq
S_t:=\E^B\left[\ind_{\{\tau>t\}}|\filF_t\right]=\Q\left[\tau>t|\filF_t\right]\;.\nonumber
\eeq

The financial interpretation of $S_t$ is a survival probability at $t$ given only observation of the default-free filtration $\FF$ up to $t$ and default monitoring $\HH$ for any name. Formally, the stochastic process $S$ is linked to the survival probability $G$ by the law of iterated expectations:
\beq
\E^B[S_t]=\E^B\left[\E^B\left[\ind_{\{\tau>t\}}\vert\filF_t\right]\right]=\E^B\left[\ind_{\{\tau>t\}}\right]=\Q[\tau>t]=G(t)\; .\label{eq:ESt}
\eeq

In many practical applications, the curve $G$ is given exogenously from market quotes (bond or credit default swaps). When this is the case, the above relationship puts constraints on the dynamics of $S$ so that the equality $\E^B[S_t]=G(t)$ is then referred to as the \textit{calibration equation}.\par

A very important result from stochastic calculus is the so-called Key Lemma (Lemma 3.1.3. in~\cite{Biel11}) that allows to get rid of the explicit default time $\tau$, focusing on the Az\'ema supermartingale instead. Applying this lemma to CVA yields the following equation, that holds whenever $V_\tau H_T$ is $\Q$-integrable and $V$ is $\FF$-predictable\footnote{In our CVA context, this second condition amounts to say that the portfolio $\Pi$ is not allowed to explicitly depend on $\tau$. For instance, it cannot contain corporate bonds whose reference entity is precisely the counterparty $C$}: 
\beq
\hbox{CVA}=(1-R)\E^B\left[\frac{V^+_\tau}{B_\tau}\ind_{\{\tau\leq T\}}\right]=-(1-R)\E^B\left[\int_{0}^T\frac{V^+_t}{B_t}dS_t\right]\label{eq:CVAGen}\;.
\eeq
The above result can be understood intuitively by localizing the default time in any possible small interval $(t,t+dt]$, for $t$ spanning the whole maturity horizon $[0,T]$. Defining $dH_t:=H_{t+dt}-H_t=\ind_{\{\tau \in (t,t+dt]\}}$ one gets
\beq
\E^B\left[\frac{V^+_\tau}{B_\tau}H_T\right] = \int_0^T \E^B\left[\frac{V^+_t}{B_t} dH_t \right]=\int_0^T \E^B\left[\frac{V^+_t}{B_t} \E^B\left[ dH_t | {\mathcal F}_t\right] \right]=-\int_0^T \E^B\left[\frac{V^+_t}{B_t} dS_t \right]\
\eeq
where we have used Fubini's theorem, the tower property and assumed that $V$ is $\FF$-adapted hence is independent from $\HH$.
%
%
\subsection{CVA in the Cox process setup}
%
An interesting specific case of Az\'ema supermartingales arises when $S$ is positive and decreasing from $S_0=1$ with zero quadratic variation.  This corresponds to the Cox setup: the process $S$ can be parametrized as
\beq
S_t=\e^{-\Lambda_t} \hbox{ where } \Lambda_t:=\int_0^t\lambda_u du\;,\nonumber
\eeq
where $\lambda:=(\lambda_t)_{t\geq 0}$ is a non-negative
, $\FF$-adapted stochastic intensity process. 

In this specific case, one can think of $S:=(S_t)_{t\geq 0}$ as a \textit{survival process} so that $\tau$ can be viewed as the (first) passage time of $S$ below a random threshold drawn from a standard uniform random variable, independent of $S$. Then, CVA (including WWR) reduces to
\beq
\boxed{\hbox{CVA}=-(1-R)\int_{0}^T\E^B\left[\frac{V^+_t}{B_t}\zeta_t\right]dG(t)\label{eq:CVAGen2}}
\eeq
where 
\beq
\zeta_t:=\frac{\lambda_t S_t}{h(t)G(t)}\;.\nonumber
\eeq
\begin{remark}
The process $\zeta$ represents the differential of the survival process ($S$) normalized with respect to the differential of its time-0 $\Q$-expectation ($G$):
\beq
\zeta_t = \left(\left.\frac{d \Q[\tau>t|\cF_s]}{dt}\right|_{s=t}\right)\Big/\left(\left.\frac{d \Q[\tau>t|\cF_s]}{dt}\right|_{s=0}\right)\;.\nonumber
\eeq
When $G$ is given exogenously from market quotes, the denominator can be considered as the prevailing market view of the default likelihood.
From that perspective, $\zeta$ is a kind of model-to-market survival rate change ratio. 
\end{remark}

In the above expression, 
$$\hbox{EPE}(t):=\E^B\left[\frac{V^+_t}{B_t}\zeta_t\right]$$ 
is the EPE under WWR : it is the deterministic profile to be integrated with respect to the survival probability curve $G$ to get the CVA (up to the constant $1-R$) including WWR, just like the EPE in the no-WWR case eq (\ref{eq:CVAind}). Moreover, from the calibration equation~(\ref{eq:ESt}),
\beq
\E^B\left[\lambda_tS_t\right]=-\E^B\left[\frac{d}{dt}S_t\right]=-\frac{d}{dt}\E^B\left[S_t\right]=-\frac{d}{dt}G(t):=-G'(t)=h(t)G(t)\nonumber
\eeq
so that $\zeta$ is a unit-$\Q$-expectation, non-negative stochastic process. In the case of independence between exposure ($V$) and risk-free rate ($r,B$) on the one hand, and credit risk ($\lambda,S$) on the other hand, the expected value in eq.~(\ref{eq:CVAGen2}) can be factorized and the equation collapses to eq.~(\ref{eq:CVAind})\footnote{Recall that eq.~(\ref{eq:CVAGen}) holds provided that the portfolio value process $V$ does not depend on the explicit default random variable. However, it may well depend on credit worthiness quantities embedded in $\mathbb{F}$, typically credit spreads $\lambda$. Consider for example the case where the default time $\tau$ is modeled as the first jump of a Cox process with a strictly positive intensity process. In that case, $\tau = \Lambda^{-1}_\xi$, with $\xi$ standard exponential independent from $\lambda$. Then, the portfolio value $V_t$ may depend on $\lambda$ up to $t$, but not on information on $\xi$.}:
\beq
\hbox{CVA}=-(1-R)\int_{0}^T\E^B\left[\frac{V^+_t}{B_t}\right]\E^B\left[\zeta_t\right]dG(t)=-(1-R)\int_{0}^T\hbox{EPE}^\perp(t)dG(t)=\hbox{CVA}^\perp\;.\nonumber
\eeq

Generally speaking however, the factorization of expectations
\beq
\E^B\left[\frac{V^+_t}{B_t}\zeta_t\right]=\E^B\left[\frac{V^+_t}{B_t}\right]\E^B\left[\zeta_t\right]\label{eq:EFact}
\eeq
is not valid. Because of WWR, we have to account for the potential statistical dependence  between market and credit risk. This is typically obtained by modeling $V,r$ and $\lambda$ using correlated risk factors. This can be achieved by modeling these processes with e.g. correlated Brownian motions. One can even think of making $\lambda$ a deterministic function of the exposure $V$, as in~\cite{Hull12}. This setup has the advantage to feature parameters that are more intuitive from a trading or risk-management perspective than an instantaneous correlation between latent risk-factors, but the calibration of the intensity process is more involved and depends on the specific portfolio composition. As the time-$t$ stochastic intensity $\lambda_t$ depends in a deterministic way on $V_t$, the survival process $S$ depends on the whole path of the exposure process up to $t$. In spite of this specificity, this approach fits in the stochastic intensity setup and hence fits in the class of methods covered here.
%
\section{The Wrong Way Measure (WWM)}
\label{sec:WWM}
%
In the general case where eq.~(\ref{eq:EFact}) does not hold, one needs to evaluate the left-hand expectation, which is much more involved than the right-hand side and is the main reason why such models are not used in practice. Nevertheless, noting that $\zeta$ is a non-negative unit-expectation process, comparing eq.~(\ref{eq:CVAind}) with eq.~(\ref{eq:CVAGen2}) suggests that the problem could be addressed using change of measure techniques. In this section, we derive a set of equivalent martingale measures allowing us to obtain such a factorization of expectations, even in presence of WWR. The main difference is the measure under which the expectations appearing on the right-hand side are taken. 
%
\subsection{Derivation of the EPE expression in the new measure}
%
We start by specifying a bit further the filtered probability space on which all stochastic processes are defined. The filtration $\FF$ is generated by a finite dimensional Brownian motion ${\bf W}$ driving exposure, rates and credit spreads. Filtration $\GG$ is thus the market filtration obtained by enlarging $\FF$ with the natural filtration of the default indicator $H$. Notice that in a Cox setup, all discounted assets that do not explicitly depend on $\tau$ (even if they depend on $\lambda$) are $\Q$-martingales under both filtrations.

With this setup at hand, we can define $C_s^t$ as the time-$s$ price of an asset protecting one unit of currency against default of the counterparty on the period $(t,t+dt]$, $t\geq s$. Using the Key lemma once again,
\beq
C_s^t:=\E^B\left[\frac{B_s}{B_t}\ind_{\{\tau\in (t,t+dt]\}}\Big\vert\filG_s\right]=\frac{\ind_{\{\tau>s\}}}{S_s}\underbrace{\E^B\left[\frac{B_s}{B_t}\lambda_tS_t\Big\vert\filF_s\right]}_{:=C^{\filF,t}_s}dt\; .\nonumber
\eeq

Because $B_s$ is $\filF_s$-measurable, $C^{\filF,t}_s=B_sM_s^t$ where
\beq
M_s^t:=\E^B\left[\frac{1}{B_t}\lambda_tS_t\Big\vert\filF_s\right]\; .\nonumber
\eeq

With this notation, $C^{\filF,t}_t=B_tM_t^t=\lambda_tS_t$. It is obvious to see that $C^{\filF,t}_s$ grows at the risk-free rate with respect to $s$ on $[0,t]$ ($t$ is fixed). Indeed, by the martingale representation theorem the positive martingale $M_s^t$ can be written on $[0,t]$ as
\beq
dM^t_s=M_s^t\pmb{\gamma}_s d{\bf W}_s\;.\nonumber
\eeq

Therefore $C^{\filF,t}_s$ can be seen as the price of a tradeable asset on $[0,t]$ computed under partial ($\FF$) information. Obviously, the corresponding expected rate of growth is equal to the risk-free rate under $\Q$:
\beq
dC^{\filF,t}_s= d(B_sM_s^t)= dB_s M_s^t + B_s dM_s^t = r_s B_s M_s^t ds + B_s dM_s^t  = r_s C^{\filF,t}_s ds + C^{\filF,t}_s \pmb{\gamma}_s d{\bf W}_s\nonumber
\eeq
or equivalently with Ito's lemma,
\beq
d\log C^{\filF,t}_s= \left(r_s -\frac{\pmb{\gamma}_s\pmb{\gamma}_s^T}{2}\right)  ds + \pmb{\gamma}_sd{\bf W}_s\;.\nonumber
\eeq

We can thus choose $C^{\filF,t}_s$ as num\'eraire for all $s,0\leq s\leq t$ and write
\beq
\E^B\left[\frac{V^+_t}{B_t}\lambda_tS_t\right]=\E^{C^{\filF,t}}\left[\frac{C^{\filF,t}_0}{C^{\filF,t}_t}\lambda_tS_t V^+_t\right]=C^{\filF,t}_0\E^{C^{\filF,t}}\left[ V^+_t\right]=\E^{C^{\filF,t}}\left[ V^+_t\right]\E^B\left[\frac{\lambda_tS_t}{ B_t}\right]\;,\nonumber
\eeq
or equivalently rescaling by $1/(h(t)G(t))$, 
\beq
\E^B\left[\frac{V^+_t}{B_t}\zeta_t\right]=\E^{C^{\filF,t}}\left[ V^+_t\right]\E^B\left[\frac{\zeta_t}{ B_t}\right]\;.\label{eq:EFact2}
\eeq

The probability measure associated to the expectation operator $\E^{C^{\cF,t}}$ is noted $\Q^{C^{\cF,t}}$ and is called the Wrong Way Measure (WWM or WW measure). This measure will be further specified from $\Q$ and the corresponding Radon-Nikod\`ym derivative process in Section~\ref{sec:RDDP}. A related measure based on a partial-information num\'eraire price had been introduced in Chapter 23 of \cite{Brigo06} to derive the CDS options market model.

Equation~(\ref{eq:EFact2}) is very similar to eq.~(\ref{eq:EFact}) except that (i) the RHS expectation of the positive exposure is taken under another measure than $\Q$ and (ii) the bank account num\'eraire $B$ does not appear in the first but in the second expectation, embedding credit risk. In contrast with (\ref{eq:EFact}), (\ref{eq:EFact2}) holds true whatever the actual dependency between all those risks. It yields another expression for the EPE under WWR:
\beq
\hbox{EPE}(t)=\E^{C^{\filF,t}}\left[ V^+_t\right]\E^B\left[\frac{\zeta_t}{ B_t}\right]\;.\label{eq:WWREPEasChangeMeasure}
\eeq
%
%
\subsection{EPE expression in the new measure under risk-free rate-credit independence}
%
It is very common to assume independence between risk-free rates and credit. Indeed, such a potential relationship has typically a very limited numerical impact (see e.g. \cite{Brigo05} for more details). With this additional assumption one gets $\E^B\left[\frac{\lambda_tS_t}{ B_t}\right]=- P^r(0,t)G'(t)$ where $P^r(s,t)$ is the time-$s$ price of a risk-free zero-coupon bond expiring at $t$, i.e.
\beq
P^r(s,t):=\E^B\left[\e^{-\int_s^tr_u du}\Big\vert\filF_s\right]\;.\nonumber
\eeq

Hence, under independence between counterparty's credit risk and the bank account num\'eraire, CVA finally reads
\beq
\boxed{\hbox{CVA}=-(1-R)\int_0^T\E^{C^{\filF,t}}\left[ V^+_t\right]P^r(0,t)dG(t)\; .\label{eq:CVAGenind}}
\eeq

The above expression looks very similar to the standard CVA expression that is revisited assuming independence between all risk factors, namely
\beq
\boxed{\hbox{CVA}^\perp=-(1-R)\int_{0}^T\E^B\left[V^+_t\right]P^r(0,t)dG(t)\; .\label{eq:CVAindind}}
\eeq

Hence, the general CVA formula~(\ref{eq:CVAGenind}) (including WWR but with the mild independence assumption between risk-free rates and counterparty credit) takes a similar form as the simple standard CVA~(\ref{eq:CVAindind})  (i.e. without WWR) where, in addition, risk-free rates are deterministic. 

In this case, the general CVA expression is given by the independent CVA expression, \textit{but} replacing $\hbox{EPE}^\perp(t)=\E^B[V^+_t]$ in eq.~(\ref{eq:CVAindind}) by $\hbox{EPE}(t)=\E^{C^{\cF,t}}\left[V^+_t\right]$. This observation suggests that changing the measure may indeed help dealing with WWR. 
%
\subsection{Radon-Nikod\`ym derivative process}
\label{sec:RDDP}
%
The num\'eraire $C^{\cF,t}=(C^{\cF,t}_s)_{0\leq s\leq t}$ is the num\'eraire associated with the WWM $\Q^{C^{\cF,t}}$.  The corresponding Radon-Nikod\`ym derivative process $Z^t$ is a $\Q$-martingale on $[0,t]$ :
\beq
Z_s^t:=\left.\frac{d\Q^{C^{\cF,t}}}{d\Q}\right|_{\cF_s}=\frac{C_s^{\cF,t}B_0}{C_0^{\cF,t}B_s}=\frac{M_s^t}{M_0^t}=\frac{\E^B\left[\frac{\lambda_tS_t}{B_t}|\cF_s\right]}{\E^B\left[\frac{\lambda_tS_t}{B_t}\right]}\label{eq:RND-Z0}\;.
\eeq

In the case of independence between rates and credit, $Z^t$ simplifies to
\beq
Z_s^t=\frac{P^{r}(s,t)\E^B\left[\zeta_t|\cF_s\right]}{B_sP^r(0,t)}\;.\nonumber\label{eq:RND-Z}
\eeq

In order for the CVA formula~(\ref{eq:CVAGenind}) to be useful in practice, we need to compute $\E^{C^{\filF,t}}\left[ V^+_t\right]$ that is, to derive the exposure dynamics under this new measure. This is the purpose of the next section. 
%
\section{Exposure's drift adjustment}
\label{sec:EDA}
%
In the sequel we restrict ourselves to the case where portfolio and credit risk are driven by a specific one-dimensional $\Q$-Brownian motion, namely $W^V$ and $W^\lambda$, respectively. These two processes can be correlated but $W^\lambda$ is independent from the short-rate drivers. These assumptions can be relaxed but help clarifying the point we want to make, which is to show how one can get rid of the link between exposure and credit by changing the pricing measure. In this setup we assume that these Brownian motions and the short-rate drivers actually generate the filtration $\FF$. 

We postulate continuous dynamics for $V$ under $\Q$,
\beq
dV_s=\alpha_s ds+\beta_s dW^V_s\label{eq:ExpDyn}
\eeq
where we assume the processes $\alpha$ and $\beta$ to be continuous and $\FF$-adapted, and derive the dynamics of $V$ under $\Q^{C^{\filF,t}}$. It is known from Girsanov's theorem that
\beq
dW^V_s=d\tilde{W}^V_s+ d\langle W^V,\log C^{\filF,t}\rangle_s\;,\nonumber
\eeq
where $\tilde{W}^V_s$ is a $(\Q^{C^{\filF,t}},\mathbb{F})$-Brownian motion. In other words, the dynamics of $V$ under $\Q^{C^{\filF,t}}$ is given by
\beq
dV_s=\left(\alpha_s + \theta_s^t\right) ds+\beta_s d\tilde{W}^V_s\;,\nonumber
\eeq
with $\theta_s^t$ is a stochastic process known as \textit{drift adjustment}. Standard results from stochastic calculus (see e.g. the change-of-num\'eraire toolkit presented in~\cite[Ch. II]{Brigo06}) yields the general form of this drift adjustment :
\beq
\theta_s^t dt = \beta_s d\langle W^V,\log C^{\filF,t}\rangle_s\; .\nonumber
\eeq

Evaluating this cross-variation requires to further specify the risk-neutral dynamics of the num\'eraire $C^{\filF,t}$ and in particular, the $\Q$-dynamics for the default intensity $\lambda$. Again, we adopt a quite general framework:
\beq
d\lambda_s = \mu^\lambda_s ds + \sigma^\lambda_s dW^\lambda_s\; ,\nonumber
\eeq
where $\mu^\lambda_s$ and $\sigma^\lambda_s$ are continuous $\FF$-adapted stochastic processes. As $\lambda$ is assumed independent from $r$, the new num\'eraire takes the form
\beq
C^{\filF,t}_s=\E^B\left[\frac{B_s}{B_t}\Big\vert\filF_s\right]\E^B\left[\lambda_tS_t\Big\vert\filF_s\right]=P^r(s,t)\E^B\left[\lambda_tS_t\Big\vert\filF_s\right]=-P^r(s,t)S_s\frac{\partial P^\lambda(s,t)}{\partial t}\label{eq:NewNum}
\eeq
where
\beqn
P^\lambda(s,t)&:=&\E^B\left[\e^{-\int_s^t \lambda_u du}\Big\vert\filF_s\right]\; .\nonumber
\eeqn

In order to proceed, we must further specify the form of $P^\lambda(s,t)$. 
%
\subsection{Affine intensity and short-rate processes}
%
In this section we derive the specific form of $\theta_s^t$ in the standard case where both risk-free rate $r$ and stochastic intensity $\lambda$ follow independent affine stochastic intensity processes, the independence assumption being justified for example in~\cite{Brigo05,Brigo06}:
\beqn
P^r(s,t)&=&A^r(s,t)\e^{-B^r(s,t)r_s}\nonumber\;,\\
P^\lambda(s,t)&=&A^\lambda(s,t)\e^{-B^\lambda(s,t)\lambda_s}\;.\nonumber
\eeqn

This in turn implies that for $x\in\{r,\lambda\}$, see for example \cite{Bjork04},
$$dx_s=\mu^x_s ds + \sigma^x_s dW^r_s$$
where $d\langle W^r, W^\lambda\rangle_t\equiv 0$ and the drift and diffusion coefficient of both processes take the specific form
\beqn
\mu^x_s&=&\mu^x_s(x_s)=a(s) + b(s) x_s\nonumber\\
\sigma^x_s&=&\sigma^x_s(x_s)=\sqrt{c(s) + d(s) x_s}\nonumber
\eeqn
for some deterministic functions $a,b,c,d$.\par

Since $P^\lambda(s,t)>0$ for all $0\leq s\leq t$, one obviously gets $A^\lambda(s,t)>0$ so we can write
\beq
P_t^\lambda(s,t):=\frac{\partial P^\lambda(s,t)}{\partial t}=\frac{\partial A^\lambda(s,t)}{\partial t}\e^{-B^\lambda(s,t)\lambda_s}-\frac{\partial B^\lambda(s,t)}{\partial t}\lambda_sP^\lambda(s,t)=:P^\lambda(s,t)\left(\frac{A^\lambda_t(s,t)}{A^\lambda(s,t)}-B^\lambda_t(s,t)\lambda_s\right)\; .\nonumber
\eeq 

Observe that the functions $A^x,B^x$ satisfy, for all $u\geq 0$ and $x\in\{r,\lambda\}$:
$$A^x(u,u)=B^x_t(u,u)=1\hbox{ and }B^x(u,u)=A^x_t(u,u)=0\;.$$

Plugging this expression $P_t^\lambda(s,t)$ in~(\ref{eq:NewNum}), one obtains
\beq
C^{\filF,t}_s=-S_sP^r(s,t)P^\lambda(s,t)\left(\frac{A^\lambda_t(s,t)}{A^\lambda(s,t)}-B^\lambda_t(s,t)\lambda_s\right)\nonumber
\eeq
and Ito's lemma yields the dynamics of the log-num\'eraire, valid for $s\in [0,t]$
\beq
d\log C^{\filF,t}_s= -\lambda_s ds + d\log P^r(s,t) + d\log P^\lambda(s,t) + d\log  \left(B^\lambda_t(s,t)\lambda_s-\frac{A^\lambda_t(s,t)}{A^\lambda(s,t)}\right)\; . \nonumber
\eeq

From the affine structure of $r$ and $\lambda$, the above equation becomes
\beqn
d\log C^{\filF,t}_s&=& (\ldots)ds - B^\lambda(s,t)d\lambda_s - B^r(s,t)dr_s + \frac{1}{B^\lambda_t(s,t)\lambda_s-\frac{A^\lambda_t(s,t)}{A^\lambda(s,t)}} d\left(B^\lambda_t(s,t)\lambda_s-\frac{A^\lambda_t(s,t)}{A^\lambda(s,t)}\right)\nonumber\\
&=& (\ldots)ds +\left(\frac{A^\lambda(s,t)B^\lambda_t(s,t)}{A^\lambda(s,t)B^\lambda_t(s,t)\lambda_s-A^\lambda_t(s,t)}-B^\lambda(s,t)\right) \sigma^\lambda_s dW^\lambda_s-B^r(s,t)\sigma^r_sdW^r_s\; .\label{eq:dlogC}
\eeqn

Finally, one gets the following relationship for the drift adjustment:
\beq
\boxed{\theta_s^t= \rho^\lambda_s\beta_s\sigma^\lambda_s\left(\frac{A^\lambda(s,t)B^\lambda_t(s,t)}{A^\lambda(s,t)B^\lambda_t(s,t)\lambda_s-A^\lambda_t(s,t)}-B^\lambda(s,t)\right)-\rho^r_s\beta_s\sigma^r_sB^r(s,t)\label{eq:DriftAdjTheta}} 
\eeq
where $\rho^\lambda_s$ represents the instantaneous correlation between the Brownian motions driving the exposure and credit risk, $\rho^\lambda_s ds:=d\langle W^V,W^\lambda\rangle_s$, and similarly $\rho^r_s ds:=d\langle W^V,W^r\rangle_s$.
%
\subsection{Deterministic approximation of the drift adjustment}
%
Our main point in this paper is to investigate the WWR impact. Therefore, we focus on deterministic risk-free rates and deterministic correlation $\rho^\lambda_s=\rho(s)$ in the sequel. This helps simplifying the framework since then $\log P^r(s,t)$ contributes zero to the quadratic variation of $\log C^{\filF,t}_s$. In such a framework, the drift adjustment simplifies to
\beq
\boxed{\theta_s^t= \rho(s)\beta_s\sigma^\lambda_s\left(\frac{A^\lambda(s,t)B^\lambda_t(s,t)}{A^\lambda(s,t)B^\lambda_t(s,t)\lambda_s-A^\lambda_t(s,t)}-B^\lambda(s,t)\right)\;.\label{eq:DriftAdjTheta}}
\eeq
\begin{remark}
The Radon-Nikod\`ym derivative process $Z^t$ derived in Section~\ref{sec:RDDP} is given as a conditional expectation of $\zeta$ rescaled by risk-free zero-coupon bond prices and the bank account num\'eraire. It is possible however to further specify the form of $Z^t$ as a function of the drift adjustment $\theta^t$ in the particular framework considered in this section. 

It is clear from eq.~(\ref{eq:RND-Z0}) that $Z_s^t=\frac{B_0}{C_0^{\cF,t}}\frac{C_s^{\cF,t}}{B_s}$ where $\frac{B_0}{C_0^{\cF,t}}=1$ and $\frac{C_s^{\cF,t}}{B_s}$ is a non-negative martingale. In the case of deterministic interest rates, the analytical expression of $C_s^{\cF,t}$ is easily obtained from the dynamics of $\log C^{\cF,t}$ given in~(\ref{eq:dlogC}) so that finally

\beq
Z_s^t=\exp\left\{\int_0^s \tilde{\theta}_u^t dW^\lambda_u-\frac{1}{2}\int_0^s \left(\tilde{\theta}_u^t\right)^2 du\right\}\nonumber
\eeq

where $\tilde{\theta}_s^t\rho(s)\beta_s:=\theta_s^t$.
\end{remark}

As the adjustment in the drift exposure features the stochastic intensity, $\theta^t=(\theta_s^t)_{0\leq s\leq t}$ is stochastic, and we cannot simplify the problem by avoiding to simulate the driver $W^\lambda$ of the intensity process. 

In order to reduce the dimensionality of the problem, one can look for deterministic approximations $\theta(s,t)$ of $\theta_s^t$. We introduce below two easy alternatives.
\subsubsection{Replace $\lambda_s$ by its expected value $\bar{\lambda}(s)$.}
%
A first method consists in replacing $\lambda_s$ by its expected value under $\mathbb{Q}$: $\bar{\lambda}(s):=\E^B[\lambda_s]$ in eq.~(\ref{eq:DriftAdjTheta}):
\beq
\theta(s,t)= \rho(s)\beta_s\sigma^\lambda_s(\boxed{\bar{\lambda}(s)})\left(\frac{A^\lambda(s,t)B^\lambda_t(s,t)}{A^\lambda(s,t)B^\lambda_t(s,t)\boxed{\bar{\lambda}(s)}-A^\lambda_t(s,t)}-B^\lambda(s,t)\right)\; ,\label{eq:DraftAdjlb}
\eeq
where we have used the notation $\sigma^\lambda_s(\lambda_s):=\sigma^\lambda_s$.
\subsubsection{Replace $\lambda_s$ by the implied hazard rate $h(s)$.}
%
A second method consists in replacing $\lambda_s$ by $h(s)$ in eq.~(\ref{eq:DriftAdjTheta}):
\beq
\theta(s,t)= \rho(s)\beta_s\sigma^\lambda_s(\boxed{h(s)})\left(\frac{A^\lambda(s,t)B^\lambda_t(s,t)}{A^\lambda(s,t)B^\lambda_t(s,t)\boxed{h(s)}-A^\lambda_t(s,t)}-B^\lambda(s,t)\right)\; .\label{eq:DraftAdjh}
\eeq

Recall that $h(t)$ is the hazard rate implied by the survival probability $G(t)=P^\lambda(0,t)$. 

\begin{remark}As calibration to market data forces the equality $\E^B\left[e^{-\int_0^t\lambda_s ds}\right]=e^{-\int_0^t h(s) ds}$, both methods are equivalent up to Jensen's effect:

$$e^{-\int_0^t h(s) ds}=\E^B\left[e^{-\int_0^t\lambda_s ds}\right]\approx e^{-\int_0^t\E^B[\lambda_s] ds}=e^{-\int_0^t\bar{\lambda}(s) ds}\;.$$

Another way to see the connections between the two approaches is to notice that $\bar{\lambda}(t)$ coincides with $h(t)$ as long as one can neglect covariance between $\lambda$ and its integrated version $\Lambda$:

\beqn
h(t)=-\frac{d}{dt}\ln G(t)=\frac{-G'(t)}{G(t)} =\frac{\E^B\left[\lambda_tS_t\right]}{\E^B\left[S_t\right]}=\bar{\lambda}_t+\frac{\cov^B\left[\lambda_t,S_t\right]}{\E^B\left[S_t\right]}\approx\bar{\lambda}(t)\;.\nonumber
\eeqn
\end{remark}
%
\subsection{Calibration and deterministically shifted affine processes}
\label{sec:subseccal}
%
The class of affine processes is quite broad; Ornstein-Uhlenbeck (OU), Cox-Ingersoll-Ross (CIR) including the case with jumps (JCIR, see \cite{Brigo10} for related calculations for CDS and CDS options) all fit in this class. Unfortunately, both only have three degrees of freedom; not enough for the calibration equation~(\ref{eq:ESt}) to hold in general. This can be circumvented by considering homogeneous affine processes that are shifted in a deterministic way. In this setup, $\lambda$ becomes a shifted version of a (latent) homogeneous affine processes $y$:
$$\lambda_t=y_t+\psi(t)\;.$$

The deterministic shift function $\psi$ is chosen to ensure that the model-implied function $P^\lambda(0,t)$ agrees with that of a given survival probability curve $G(t)$ provided externally. This is exactly eq~(\ref{eq:ESt}):
$$G(t)=P^\lambda(0,t)=\E^B\left[e^{-\int_0^t \lambda_u du}\right]=\E^B\left[e^{-\int_0^t y_u du}\right]e^{-\int_0^t \psi(u) du}=P^y(0,t)e^{-\int_0^t \psi(u) du}\; .$$

Setting $\Psi(t):=\int_0^t\psi(u)du$, $\Psi(s,t):=\Psi(t)-\Psi(s)$ and using the affine property of $y$,
$$P^\lambda(s,t)=A^y(s,t)e^{-B^y(s,t)y_s}e^{-(\Psi(t)-\Psi(s))}=A^y(s,t)e^{B^y(s,t)\psi(s)-\Psi(s,t)}e^{-B^y(s,t)\lambda_s}\; .$$

Hence, the shifted process $\lambda$ is affine too, with 
\beqn
A^\lambda(s,t)&=&A^y(s,t)e^{B^y(s,t)\psi(s)-\Psi(s,t)}\; ,\nonumber\\
B^\lambda(s,t)&=&B^y(s,t)\; .\nonumber
\eeqn

These $A^\lambda,B^\lambda$ are the $A,B$ functions to be used in the drift adjustment~(\ref{eq:DraftAdjh}). In particular,
\beqn
A^\lambda_t(s,t)&=&A^y_t(s,t)e^{B^y(s,t)\psi(s)-\Psi(s,t)}+A^\lambda(s,t)\left(B^y_t(s,t)\psi(s)-\psi(t)\right)\nonumber\\
&=&A^\lambda(s,t)\left(\frac{A^y_t(s,t)}{A^y(s,t)}+B^y_t(s,t)\psi(s)-\psi(t)\right)\; ,\nonumber\\
B^\lambda_t(s,t)&=&B^y_t(s,t)\; .\nonumber
\eeqn

Hence, the functions $A^\lambda$ and $B^\lambda$ as well as their derivatives can be easily computed from the functions $A^y,B^y$ of the underlying process $y$ and the survival probability $G$. For the sake of completeness we give the explicit formulae in the appendix for $y$ being OU, CIR and JCIR. The shifted versions $\lambda_t=y_t+\psi(t)$ are called Hull-White, CIR$^{++}$ and JCIR$^{++}$, respectively.
%
\section{Numerical experiments}
\label{sec:numexp}
%
In this section we compare the WW measure approach with deterministic approximation of the drift adjustment to the standard Monte-Carlo setup featuring a 2D Euler scheme of the bivariate SDE driving the exposure and intensity. We assume various Gaussian exposures and CIR$^{++}$ stochastic intensity and disregard the impact of discounting to put the focus and the treatment of the credit-exposure dependency.\footnote{Equivalently, the processes $V$ considered below can be seen as the stochastically discounted exposure $V/B$ and then set $r\equiv 0$ ($B\equiv 1$).} 
%
\subsection{Exposure processes, EPE and WWR-EPE}
%
Assuming Gaussian exposures has the key advantage of leading to analytical expressions of EPEs. For instance, let $\cN(\mu,\sigma)$ stands for the Normal distribution with mean $\mu$ and standard deviation $\sigma$. Then, assuming 
\beq
V_t\stackrel{(\Pr)}{\sim}\cN\left(\mu(t),\sigma(t)\right)\label{eq:dVGaussExp}
\eeq
for some deterministic functions $\mu(t)$, $\sigma(t)>0$ and probability measure $\Pr$,
\beq
\E^{\Pr}\left[V_t^+\right]=\sigma(t)\phi\left(\frac{\mu(t)}{\sigma(t)}\right)+\mu(t)\Phi\left(\frac{\mu(t)}{\sigma(t)}\right)\;,\label{eq:AnEPE}
\eeq
where $\phi$ is the standard Normal density and $\Phi$ the corresponding cumulative distribution function.

Hence, the $\Q$-EPE is analytically tractable if the exposure dynamics under $\Q$ features an affine drift and a deterministic diffusion coefficient i.e. when $\alpha_s=\tilde{\alpha}(s)+\alpha(s)V_s$ and $\beta_s=\beta(s)$ in~(\ref{eq:ExpDyn}). This includes the special cases where the exposure is modeled via an arithmetic Brownian motion, as in the Bachelier model, or as a mean-reverting Ornstein-Uhlenbeck process, as in Vasicek's models. In case the exact exposure dynamics is not Gaussian, one might consider the Gaussian assumption as an approximation, possibly obtained via moment-matching or drift-freezing techniques, see for example \cite{Rapisarda} for the lognormal case applied to basket options and \cite{Brigo13} Chapters 4.4 and 4.5 for swap portfolios.

Moreover, the profiles can take various forms, and can successfully depict the behavior of exposure profiles of equity return swaps or forward contracts (in the simple Brownian case), or exposure profiles of interest rates swap (IRS, if drifted Brownian bridges are used instead). 

We focus below on specific exposures and stochastic intensity processes that lead to analytical tractability.

%
\subsubsection{Forward-type Gaussian exposure}
%
We choose for the coefficients of exposure dynamics~(\ref{eq:ExpDyn}) $\alpha_s\equiv 0$ and $\beta_s\equiv\nu$ so that the exposure is a rescaled Brownian motion, implying
$$V_t\stackrel{(\Q)}{\sim} \cN\left(0,\nu\sqrt{t}\right)$$
Hence, the EPE collapses to the RHS of~(\ref{eq:AnEPE}) with $\mu(t)=0$ and $\sigma(t)=\nu\sqrt{t}$:
$$\hbox{EPE}^\perp(t)=\nu\sqrt{t}\phi(0)\;.$$

In order to compute the EPE under WWR for a given survival probability curve $G$, we consider an affine stochastic intensity process $\lambda$ and imply the drift function $\psi$ from the calibration equation $P^\lambda(0,t)=G(t)$ as in Section~\ref{sec:subseccal}:
\beq
\psi(t)=\frac{d}{dt}\ln\frac{P^y(0,t)}{G(t)}\nonumber\;.
\eeq

One can then easily estimate the EPE (and thus CVA) $\E^B[V^+_t\zeta_t]$ by jointly simulating $\lambda$ and $V$ in a standard Monte Carlo framework. In this specific case however, the change-of-measure technique proves to be very useful. Indeed, under the deterministic drift adjustment approximation or under drift freezing, $V_t$ is interestingly again normally distributed under $\Q^{C^{\filF,t}}$ 
with mean $\mu(t)=\Theta(t):=\int_0^t\theta(u,t)du$ and standard deviation $\sigma(t)=\nu\sqrt{t}$. Applying eq.~(\ref{eq:AnEPE}) yields
$$\hbox{EPE}(t)\approx\nu\sqrt{t}\phi\left(\frac{\Theta(t)}{\nu\sqrt{t}}\right)+\Theta(t)\Phi\left(\frac{\Theta(t)}{\nu\sqrt{t}}\right)\;.$$
\subsubsection{Swap-type, mean-reverting Gaussian exposure.}
%
In this section we adopt an exposure profile that mimics that of an interest rate swap in the sense that there is a pull-to-parity effect towards maturity. To that end, we use a drifted Brownian bridge. More explicitly, we follow~\cite[Th. 4.7.6]{Shrev04} and set the coefficients in~(\ref{eq:ExpDyn}) to $\alpha_s=\gamma(T-s)-\frac{V_s}{T-s}$ and $\sigma_s=\nu$ so that 
$$V_t=\gamma t(T-t) + \nu(T-t)\int_0^t \frac{1}{T-s} dW_s^V$$
leading to
$$V_t\stackrel{(\Q)}{\sim} \cN\left(\gamma t(T-t),\nu\sqrt{t(1-t/T)}\right)\; .$$

In this model, $\gamma$ governs the future expected moneyness of the swap implied by the forward curve and $\nu$ drives the volatility. Because the diffusion part is the same in both forward-type and swap-type SDEs of $V$, the drift adjustment process $\theta^t$ takes the same form in either cases. However, the marginal distributions of the WWR exposure change. We compute them below. 

Recall that the dynamics of a OU process with time-dependent coefficients takes the form
\beq
dX_s=\kappa(s)(\eta(s)-X_s)ds+\epsilon(s) dW_s\; .\label{eq:GenOU}
\eeq

The solution to this SDE is a Gaussian process whose solution can be easily found to be
$$X_t|X_s=G(s,t) (X_s + I(s,t)+ J(s,t))\;,$$
with
\beqn
G(s,t)&:=&\e^{-\int_{s}^t\kappa(u)du}\nonumber\\
I(s,t)&:=&\int_{s}^t \kappa(u)\eta(u)\e^{\int_{s}^u\kappa(v)dv}du\nonumber\\
J(s,t)&:=&\int_{s}^t \epsilon(u)\e^{\int_{s}^u\kappa(v)dv}dW_u\;.\nonumber
\eeqn

In particular, $X_t$ is Normally distributed with mean $m(t)$ and variance $v(t)$:
$$m(t)=G(s,t)\left(m(s)+I(s,t)\right)~\hbox{ and }~v(t)=G^2(s,t)\left(v(s)+\int_{s}^t \left(\frac{\epsilon(u)}{G(s,u)}\right)^2 du\right)$$

In the simplest case of constant volatility $\epsilon(s)=\epsilon$, $X_t$ is distributed as
$$X_t\stackrel{(\Q)}{\sim} \cN\left(e^{-\int_0^t \kappa(u)du}\int_0^t \kappa(u)\eta(u)e^{-\int_0^u \kappa(v)dv}du ,e^{-2\int_0^t \kappa(u)du}\int_0^t \epsilon^2e^{2\int_0^u \kappa(v)dv}du\right)\;.$$

Basic algebra shows that under $\Q^{C^{\filF,t}}$ the exposure $V_t$ takes the form~(\ref{eq:GenOU}) with
\beqn
\kappa(s)&=&(T-s)^{-1}\nonumber\\
\eta(s)&=&(T-s)(\gamma(T-s)+\theta_s^t)\nonumber\\
\epsilon&=&\nu\;.\nonumber
\eeqn

Unfortunately, $\eta(s)$ is not a deterministic function; it features the intensity process via $\theta_s^t$. However, it becomes deterministic if one replaces $\theta_s^t$ by its deterministic proxy $\theta(s,t)$:
$$\eta(s)\approx (T-s)(\gamma(T-s)+\theta(s,t))\; .$$ 

Under this approximation, the exposure becomes a generalized OU process and
$$V_t\stackrel{\left(\Q^{C^{\filF,t}}\right)}{\sim} \cN\left(\gamma t (T-t) + (t-T)\int_0^t \frac{\theta(u,t)}{u-T} du,\nu\sqrt{t(1-t/T)}\right)$$
providing a closed form expression for the EPE under WWR given by eq.~(\ref{eq:AnEPE}) with $\Pr=\Q^{C^{\cF,t}}$ and
\beqn
\mu(t)&=&\gamma t (T-t) + (t-T)\int_0^t \frac{\theta(u,t)}{u-T} du\;,\nonumber\\
\sigma(t)&=&\nu\sqrt{t(1-t/T)}\;.\nonumber
\eeqn

\begin{remark} Notice that some non-Gaussian exposures are analytically tractable too with the deterministic drift approximation. For instance, if one knows beforehand that the exposure will be positive, one can consider the coefficients in the exposure process dynamics~(\ref{eq:ExpDyn}) to be $\alpha_s=\alpha(s)V_s$ and $\beta_s=\beta(s)V_s$, with $\alpha(\cdot)$ and $\beta(\cdot)$ deterministic functions of time:
$$V_t=V_0\exp\left\{\int_0^t\left(\alpha(s)-\frac{\beta^2(s)}{2}\right)ds+\int_0^t\beta(s)dW^V_s\right\}\;,$$
leading to

$$\hbox{EPE}^\perp(t)=\E^B[V^+_t]=\E^B[V_t]=V_0e^{\int_0^t\alpha(s)ds}\;.$$
Using Girsanov's theorem, the solution $V$ can be written as a function of the $\Q^{C^{\cF,t}}$-Brownian motion
$$V_t=V_0\exp\left\{\int_0^t\left(\alpha(s)+\theta_s^t-\frac{\beta^2(s)}{2}\right)ds+\int_0^t\beta(s)d\tilde{W}^V_s\right\}\;,$$
where $\theta_s^t$ is given in~(\ref{eq:DriftAdjTheta}) after replacing $\beta_s$ by $\beta(s)$.

Using the deterministic approximation, one obtains
$$\hbox{EPE}(t)=\E^{C^{\cF,t}}\left[V^+_t\right]=\E^{C^{\cF,t}}\left[V_t\right]=V_0e^{\int_0^t\alpha(s)ds}\E^{C^{\cF,t}}\left[e^{\int_0^t\theta_s^t ds}\right]\approx V_0e^{\int_0^t(\alpha(s)+\theta(s,t))ds}=\hbox{EPE}^\perp(t)\e^{\Theta(t)}\;.$$
Note that the geometric Browian motion assumption for $V$ here can be seen as an approximation stemming from moment matching, see for example \cite{Rapisarda} for the case of geometric Brownian motion.
%
%
\end{remark}

%
\subsection{Discretization schemes}
%
The deterministic approximation to the stochastic drift-adjustment resulting from the change-of-measure has the appealing feature to avoid having to simulate the default intensity. In the case where the exposures are normally or lognormally distributed, this leads to semi-closed form expressions for CVA, where only two numerical integrations are required (one to compute $\Theta(t)=\int_0^t\theta(u,t)du$ and the other one to integrate the EPE profile with respect to the survival probability curve $G$ to get the final CVA).\par

This contrasts with the standard method that consists in the joint simulation of the exposure and default intensity. To do so, one must rely on 2D Monte Carlo scheme. In two dimensions, it is hard to avoid using small time steps as the joint distribution of $(V,\lambda)$ is unavailable when $\lambda$ is CIR and $V$ is Gaussian or lognormal, for example, under non-zero quadratic covariation (or ``correlation'') between the driving Brownian motions. Several schemes have been tested for simulating the CIR process for $\lambda$. Most of them are comparable when the Feller condition is satisfied and when volatility is small. However, it is known that simulating such processes is typically difficult otherwise. In particular, the performances deteriorate when the volatility is large.

The CIR schemes can be divided in two classes.\footnote{Notice that many schemes are available for CIR. We restrict ourselves to present two of them who exhibit decent performances in our CVA application and able to deal with the non-Feller case.} 
%


\subsubsection{Non-negative schemes}
%
A first class of schemes avoid negative samples whatever the time step $\delta$ and the volatility. This is the case for instance of the ``reflected scheme'' originally introduced in \cite{Diop03}:
\beq
y_{(i+1)\delta}=\left|y_{i\delta}+\kappa(\theta-y_{i\delta})\delta +\sigma\sqrt{\delta y_{i\delta}}z_i\right|\;,\label{eq:CIR:Reflex}
\eeq
where $z_i$ are iid standard Normal samples. 

\begin{figure}[h!]
\centering
\subfigure[$\delta=1\hbox{E}^{-2}$]{\includegraphics[width=0.99\columnwidth]{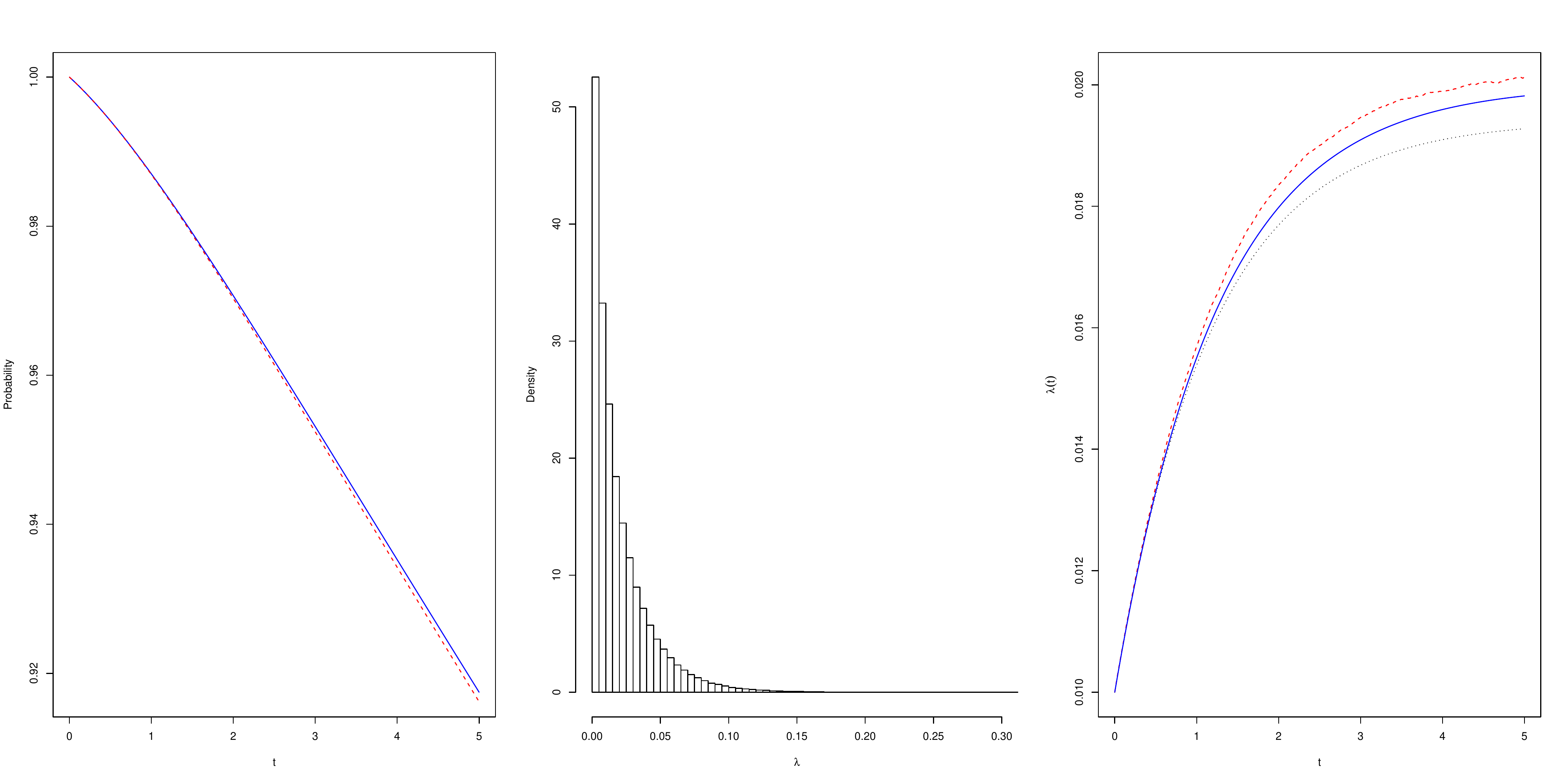}}\\
\subfigure[$\delta=1\hbox{E}^{-3}$]{\includegraphics[width=0.99\columnwidth]{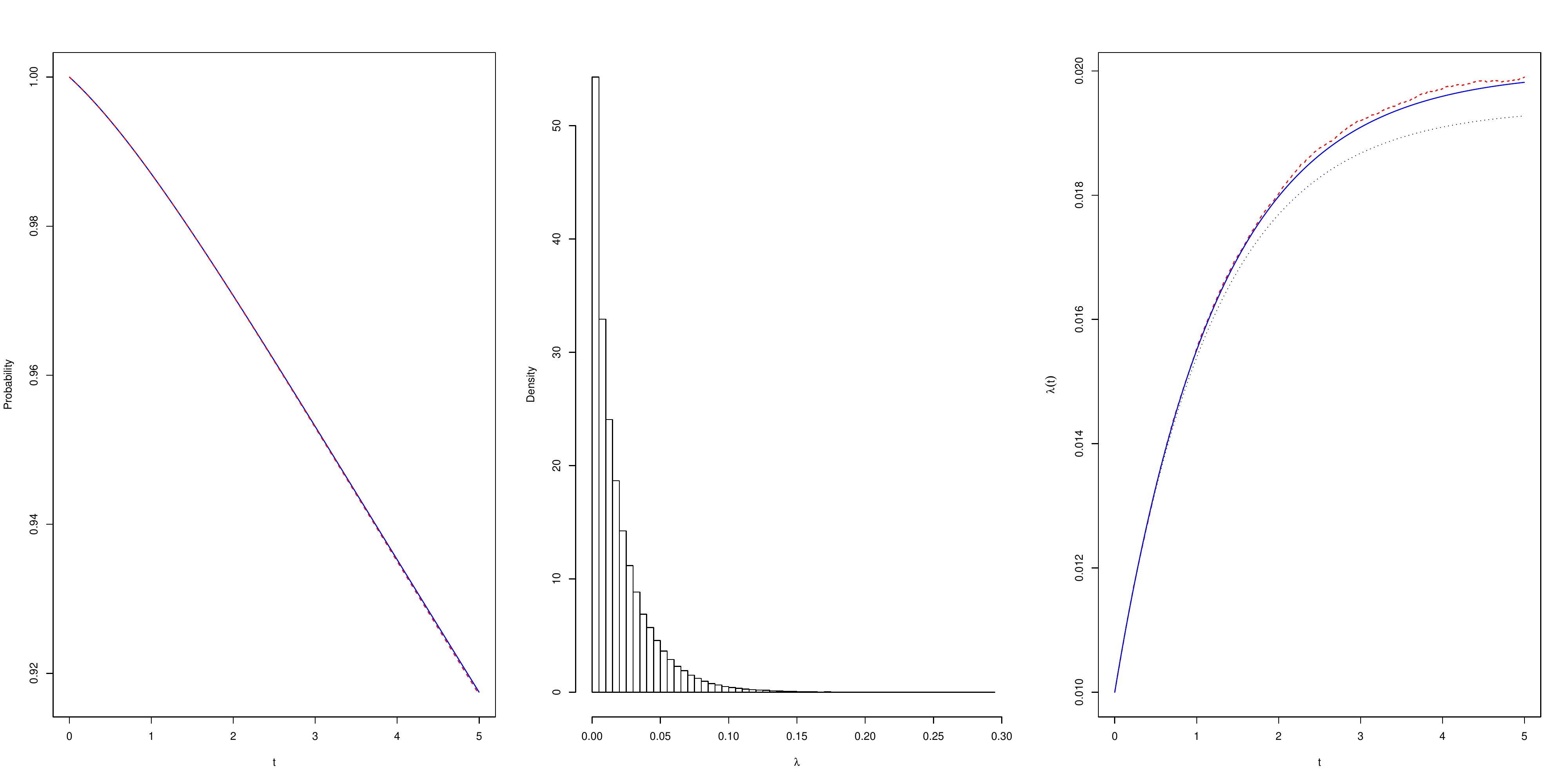}}
\caption{Statistics of samples generated via scheme (\ref{eq:CIR:Reflex}) based on 300k paths with time step $\delta$ for 5Y maturity with CIR parameters given by Set 2 in Table~\ref{tab:cirparams}. Survival probabilities (theoretical, blue solid and empirical dotted, red), histograms of $\lambda_t$ (middle) and proxies $\lambda(t)$ being either the expectation of $\lambda_t$ (theoretical, blue solid and empirical, red dotted) and $h(t)$ (black, dotted). One can see that all samples are non-negative (as expected) but the fit between theoretical and empirical survival probability curves is quite poor. }\label{fig:CIR-Scheme1}
\end{figure}

The positivity can also be imposed using implicit schemes, see e.g.~\cite{Alf05} and~\cite{Brigo05} for a discussion on the performances. The issue is that the convergence is rather disappointing : depending on the data, even a very small $\delta$ of $1\hbox{E}-5$ may not be enough in order for the empirical expectation $\hat{\E}^B[S_t]$ to fit reasonably well the theoretical expression $P^\lambda(0,t)$. This is of course a major obstacle as many (bivariate) sample paths have to be drawn, potentially for large maturities. In fact, even in this simple framework, using these positive schemes becomes quickly unmanageable on a standard computer because the time step required to ensure the above fit is too small. Fig.~\ref{fig:CIR-Scheme1} illustrates this. We have simulated $N=300k$ sample paths from~(\ref{eq:CIR:Reflex}). The left chart provides the sample mean of the Az\'ema supermartingale $\hat{\E}^B[S_t]:=N^{-1}\sum_{n=1}^N S_t(\omega_n)$ (dashed) with the theoretical expectation $P^\lambda(0,t)$ (solid). The middle plot exhibits the histogram of $\lambda$; one can check that obviously, there is no negative samples. Finally, the right plot provides a comparison of $\hat{\lambda}_t^n:=\hat{\E}^B[\lambda_t]:=\sum_{n=1}^N \lambda_t(\omega_n)$ (dashed), the analytical counterpart  $\bar{\lambda}_t=\E^B[\lambda_t]$ (solid) as well as $h(t)=-\left(\frac{d}{dt}\ln(P^\lambda(0,t))\right)/P^\lambda(0,t)$ (dotted). One can see by visual inspection, that the approximations $\hat{\E}^B[S_t]\approx P^\lambda(0,t)$ and $\hat{\E}^B[\lambda_t]\approx\bar{\lambda}_t$ are relatively poor for $\delta=1\hbox{E}^{-2}$ (top row). The bottom row provides the same plots but for $\delta=1\hbox{E}^{3}$. As expected, the fit improves when decreasing the time step, but the computation time explodes from 36 s ($\delta=1\hbox{E}^{-2}$, top) to  about 6 minutes ($\delta=1\hbox{E}^{-3}$, bottom) on a standard laptop computer.
\subsubsection{Relaxing the non-negativity constraint}
%
Alternatively, the scheme proposed by~\cite{Diop03} and discussed in~\cite{Lord10} seems to work well. It consists in the following discretization scheme
\beq
y_{(i+1)\delta}=y_{i\delta}+\kappa(\theta-y^+_{i\delta})\delta t+\sigma\sqrt{\delta y^+_{i\delta}}z_i\;.\label{eq:CIR:Lord}
\eeq

\begin{figure}[h!]
\centering
\subfigure[$\delta=1\hbox{E}^{-2}$]{\includegraphics[width=0.99\columnwidth]{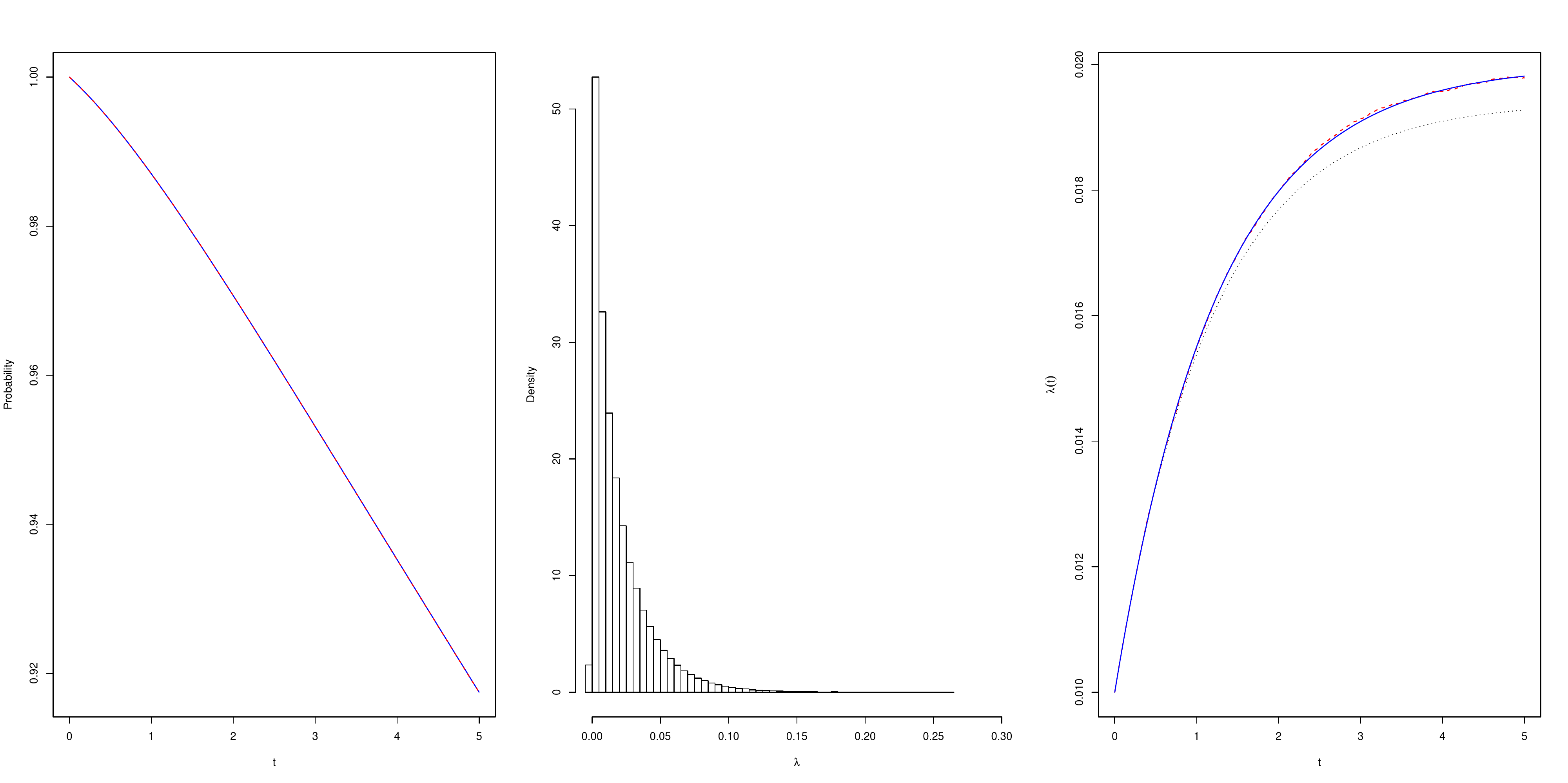}}\\
\subfigure[$\delta=1\hbox{E}^{-3}$]{\includegraphics[width=0.99\columnwidth]{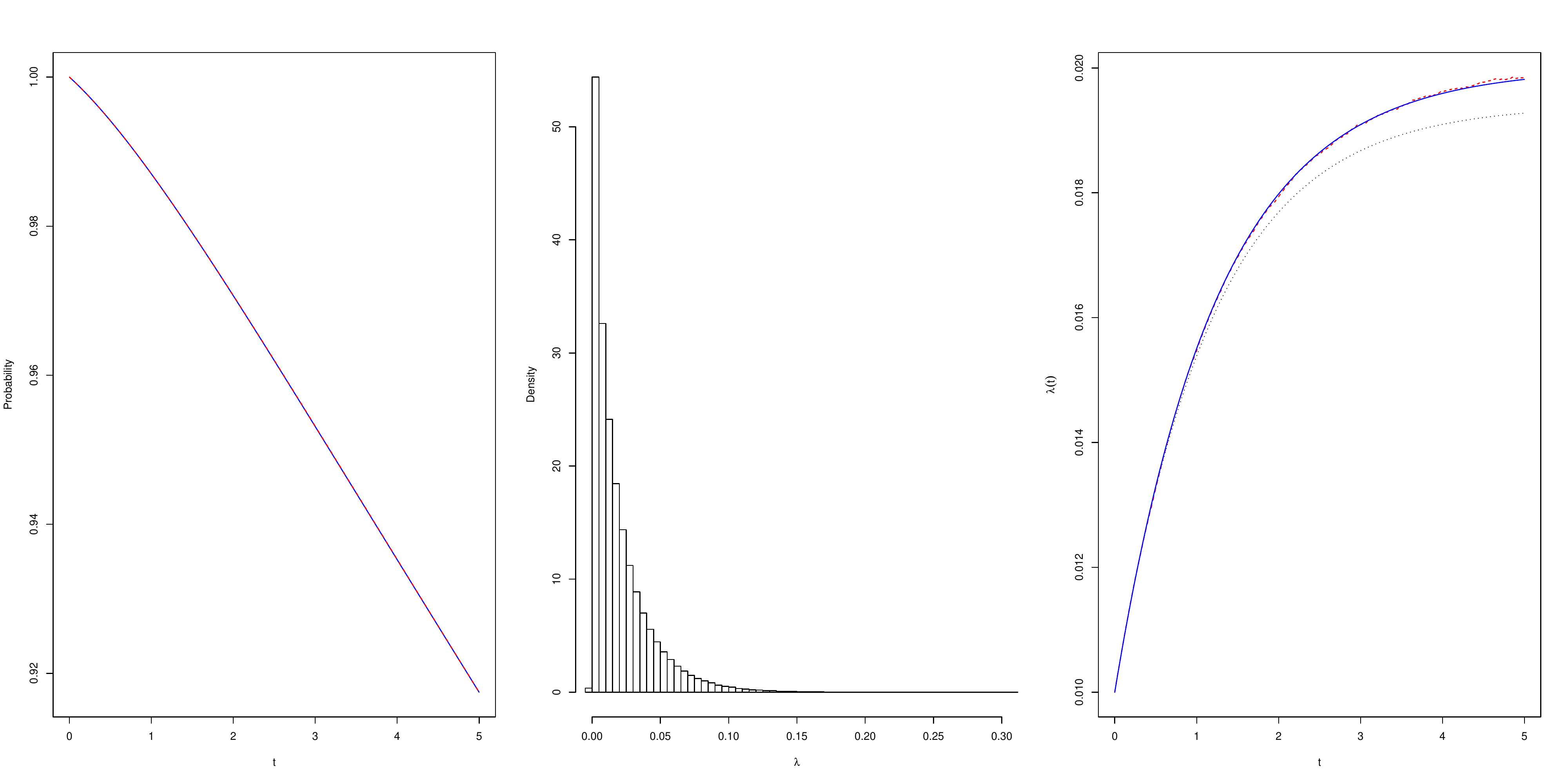}}
\caption{Statistics of samples generated via scheme (\ref{eq:CIR:Lord}) based on 300k paths with time step $\delta$ for 5Y maturity with CIR parameters given by Set 2 in Table~\ref{tab:cirparams}. Survival probabilities (theoretical, blue solid and empirical dotted, red), histograms of $\lambda_t$ (middle) and proxies $\lambda(t)$ being either the expectation of $\lambda_t$ (theoretical, blue solid and empirical, red dotted) and $h(t)$ (black, dotted). One can see by visual inspection that the fit between theoretical and empirical survival probability curves is quite good even for $\delta=1\hbox{E}^{-2}$.}\label{fig:CIR-Scheme2}
\end{figure}

As clearly visible from the histogram in Fig.~\ref{fig:CIR-Scheme2}, this scheme has the major drawback of not preventing negative samples for the intensity for a finite $\delta$ (especially when volatility is large). However, the fit between $\hat{\E}^B[S_t]$ and $P^\lambda(0,t)$ (and similarly $\bar{\lambda}_t\approx\hat{\lambda}_t^n$) is already pretty good when $\delta=1\hbox{E}^{-2}$. As expected, the proportion of negative samples decreases with decreasing time step but the computation time explodes (the computation times are comparable to those of the previous scheme).

The choice of the discretization scheme will be shown to have little impact on CVA figures  in Section~\ref{sec:CVADiscreteScheme}. Hence, one can opt indifferently for any standard scheme provided that it can deal with cases where Feller's condition is violated. This violation often happes in real market cases when the large credit volatility tends to push trajectories up and down in a way that is not compatible with Feller's condition. We choose the scheme~(\ref{eq:CIR:Lord}) which is rather standard in practice. 
\subsubsection{Discretization scheme for the JCIR}
%
The JCIR process can easily be simulated by adjusting any of the above method for sampling a CIR process for the jumps, path-by-path and period-by-period. Sample paths of the compound Poisson process are simulated independently, and at the end of each period featuring a jump, the corresponding CIR paths are adjusted by the associated jump size. Because of discretization errors, the scheme is of course satisfactory only for time step $\delta$ being small enough. It is worth mentioning that Giesecke and Smelov recently proposed in~\cite{Giese13} an exact scheme to sample jump diffusions: the standard error look comparable to a naive discretization but the computational time is cut by more than 75\% and more interestingly, the bias is killed. We rely on the standard discretization algorithm in this paper as our focus is precisely to propose a method allowing one to get rid of intensity simulation. 
%
\subsection{Wrong-way EPE profiles}
%
From the counterparty risk pricing point of view only CVA that is, the integral of the EPE profile with respect to the survival probability curve matters. Yet, it is interesting to first have a look at the EPE profiles under wrong-way risk, i.e. at $\hbox{EPE}(t)=\E^B\left[\frac{V_t^+}{B_t}\zeta_t\right]$ as deterministic functions of time. This helps getting an idea of how good the change-of-measure technique is (combined with the deterministic approximation of the drift adjustment) not at the aggregate level, but for the exposure at a specific time. This is important for analysts monitoring counterparty exposure, and more generally for risk-management purposes.
 
Therefore in this section we provide EPE profiles for specific parameter values of the exposure and stochastic intensity processes.
The parameter values are chosen such that specific EPE shapes are generated (e.g. exposure profiles being not a monotonic function of correlation). This proves particularly interesting as it allows us to analyze whether the drift-adjustment method is able to reproduce the subtleties of these profiles, like asymmetry and crossings for example.

\begin{landscape}
\begin{figure}
\centering
\subfigure[$T=5Y$, $h(t)=15\%$]{\includegraphics[width=0.28\columnwidth]{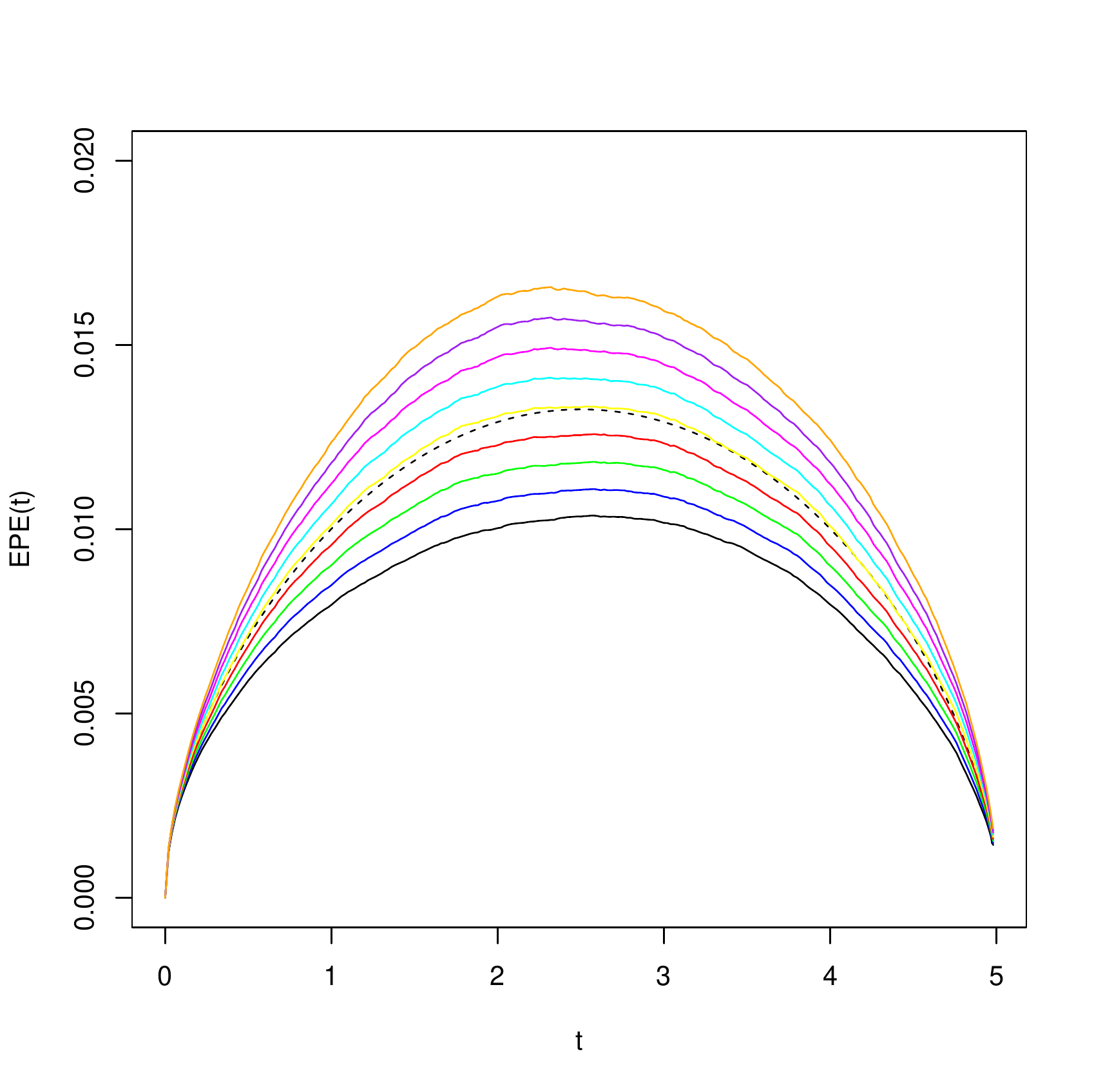}}\hspace{0.8cm}
\subfigure[$T=15Y$, $h(t)=15\%$]{\includegraphics[width=0.28\columnwidth]{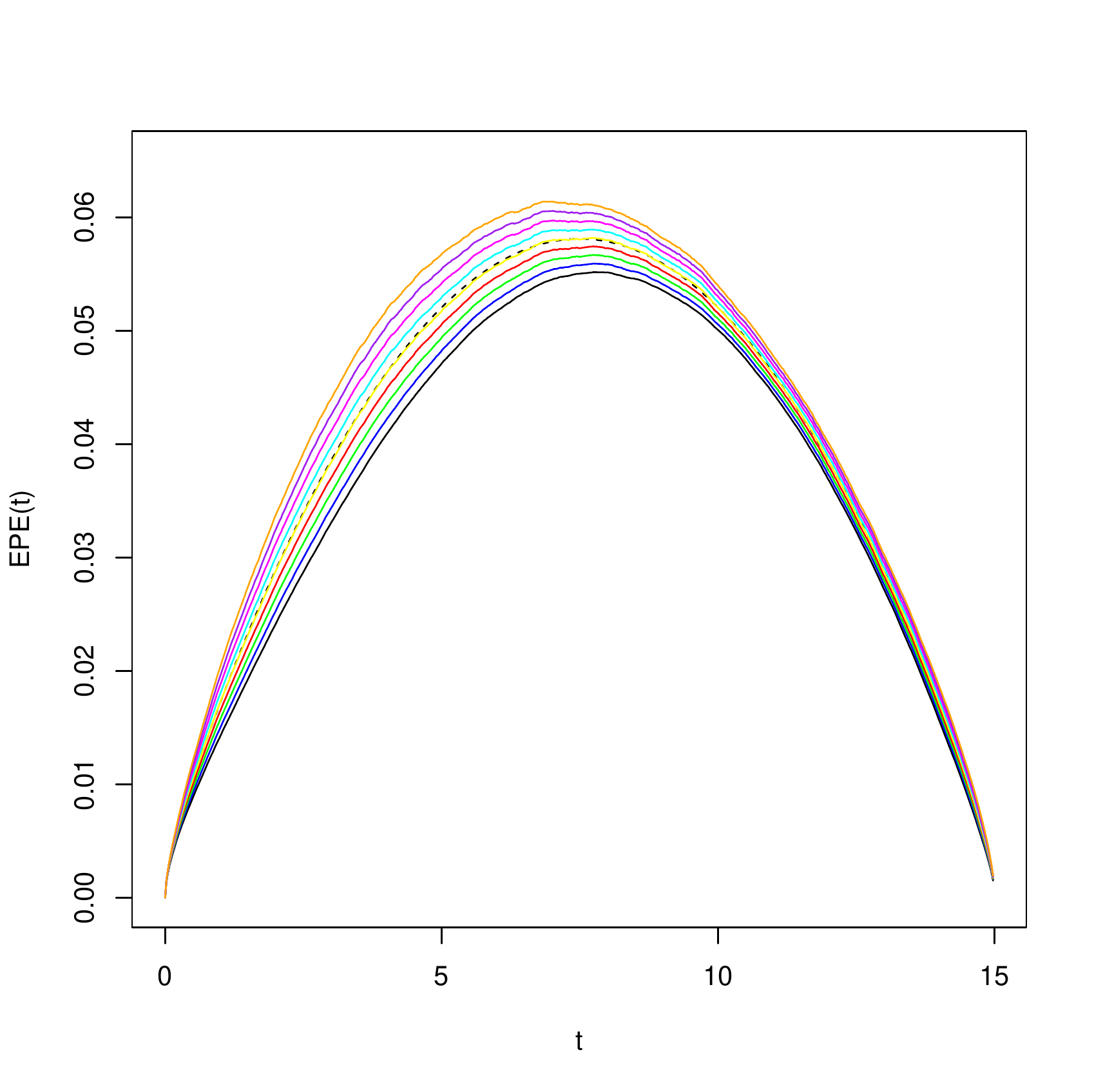}}\hspace{0.8cm}
\subfigure[$T=15Y$, $h(t)=30\%$]{\includegraphics[width=0.28\columnwidth]{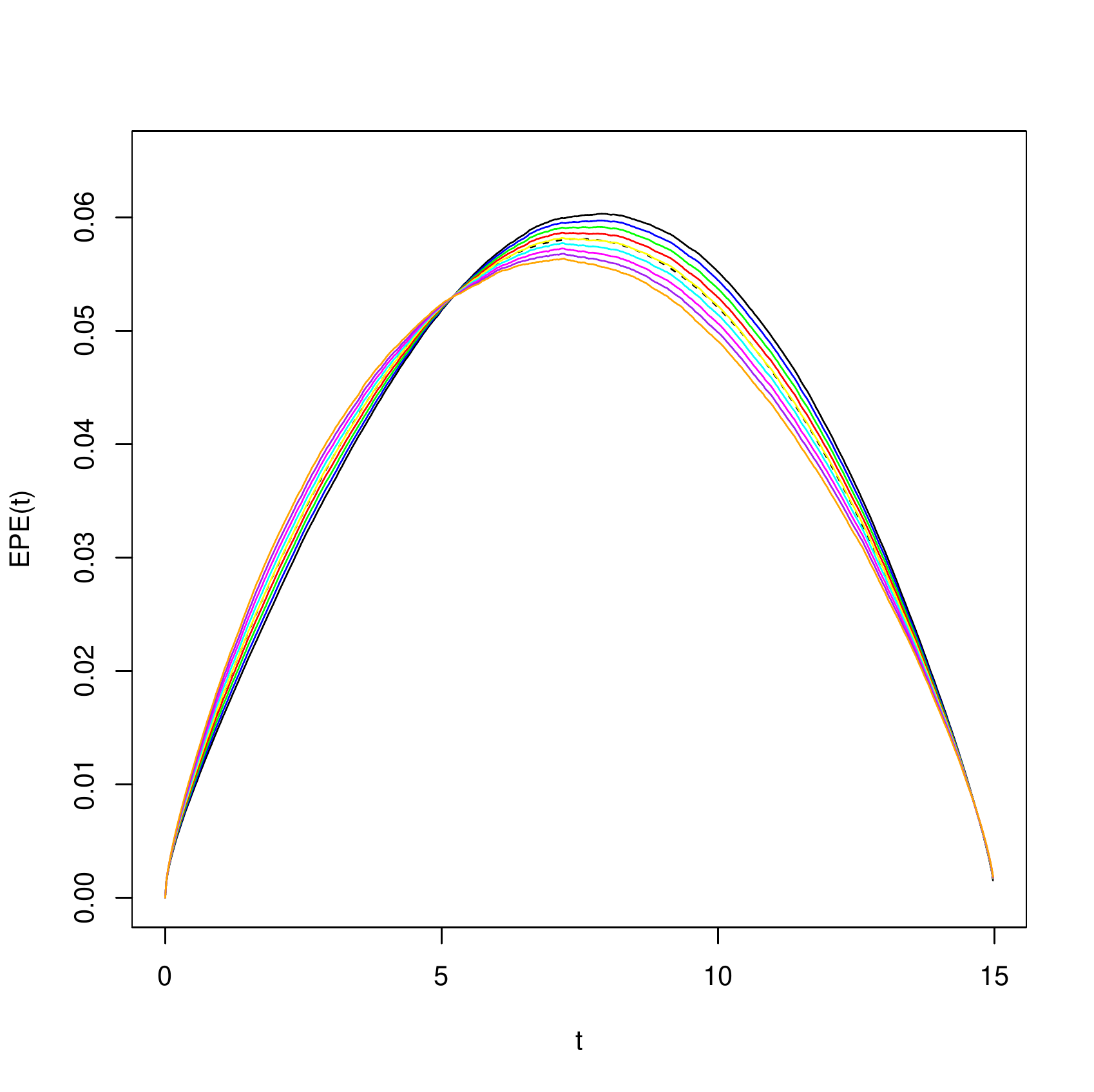}}\\
\subfigure[$T=5Y$, $h(t)=15\%$]{\includegraphics[width=0.28\columnwidth]{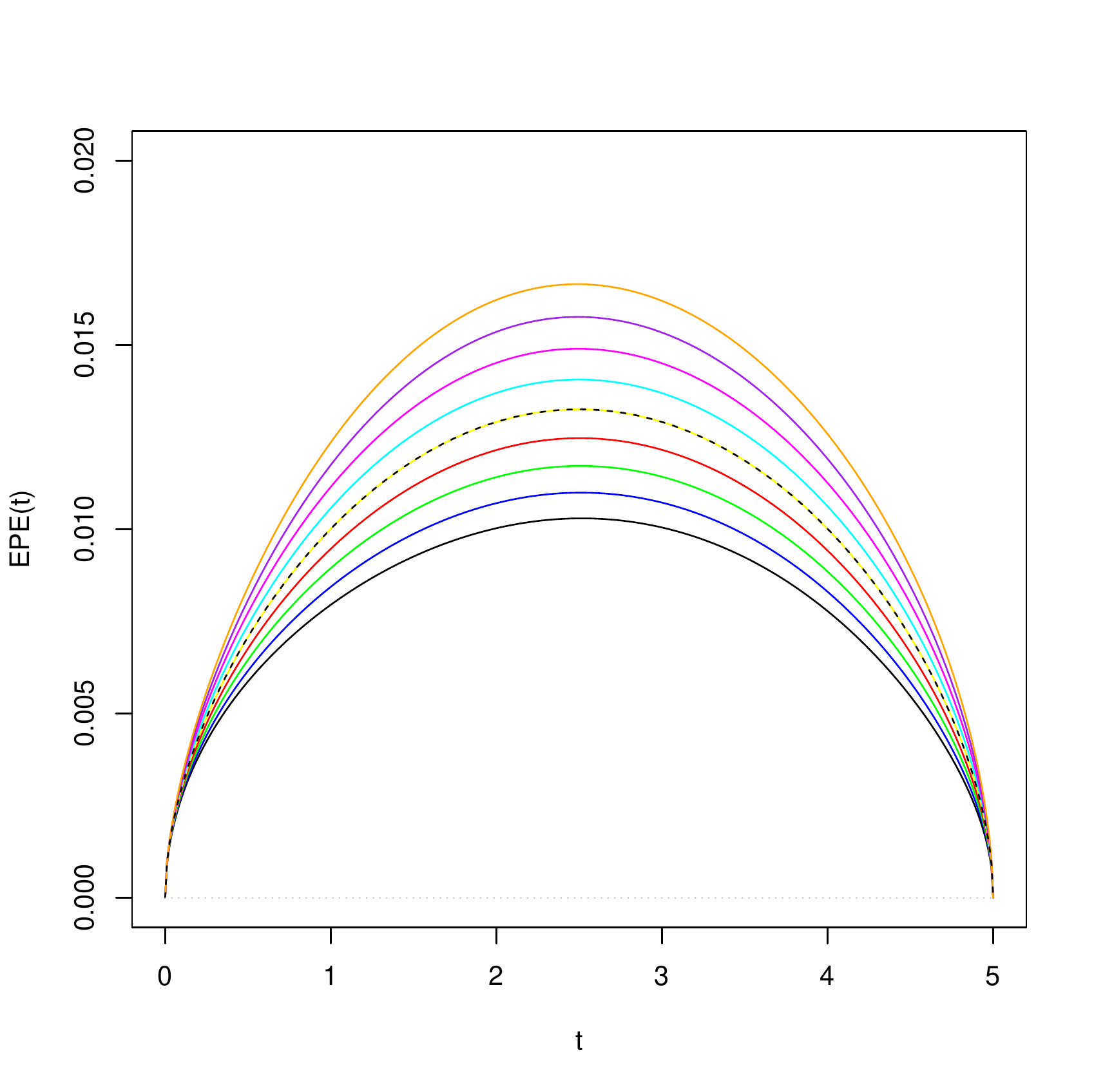}}\hspace{0.8cm}
\subfigure[$T=15Y$, $h(t)=15\%$]{\includegraphics[width=0.28\columnwidth]{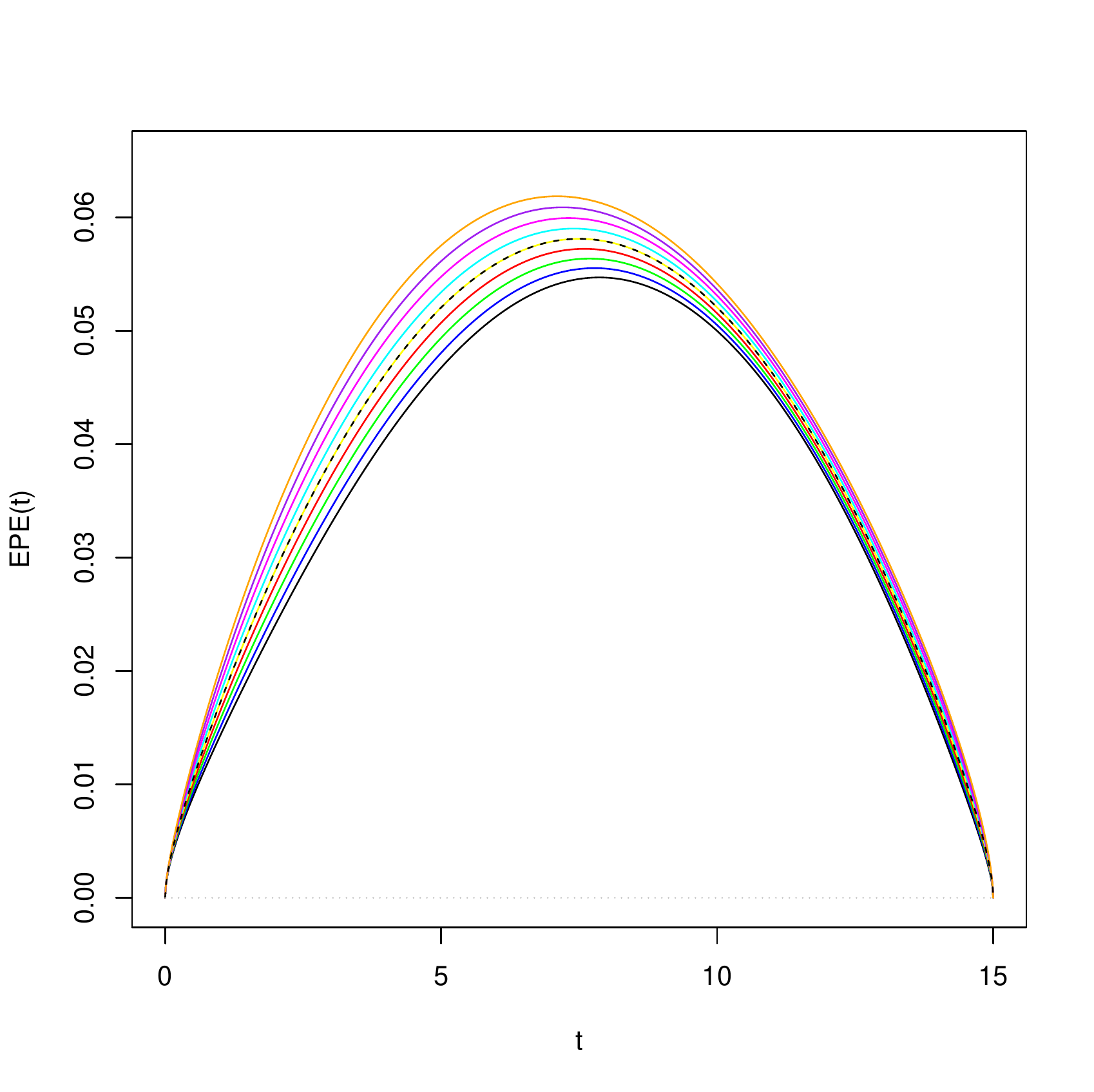}}\hspace{0.8cm}
\subfigure[$T=15Y$, $h(t)=30\%$]{\includegraphics[width=0.28\columnwidth]{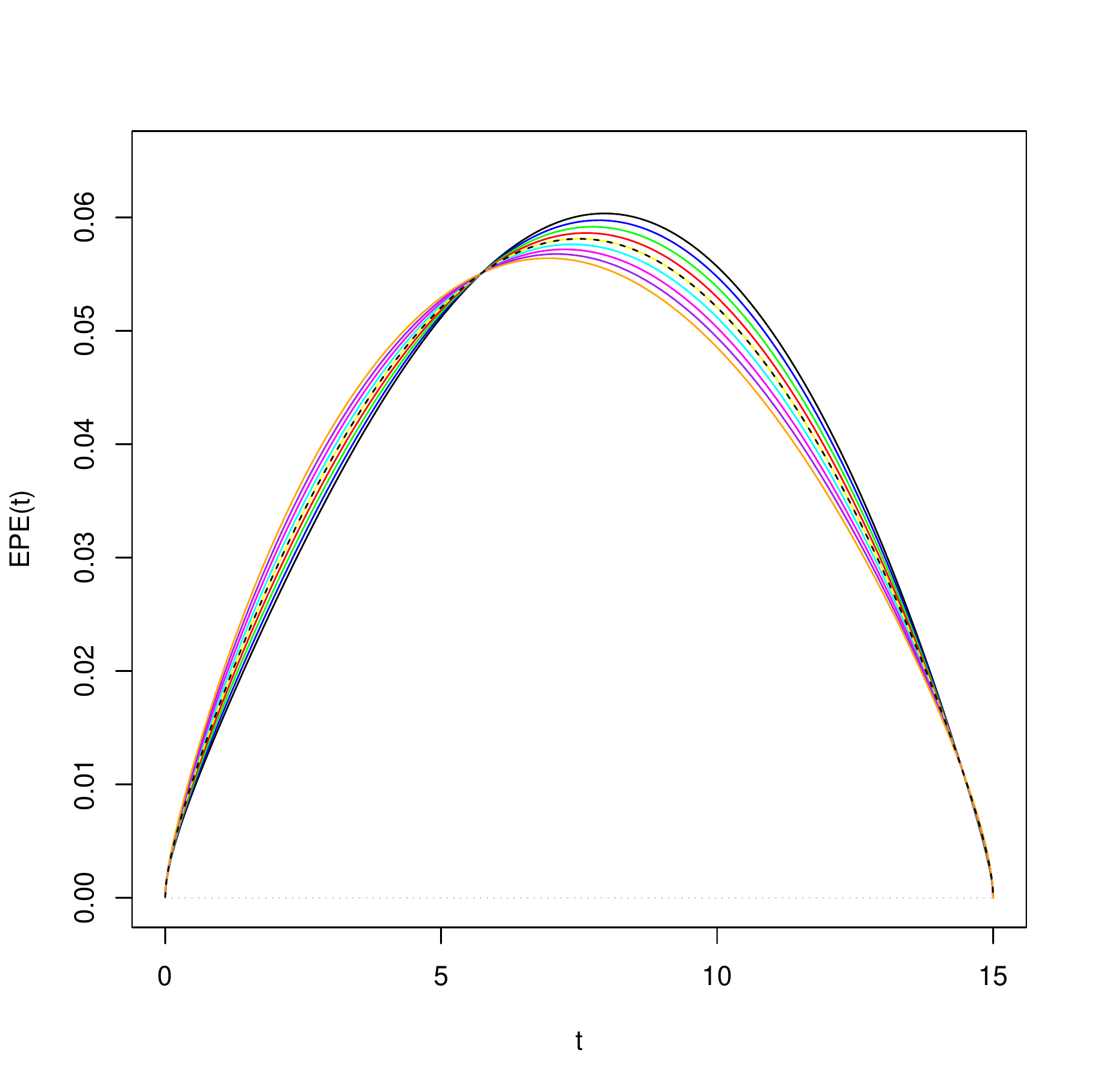}}\\
\caption{EPE$^\perp$ (dashed) and EPE (solid) for various correlation levels, from $\rho=80\%$ (orange) to $\rho=-80\%$ (black) by steps of 20\%. Full 2D Monte Carlo (top, 30k paths, $\delta=2\hbox{E}^{-2}$) and deterministic drift adjustment (bottom). Parameters: $\gamma=0.1\%,\nu=2.2\%$ (exposure) and $y_0=h(0)$, $\kappa=35\%$, $\theta=12\%$, $\sigma=12\%$ (intensity).}\label{Fig.IRS.CIR}
\end{figure}
\end{landscape}

Figure~\ref{fig:FRA-CIR} below shows EPE profiles for Gaussian (forward-type or equity return) exposures for the CIR$^{++}$ ($\mu^\lambda_s=\kappa(\theta-\lambda_s)$ and $\sigma^\lambda_s=\sigma^\lambda(\lambda_s)$ with $\sigma^\lambda(x)=\sigma^\lambda\sqrt{x}$). Top panels show the EPE obtained by the full (2D) Monte Carlo simulation. They prove to be extremely close to the corresponding panels at the bottom, obtained semi-analytically with the measure-change under deterministic drift adjustment approximation~(\ref{eq:DraftAdjh}). 
\begin{figure}[h!]
\centering
\subfigure[$T=5Y$, $h(t)=15\%$]{\includegraphics[width=0.49\columnwidth]{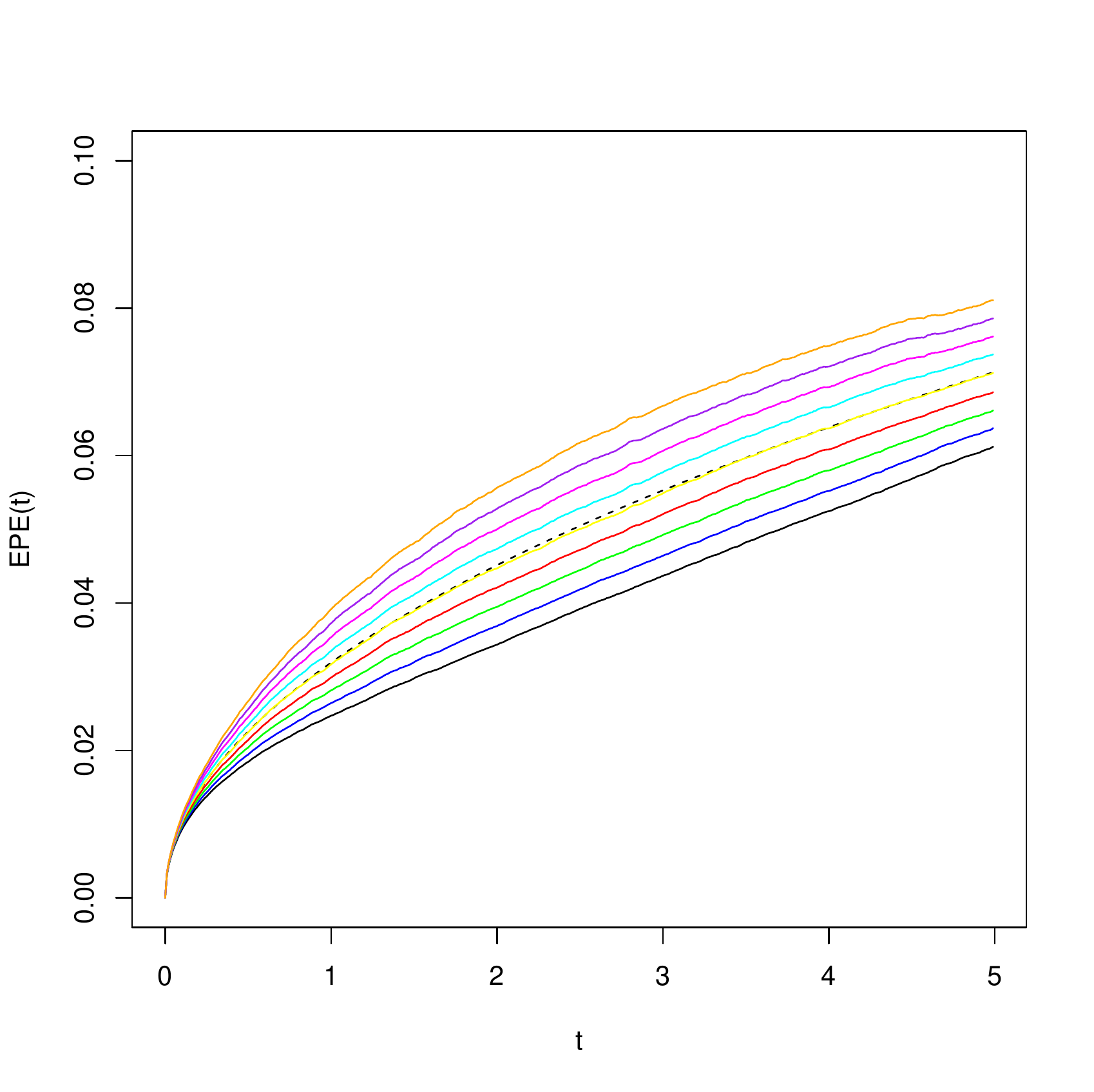}}\hspace{0.02cm}
\subfigure[$T=10Y$, $h(t)=30\%$]{\includegraphics[width=0.49\columnwidth]{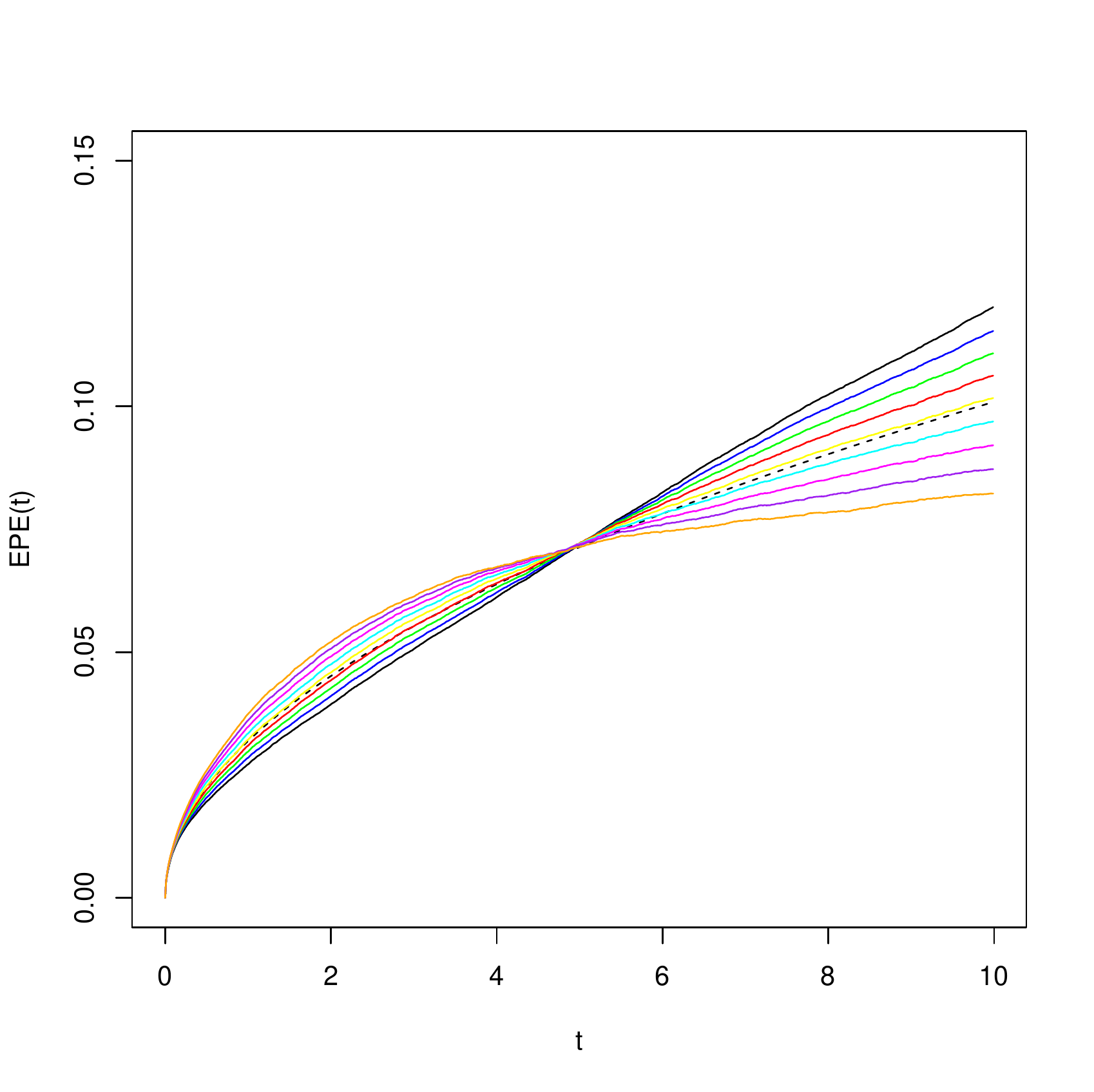}}\\
\subfigure[$T=5Y$, $h(t)=15\%$]{\includegraphics[width=0.49\columnwidth]{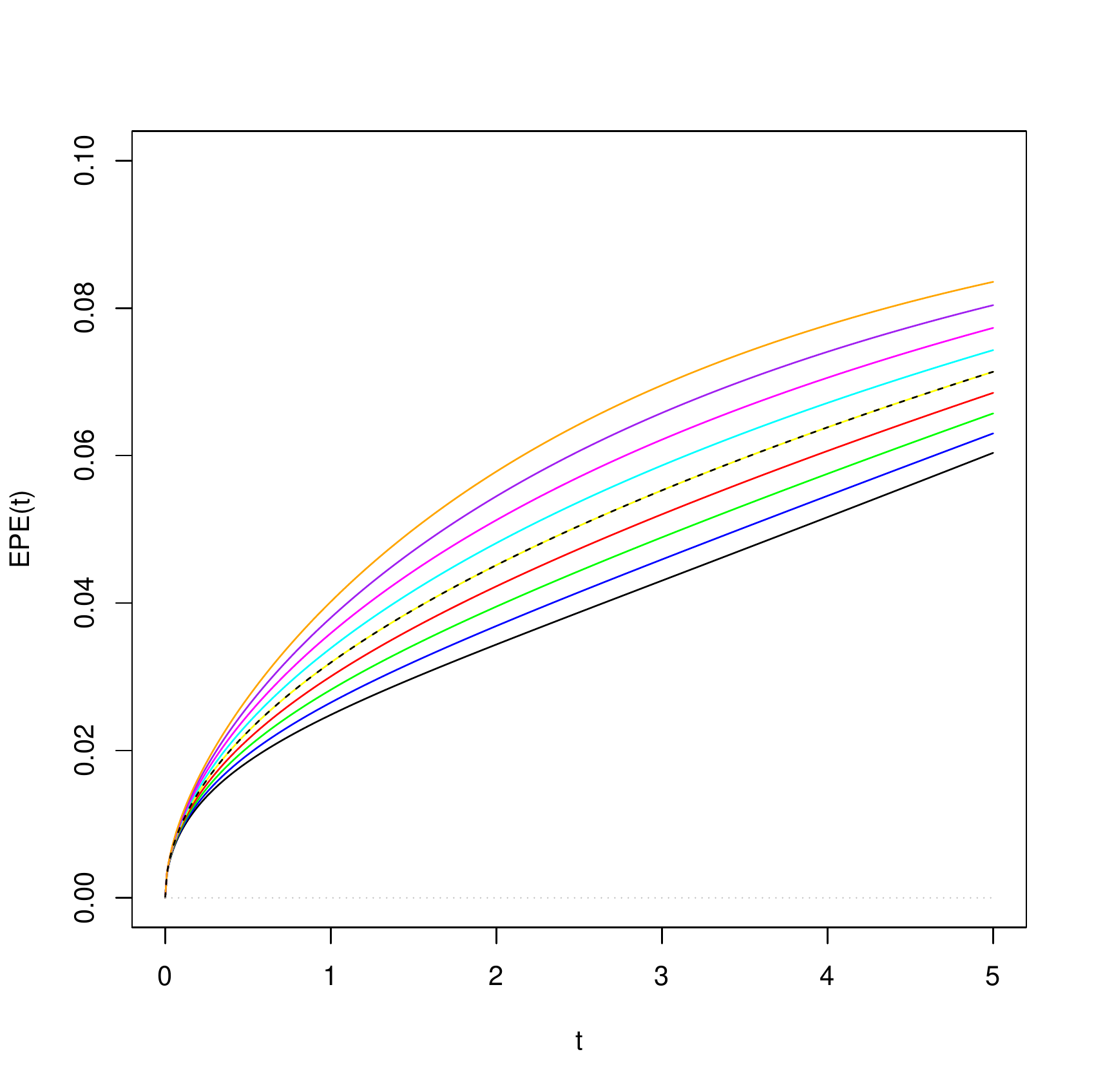}}\hspace{0.02cm}
\subfigure[$T=10Y$, $h(t)=30\%$]{\includegraphics[width=0.49\columnwidth]{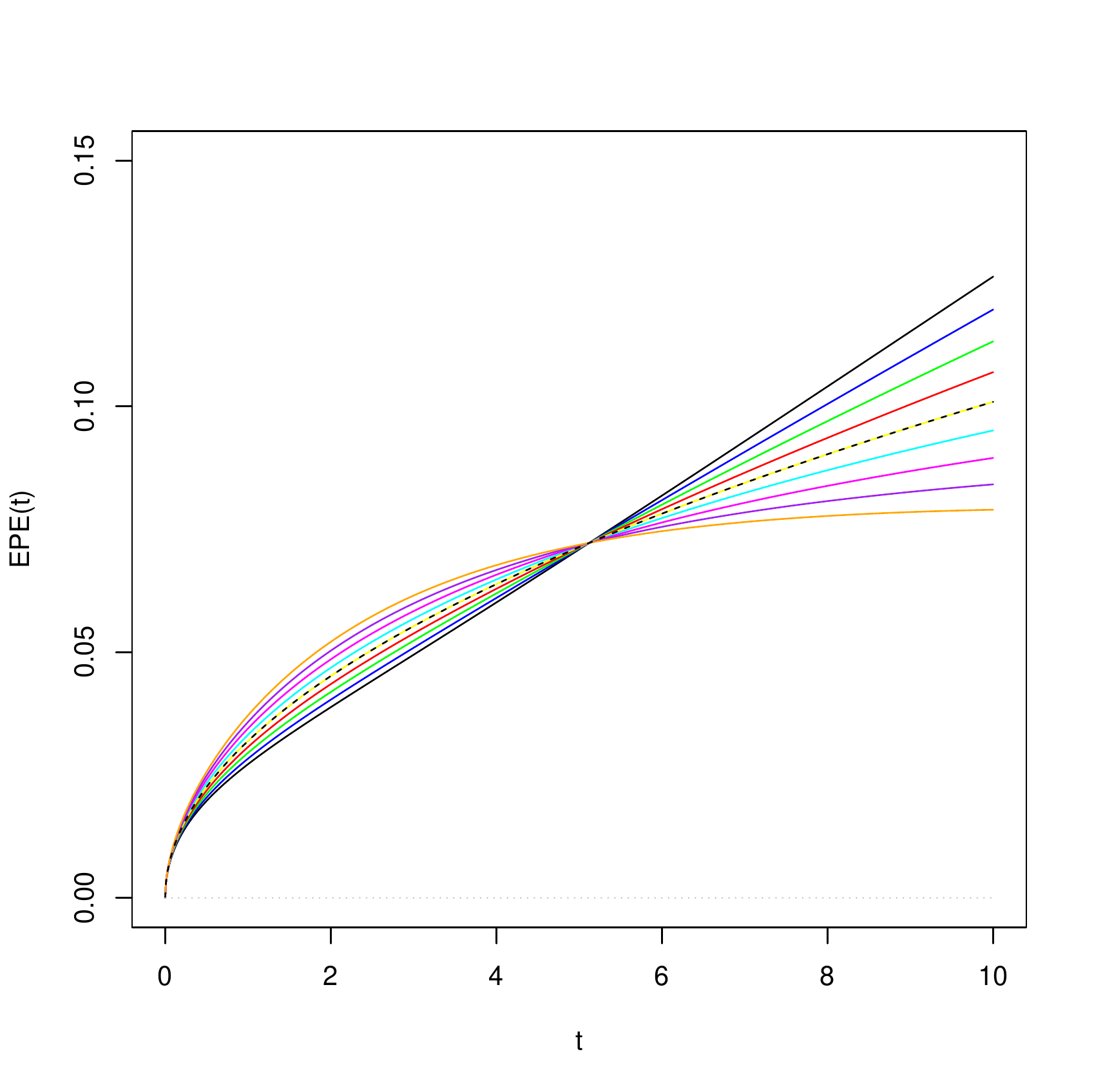}}\\
\caption{EPE$^\perp$ (dashed) and EPE (solid) with CIR$^{++}$ intensity for various correlation levels, from $\rho=80\%$ (orange) to $\rho=-80\%$ (black) by steps of 20\%. Full 2D Monte Carlo (top, 30k paths, $\delta=1\hbox{E}^{-2}$) and WWR measure with deterministic drift adjustment (bottom). Parameters: $\nu=8\%$ (exposure) and $y_0=h(0)$, $\sigma=12\%$, $\kappa=35\%$, $\theta=12\%$ (intensity).}\label{fig:FRA-CIR}
\end{figure}
The approximation performs very well for the CIR$^{++}$ intensity in swap-type profiles too, as one can see from Fig.~\ref{Fig.IRS.CIR}. 
%
%
%
%

Comparing top and bottom rows of figures \ref{fig:FRA-CIR} and \ref{Fig.IRS.CIR} suggests that the deterministic approximation of the drift adjustment preserves the ability of the change-of-measure approach to reproduce specific properties of the EPE profiles, including crossings and asymmetry. 

We have considered CIR and CIR$^{++}$ here for intensities, but one might also wish to consider other affine models. Shifted OU (also known as Hull-White) is one of them. This extremely tractable model is very popular for interest-rates modeling. However, as this is a Gaussian process, it is not appropriate for default intensities or positive exposures given the possibility of negative values. More specifically, $\Q$-EPE can be computed semi-analytically in the case of Gaussian exposures and OU dynamics for ``intensities'' in which case CVA can indeed become negative, which is clearly wrong  (see e.g. \cite{Vrins16a}). By contrast, expressions of the form $\E^{\Pr}[V_t^+]$ will of course always be non-negative, whatever the measure $\Pr$ so that there is no hope that the WWM approach will agree with the results found by computing the EPE under $\Q$ directly. The reason for this mismatch is of course that the choice of the num\'eraire is not valid in this case as it is not guaranteed to be positive. However, the change-of-measure technique is acceptable when the process parameters are such that $\lambda$ takes negative values with very small probability, at least when $V_t$ is positive (positive $\rho$). We do not discuss further the results corresponding to OU ``intensities''. Another possible model is JCIR (or its shifted version JCIR$^{++}$). For sake of brevity, we will analyze CVA figures directly in Section~\ref{sec:CVAJCIR}. 
%
%
%
\subsection{CVA figures}
%
The above section emphasizes that the change-of-num\'eraire technique, in spite of the deterministic approximation of the drift adjustment, allows to adequately represent the functional form of the EPE profiles under WWR. In this section we focus on CVA figures and compare the results obtained by using either the full Monte Carlo simulation or the semi-analytical results using the deterministic drift adjustment. Instead of specifying a given survival probability curve, we start from the CIR parameters and take $P^y(0,t)$ as $G(t)$ so that no shift is needed, i.e. $\lambda\equiv y$. This way of proceeding rules out potential problems of getting negative intensities as a result of a negative shift and yields a large degree of freedom to play with the parameters.
%
\subsubsection{Effect of the long-term mean}
%
We ha fix the CIR parameters and play with four different values of the long term mean (driving the slope of the CDS curve, i.e. contango or backwardation) as well as with the maturity, the type and the volatility of the exposure process. 

\begin{landscape}
\begin{figure}
\centering
\subfigure[$\alpha=0.5$]{\includegraphics[width=0.23\columnwidth]{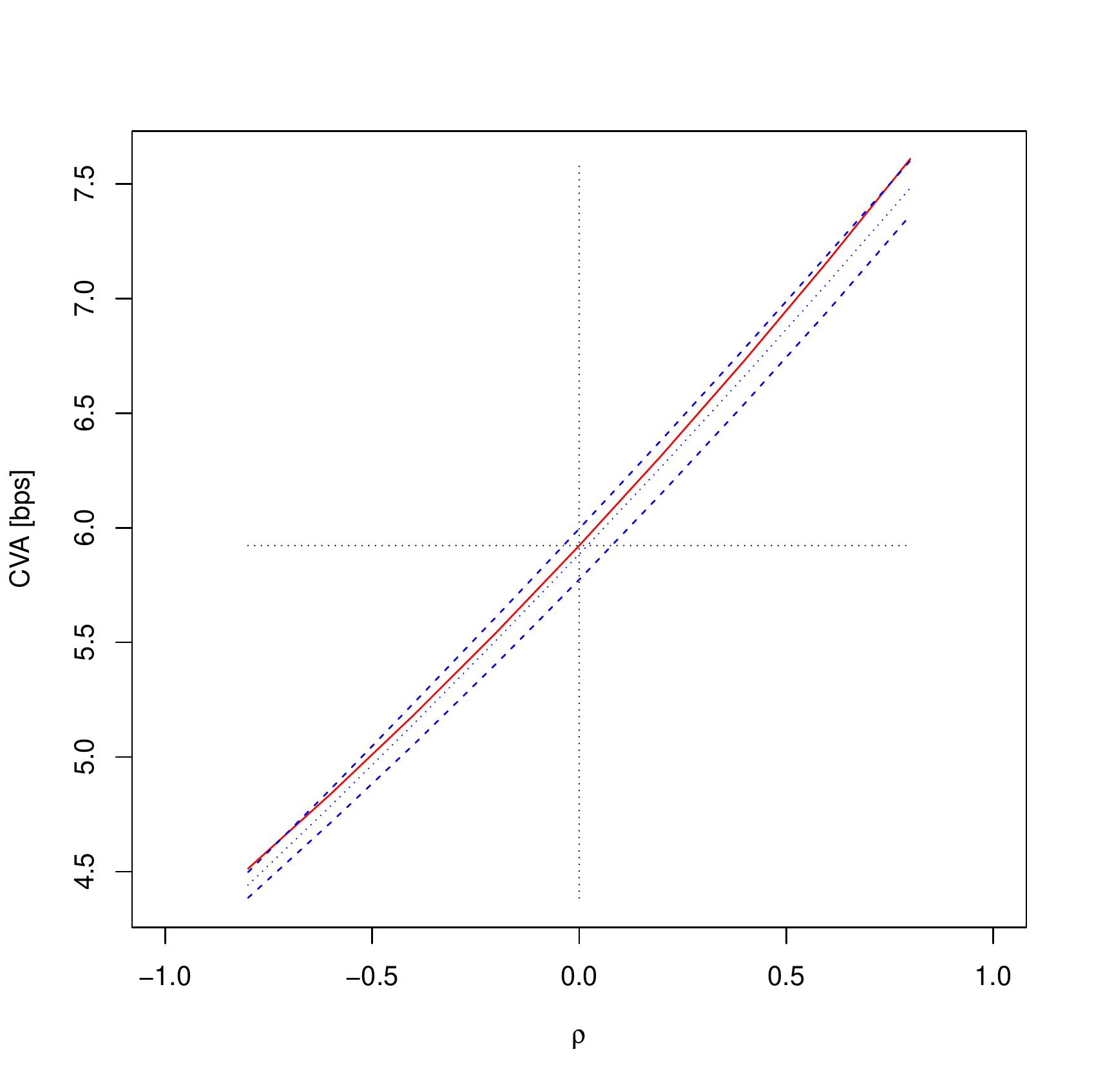}}\hspace{0.02cm}
\subfigure[$\alpha=1.5$]{\includegraphics[width=0.23\columnwidth]{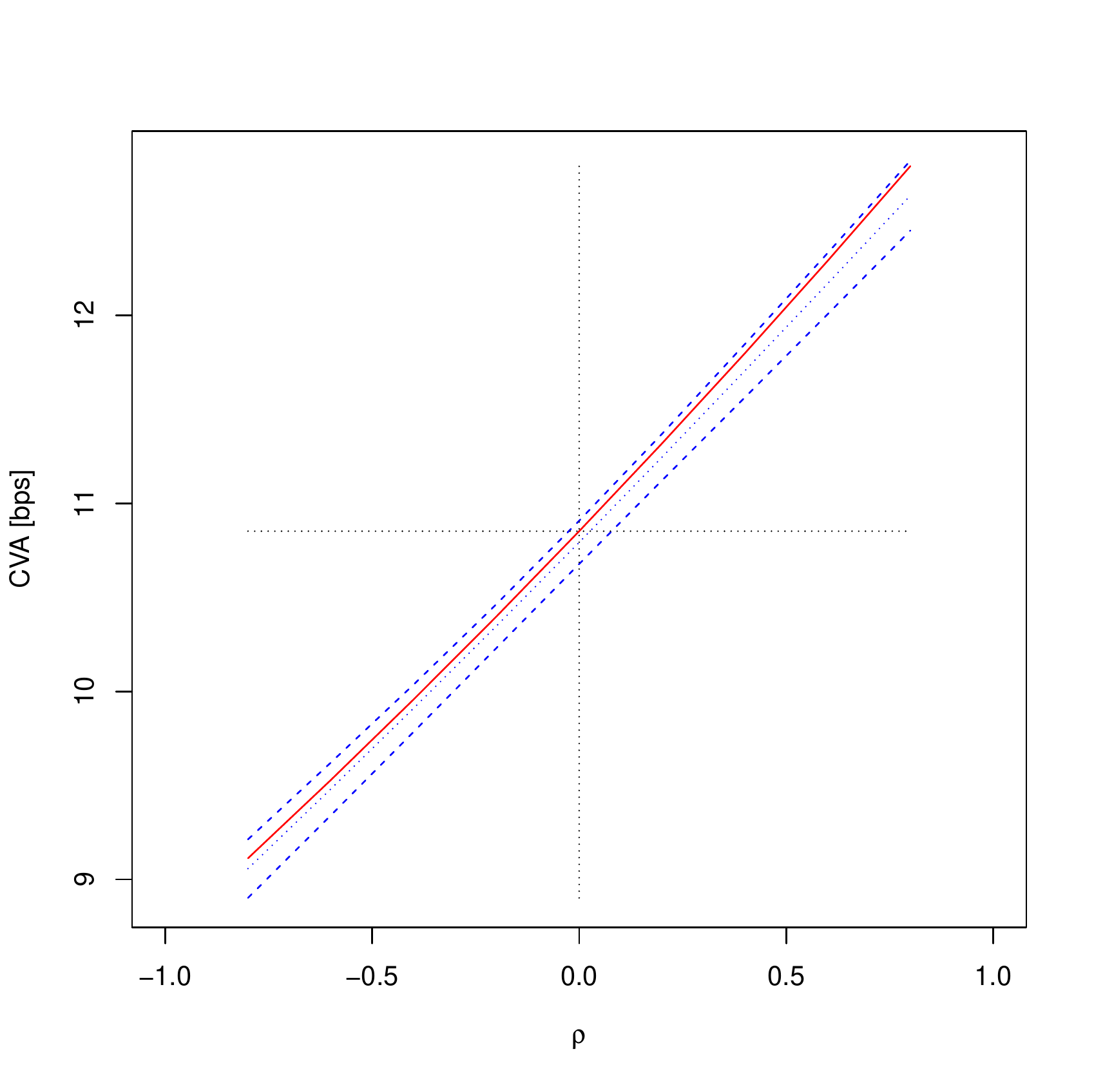}}\hspace{0.02cm}
\subfigure[$\alpha=5$]{\includegraphics[width=0.23\columnwidth]{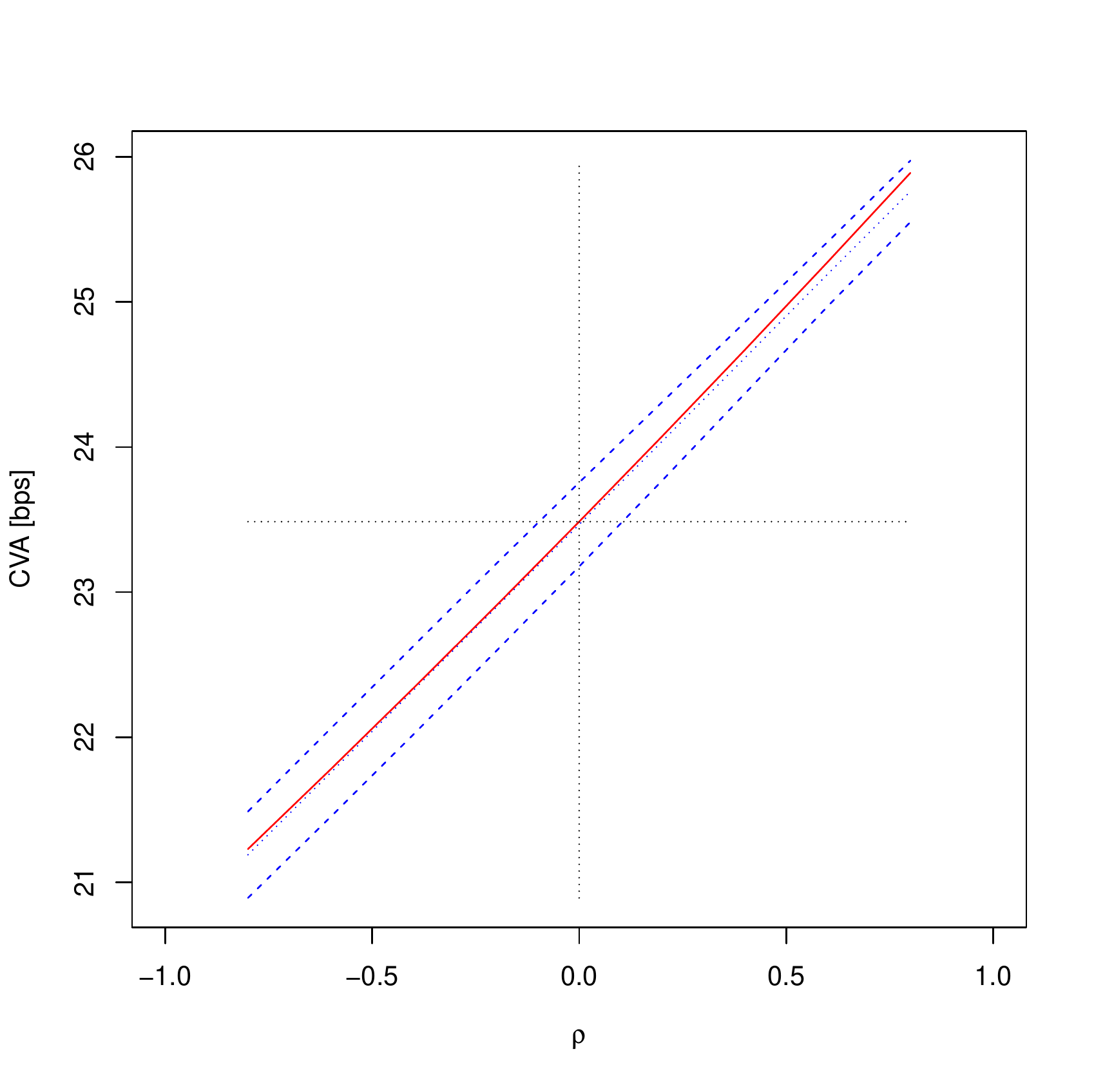}}\hspace{0.02cm}
\subfigure[$\alpha=10$]{\includegraphics[width=0.23\columnwidth]{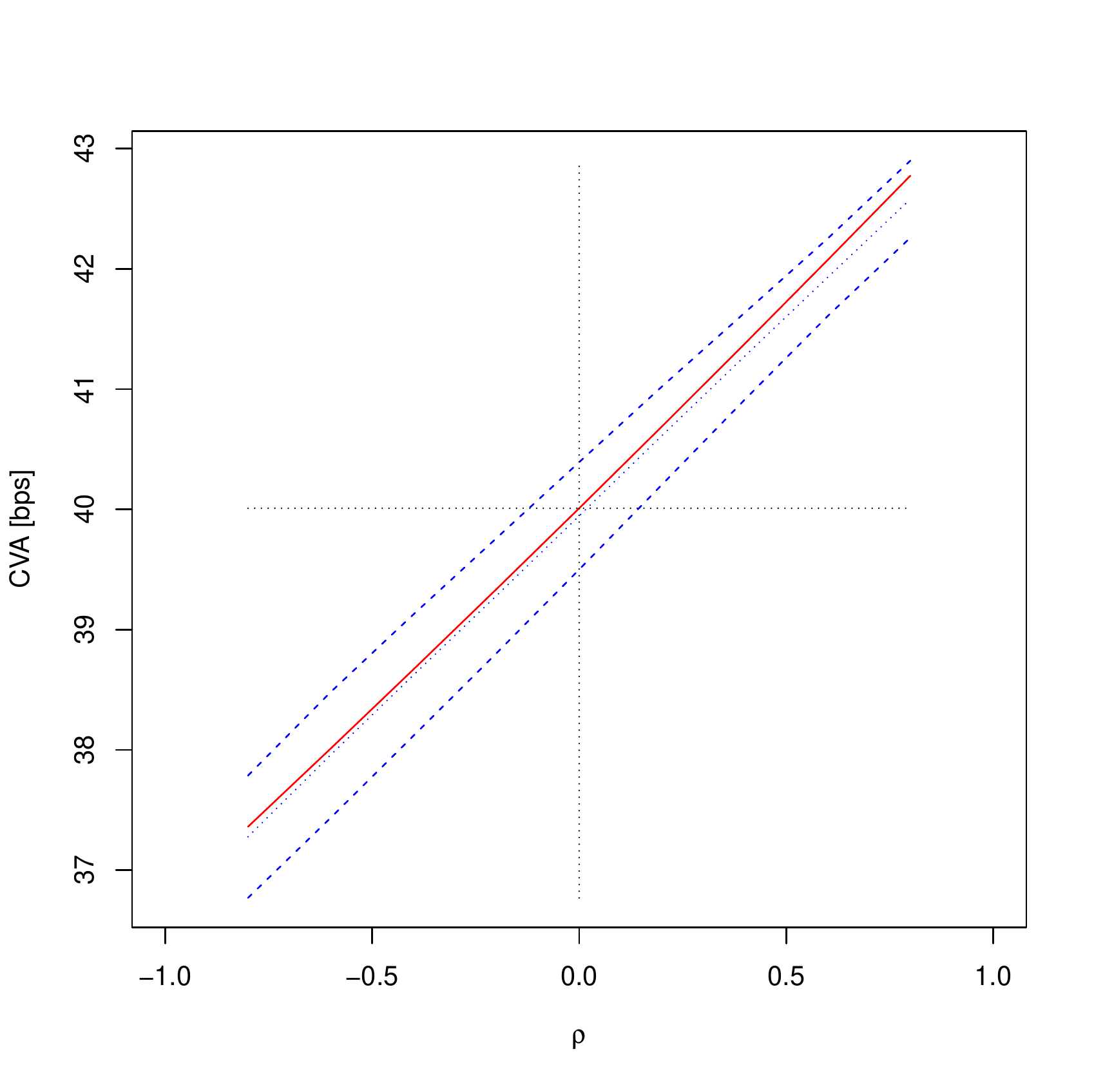}}\\
\subfigure[$\alpha=0.5$]{\includegraphics[width=0.23\columnwidth]{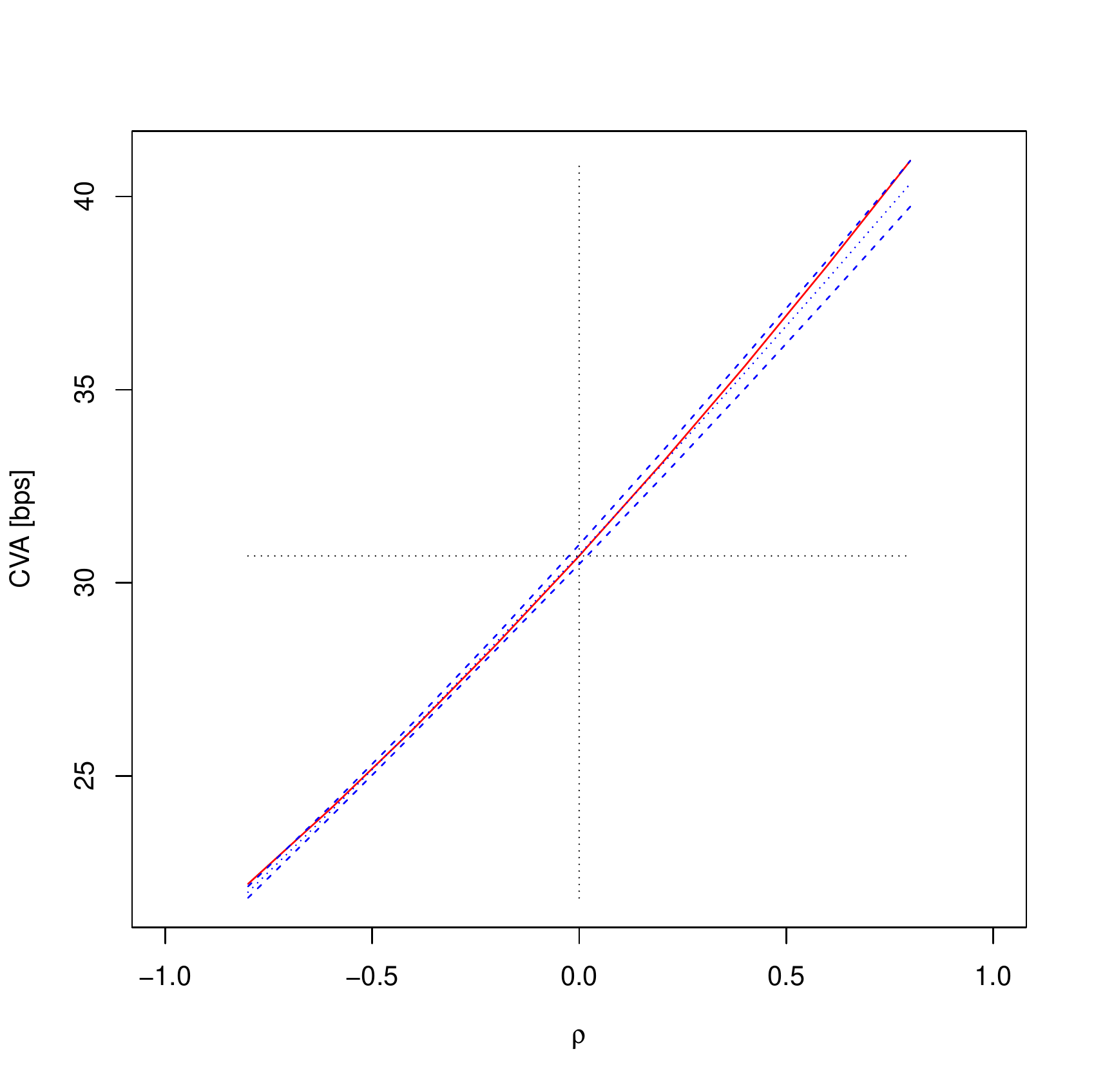}}\hspace{0.02cm}
\subfigure[$\alpha=1.5$]{\includegraphics[width=0.23\columnwidth]{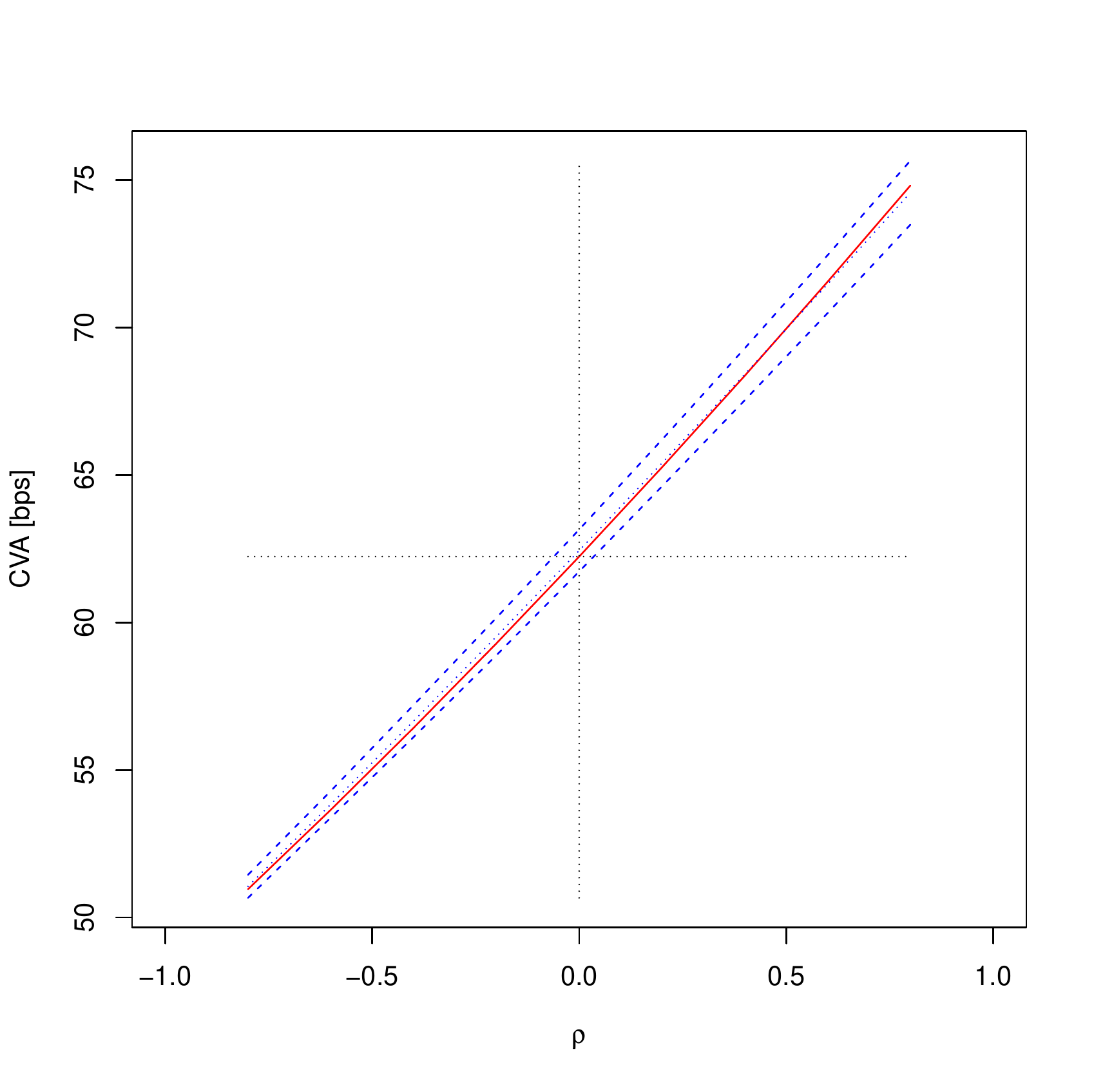}}\hspace{0.02cm}
\subfigure[$\alpha=5$]{\includegraphics[width=0.23\columnwidth]{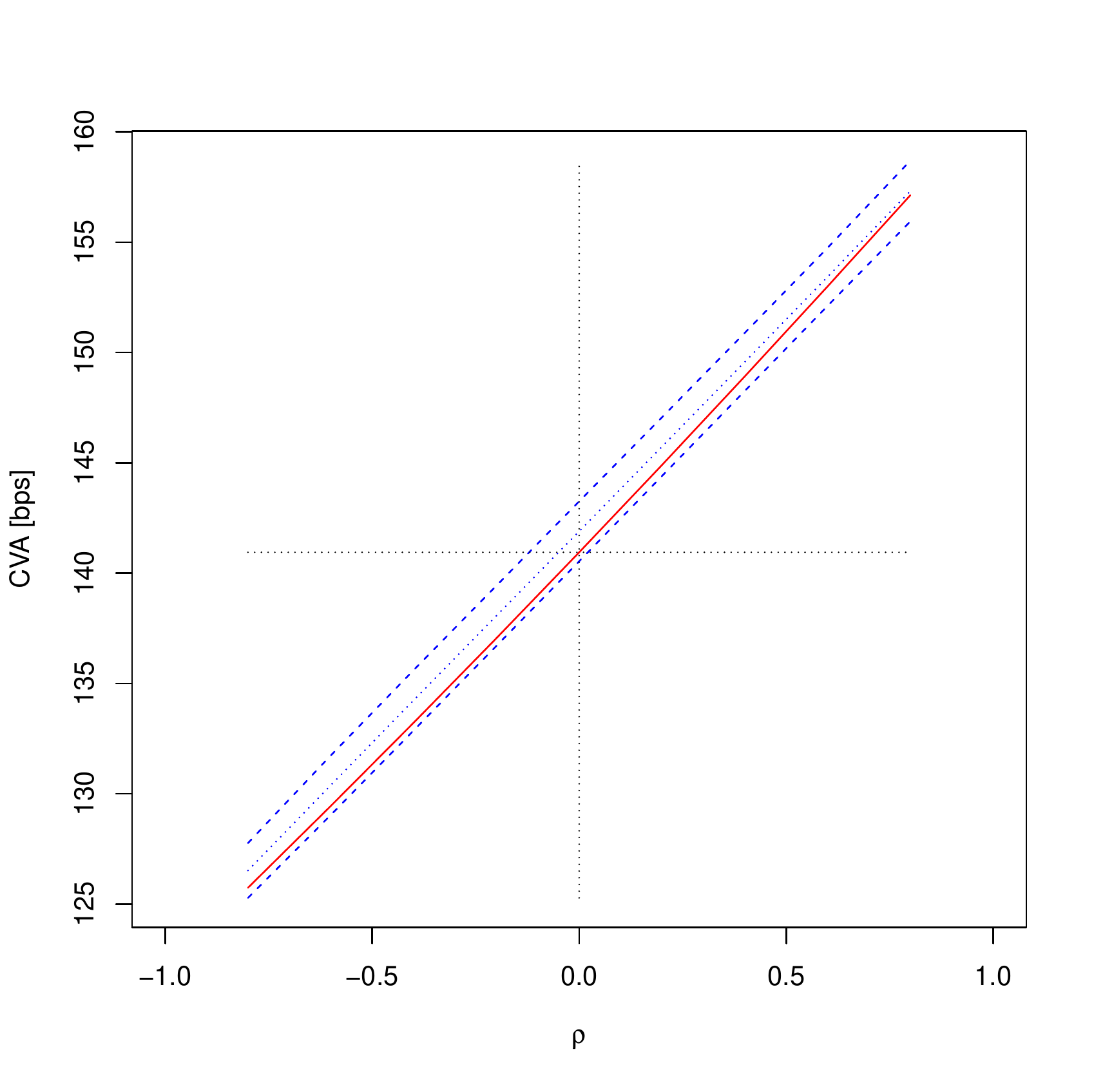}}\hspace{0.02cm}
\subfigure[$\alpha=10$]{\includegraphics[width=0.23\columnwidth]{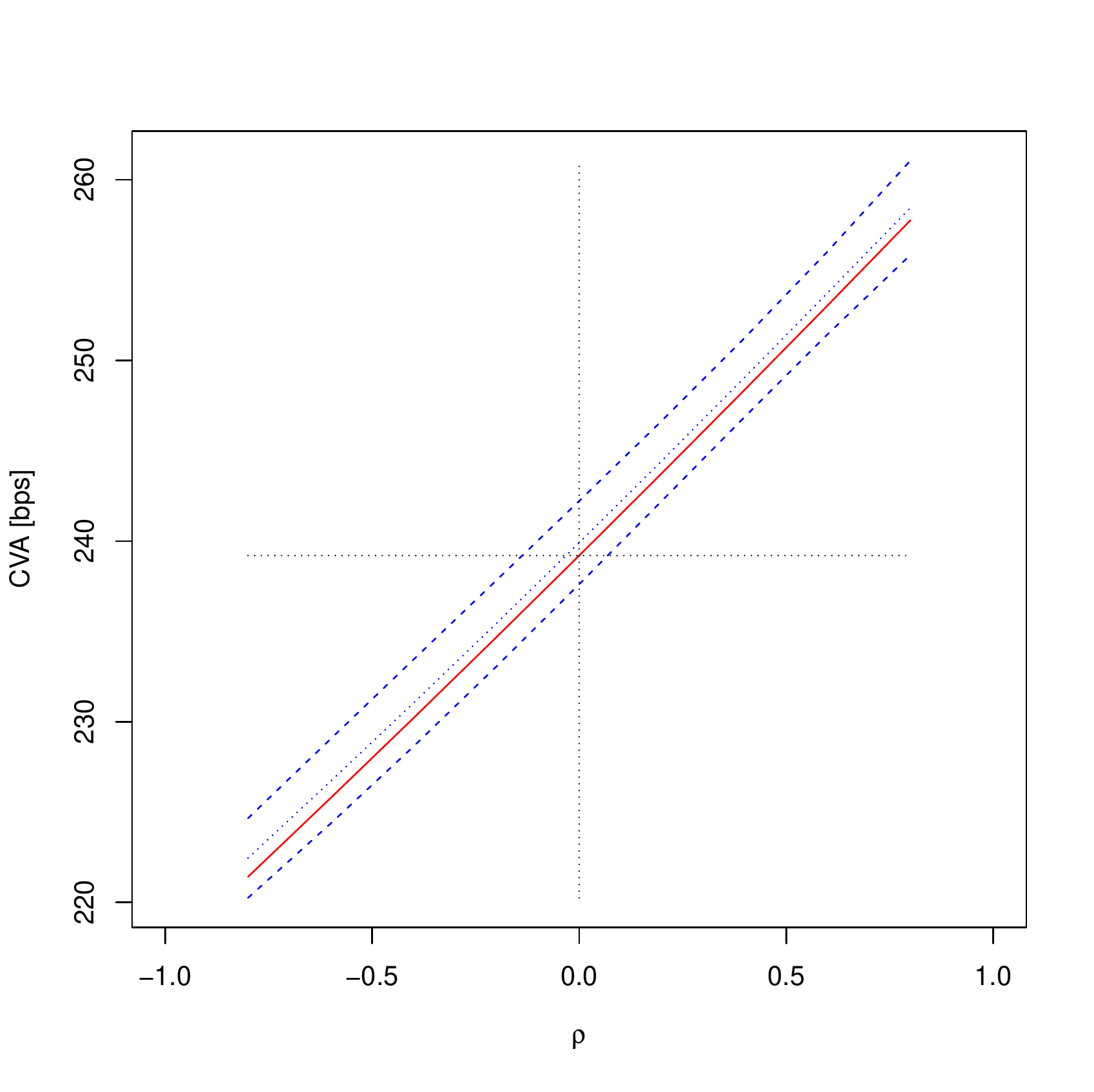}}
\caption{CVA Figures with $\kappa=10\%$, $\theta=\alpha y_0\%$, $y_0=50$ bps. Top row: Brownian exposure $T=10Y$ with $\nu=2.2\%$, $\sigma=0.8\%$, Bottom row drift-inclusive Brownian bridge exposure $T=15Y$ with $\nu=8\%$, $\sigma=1\%$. Legend: CVA with change-of-measure technique (solid red), average of 10 Full 2D Monte Carlo runs with 30k each (dotted blue) and corresponding confidence interval (2 times standard deviation estimated from the sets of runs).}\label{Fig.CVA.CIR}
\end{figure}
\end{landscape}

The corresponding CVA figures are given in Figure~\ref{Fig.CVA.CIR}. Notice that the CVA is quoted in basis points upfront. They can be converted in a running premium Chapter 21.3 in~\cite{Brigo06} and \cite{Vrins12}.
\subsubsection{Comparison of performances for 4 sets of CIR parameters}
%
Some possible sets for the CIR parameters are given in Table~\ref{tab:cirparams}. Set 1 has been chosen exogeneously, Set 2 is taken from~\cite{brigoccdsrisk} while Set 3 \& Set 4 come from~\cite{bpp10}. We refer to these works for CDS implied volatilities and other market pattern implied by these parameters. Notice that Set 4 looks relatively extreme in that the volatility parameter is quite large and Feller condition is strongly violated. 
\begin{table}
\centering
\begin{tabular}{|c|c|c|c|c|c|}
\hline
Set&$y_0$ (bps)&	$\kappa$ &	$\theta$ (bps) & $\sigma$ & $2\kappa\theta-\sigma^2$\\
\hline
1&300&2\% & 1610 & 8\% &4E$^{-5}$\\
2&350&35\% & 450 & 15\% &0.9\%\\
3&100&80\% & 200 & 20\% &-0.8\%\\
4&300&50\% & 500 & 50\% &-20\%\\
\hline
\end{tabular}
\caption{Feller condition is violated in some cases, specifically in Set 4.}\label{tab:cirparams}
\end{table}

\begin{landscape}
\begin{figure}
\centering
\subfigure[Set 1, $\nu=2.2\%$]{\includegraphics[width=0.23\columnwidth]{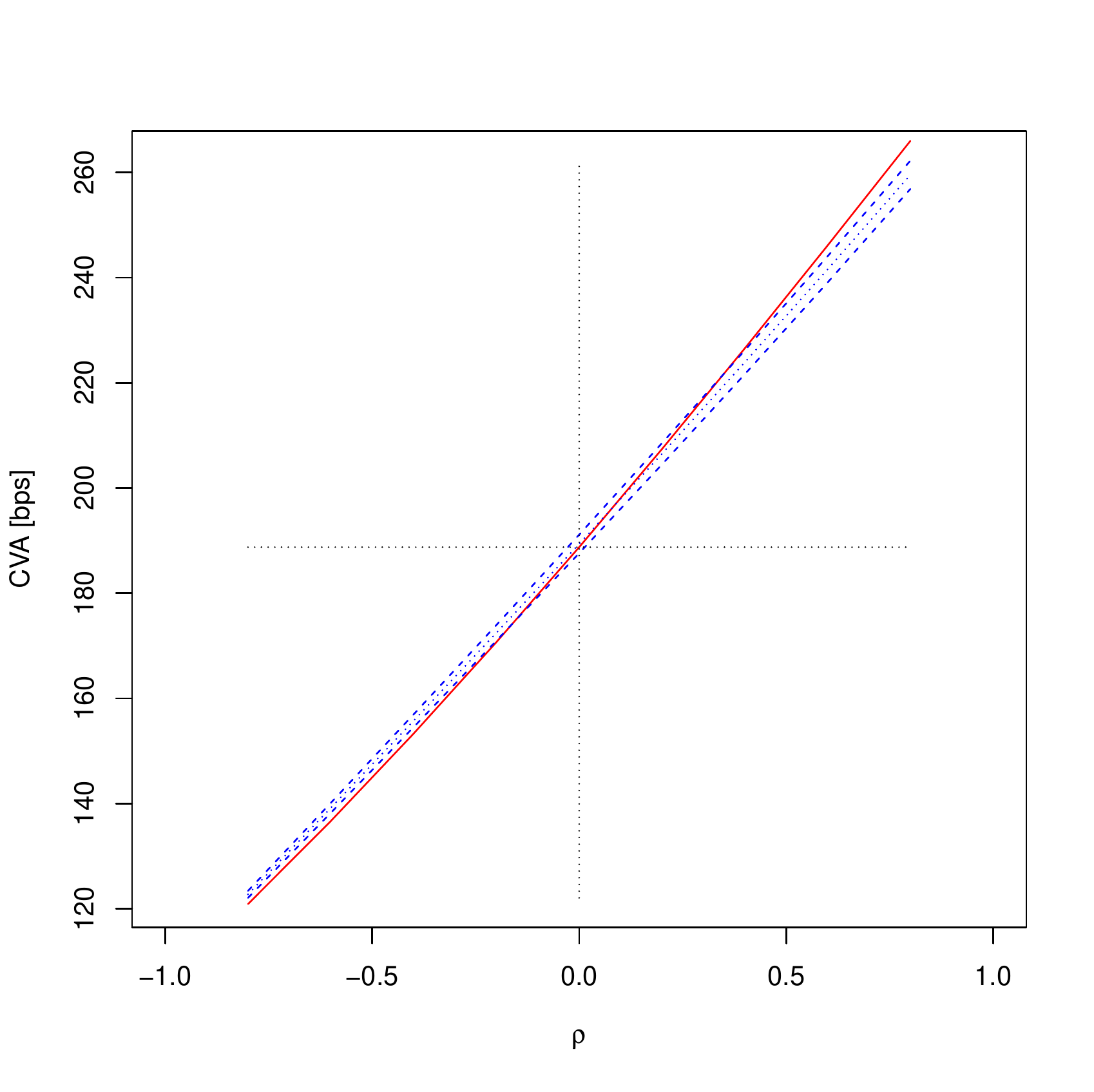}}\hspace{0.02cm}
\subfigure[Set 2, $\nu=2.2\%$]{\includegraphics[width=0.23\columnwidth]{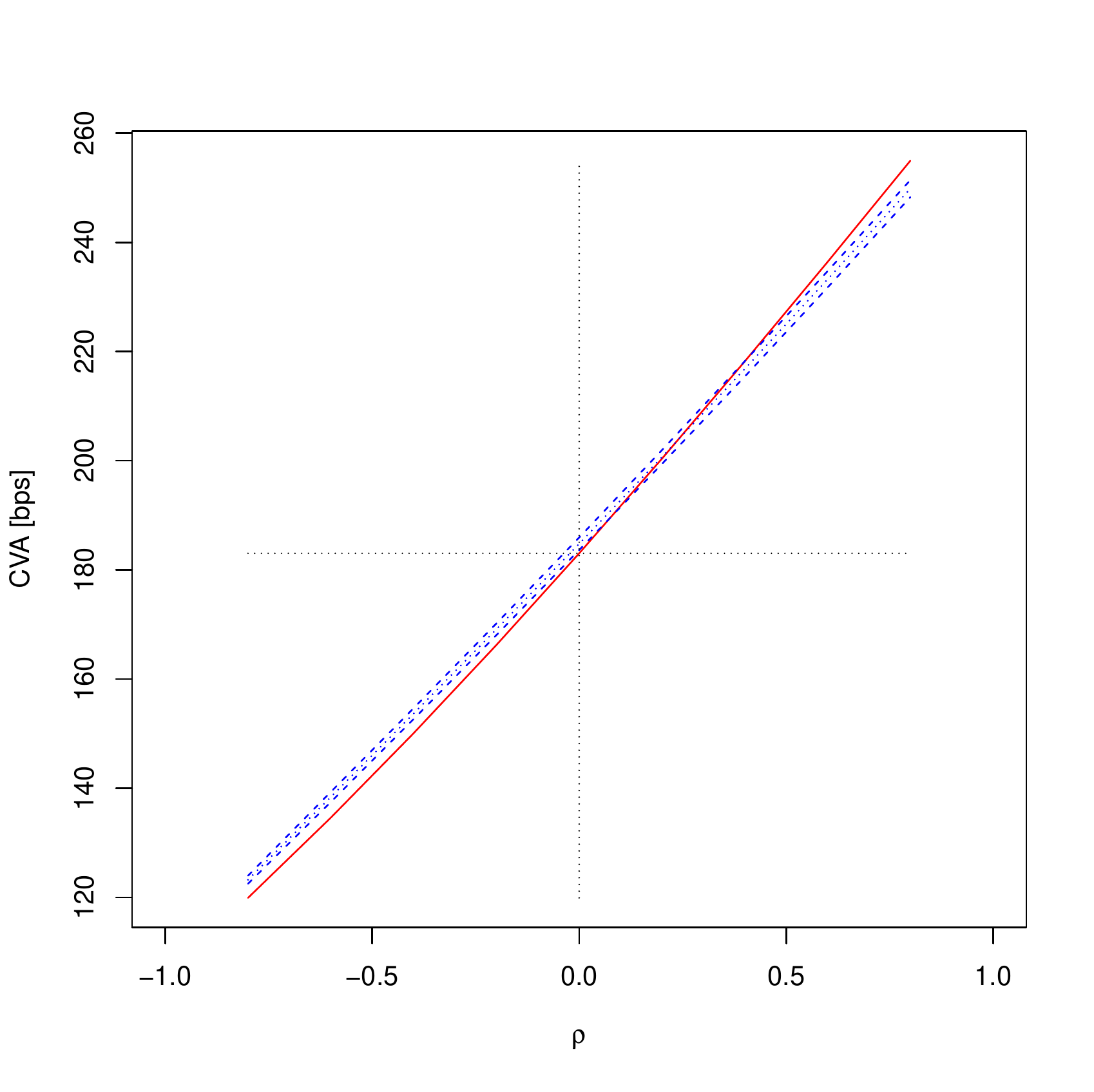}}\hspace{0.02cm}
\subfigure[Set 3, $\nu=2.2\%$]{\includegraphics[width=0.23\columnwidth]{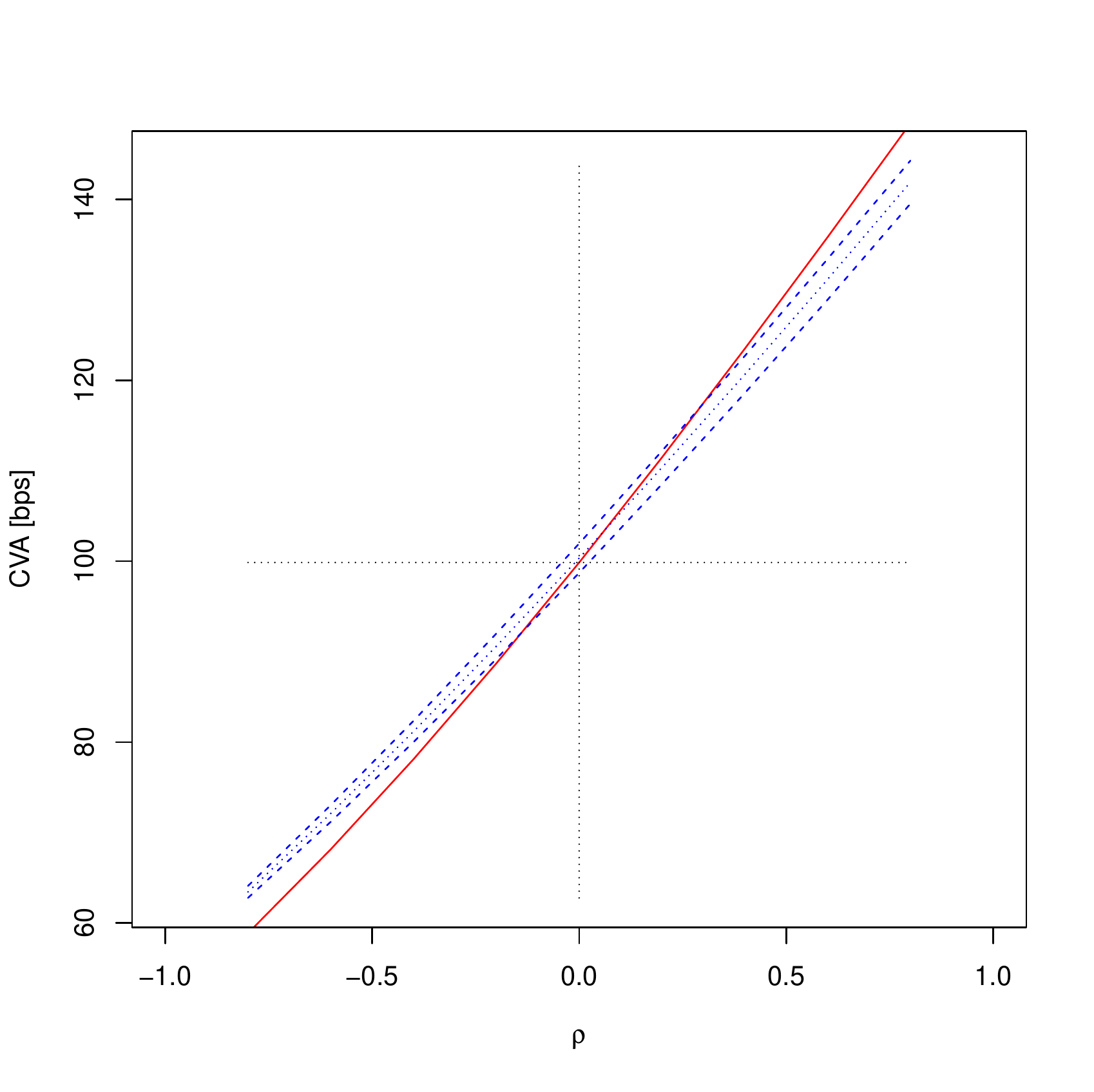}}\hspace{0.02cm}
\subfigure[Set 4, $\nu=2.2\%$]{\includegraphics[width=0.23\columnwidth]{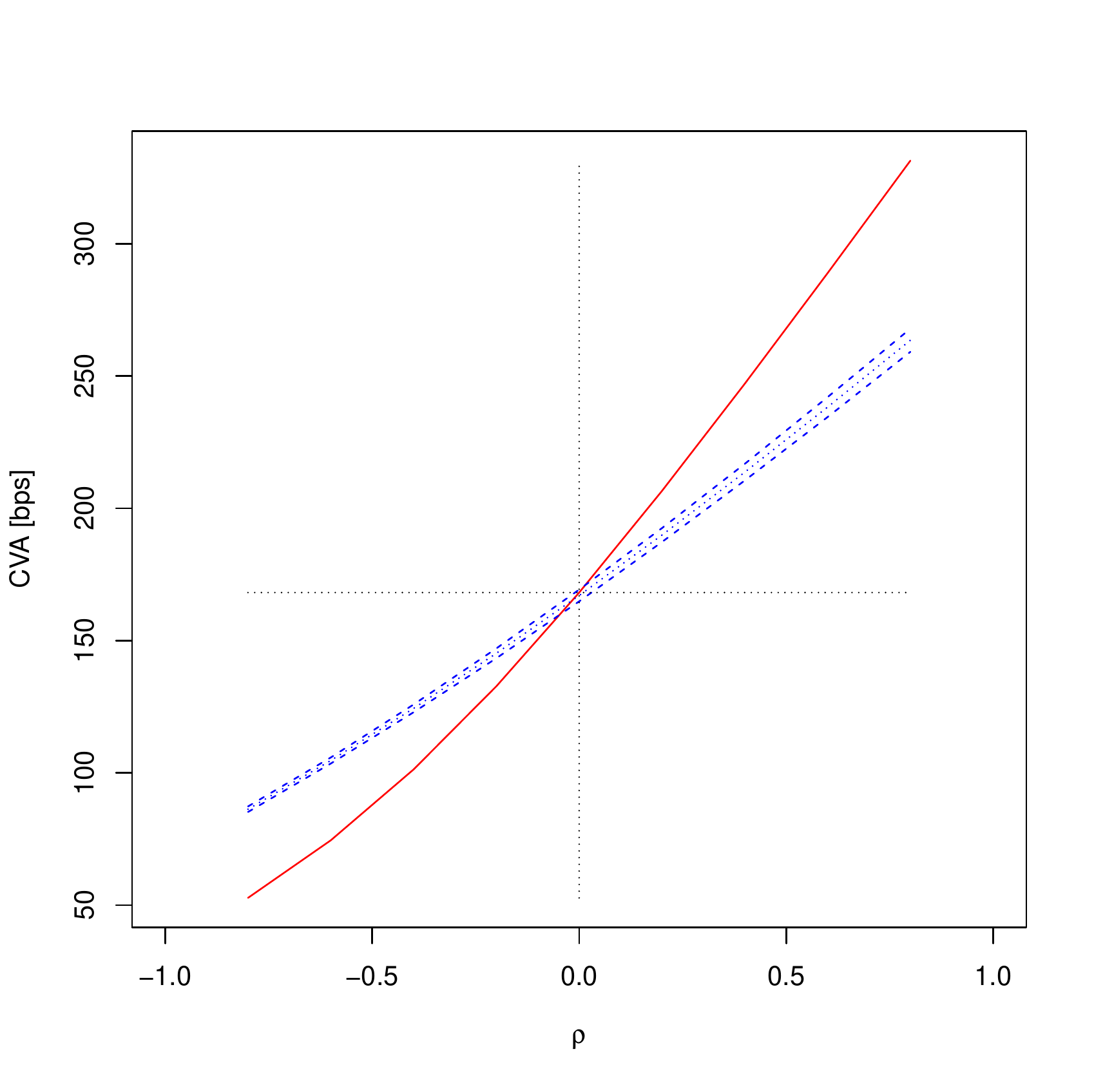}}\\
\subfigure[Set 1, $\nu=8\%$]{\includegraphics[width=0.23\columnwidth]{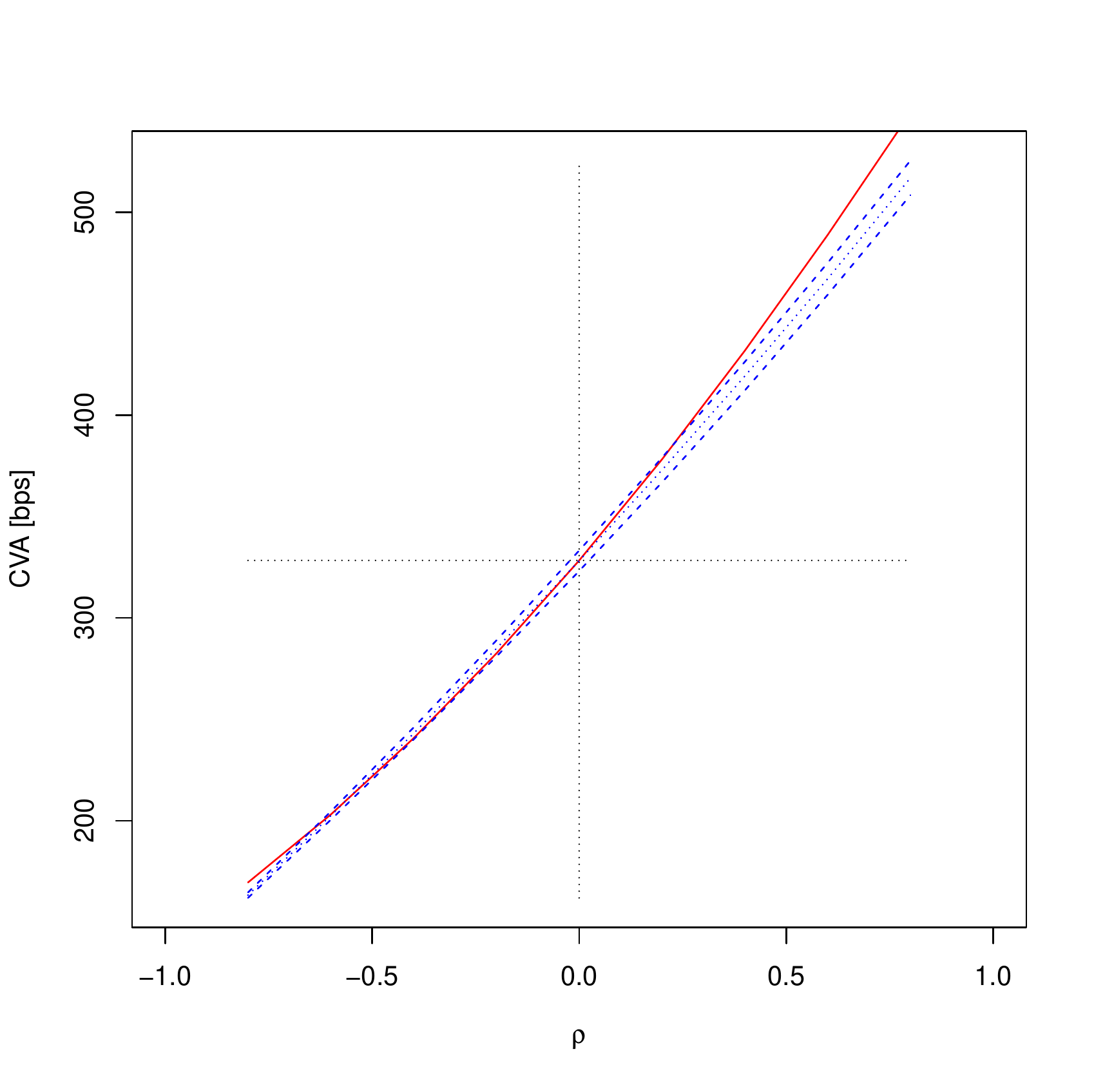}}\hspace{0.02cm}
\subfigure[Set 2, $\nu=8\%$]{\includegraphics[width=0.23\columnwidth]{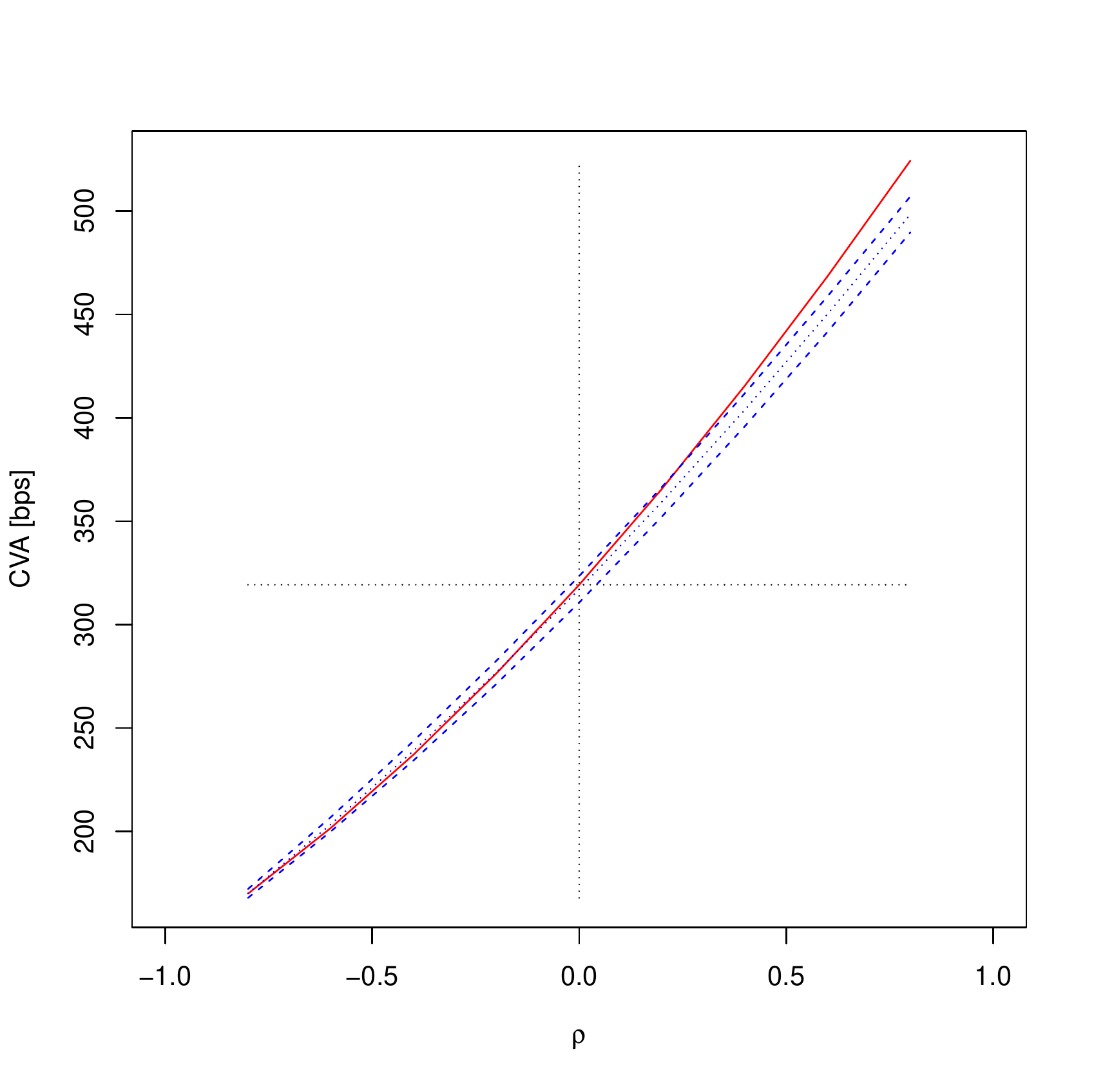}}\hspace{0.02cm}
\subfigure[Set 3, $\nu=8\%$]{\includegraphics[width=0.23\columnwidth]{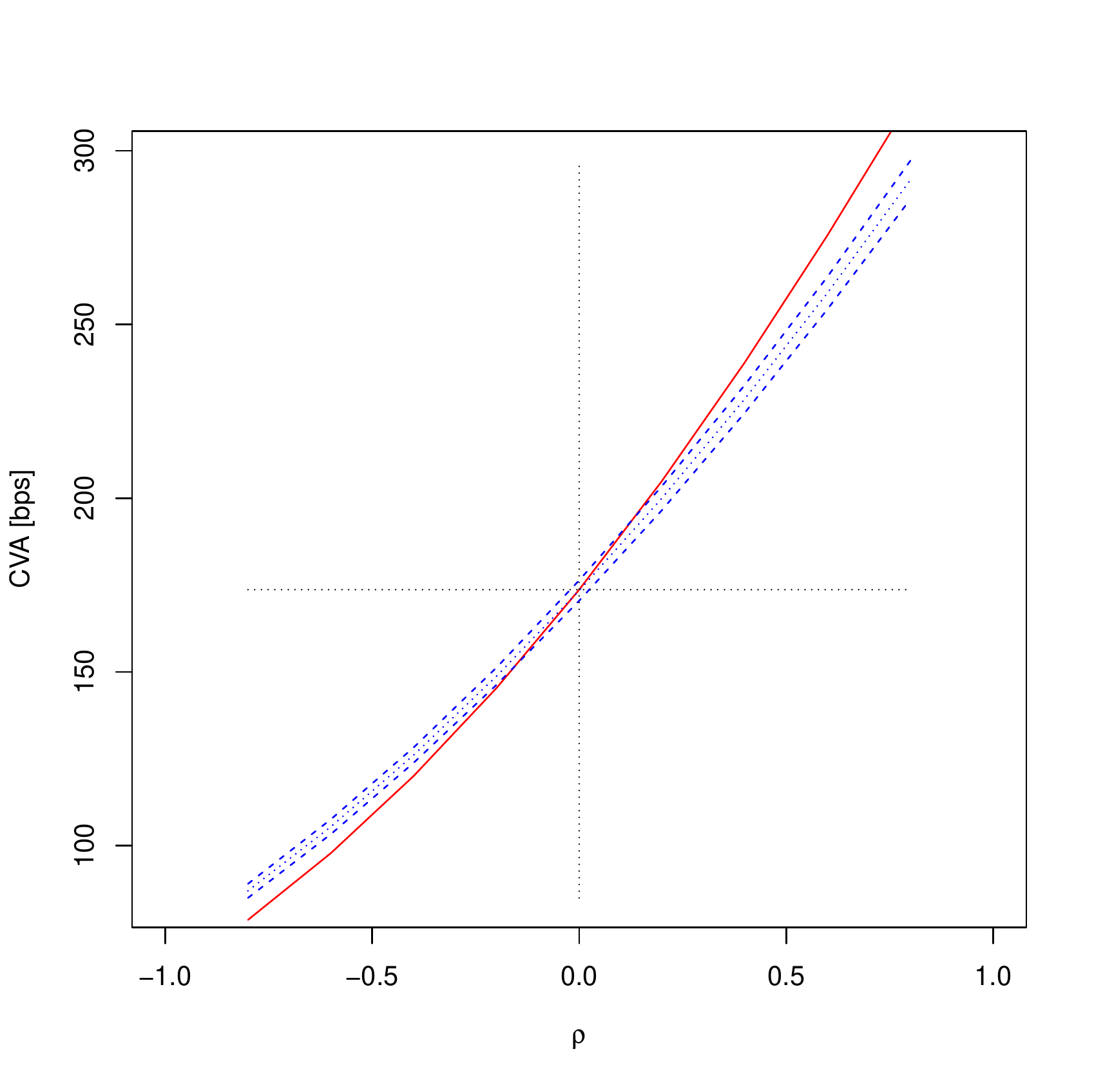}}\hspace{0.02cm}
\subfigure[Set 4, $\nu=8\%$]{\includegraphics[width=0.23\columnwidth]{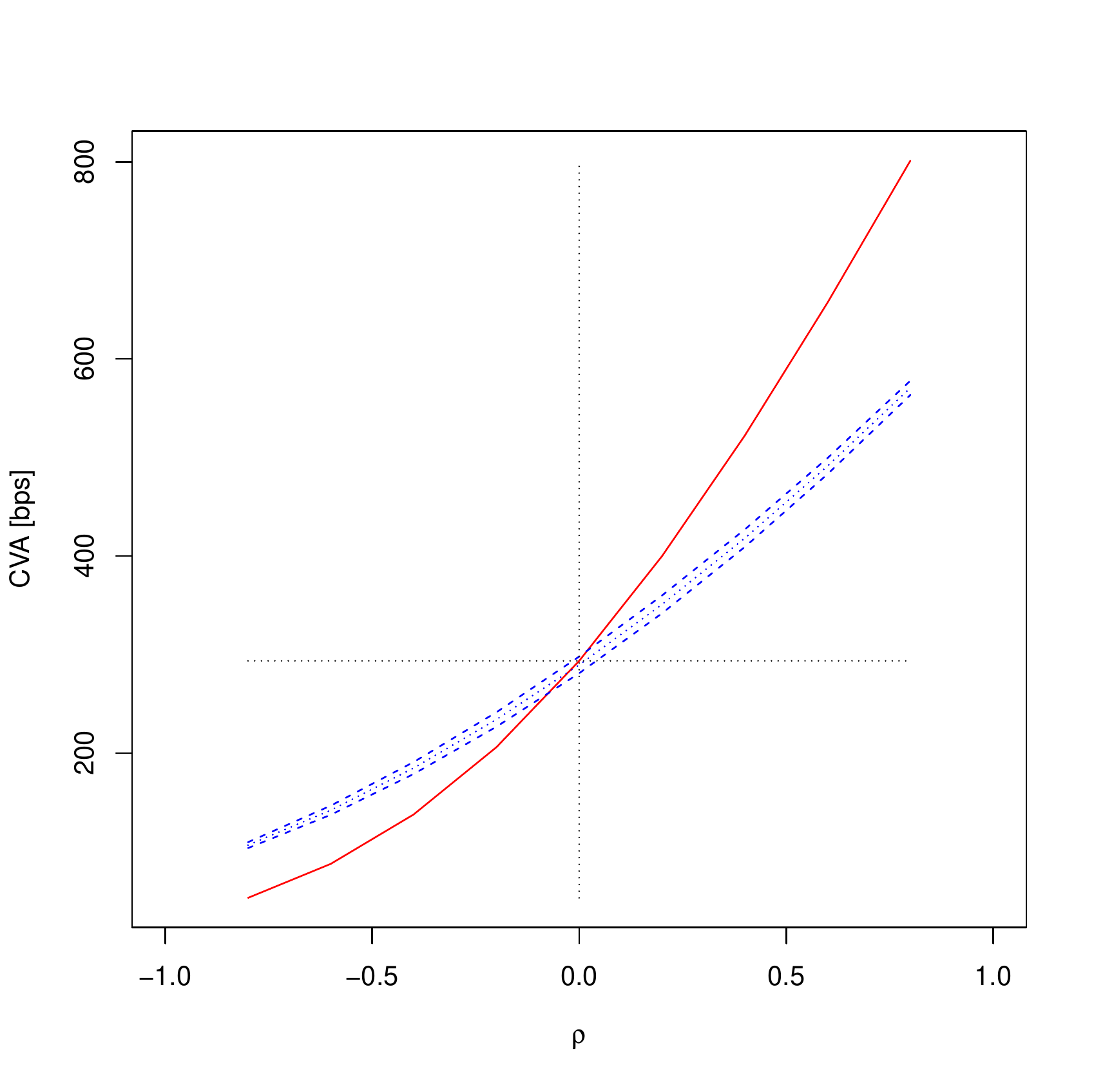}}\\
\caption{CVA Figures for both change-of-measure and Monte Carlo methods (30k paths) for a 15Y swap-type exposures for various exposure volatility and CIR parameters 
.}\label{Fig.CVA.CIR.DB}
\end{figure}
\end{landscape}

In this section we stress the impact of the volatility on the quality of the deterministic approximation of the drift adjustment. The CVA figures are shown with respect to correlation on Fig.~\ref{Fig.CVA.CIR.DB}.
%
%
%
%
%
\subsubsection{Comparison between CIR and JCIR}
\label{sec:CVAJCIR}
\begin{figure}
\centering
\subfigure[CIR]{\includegraphics[width=0.48\columnwidth]{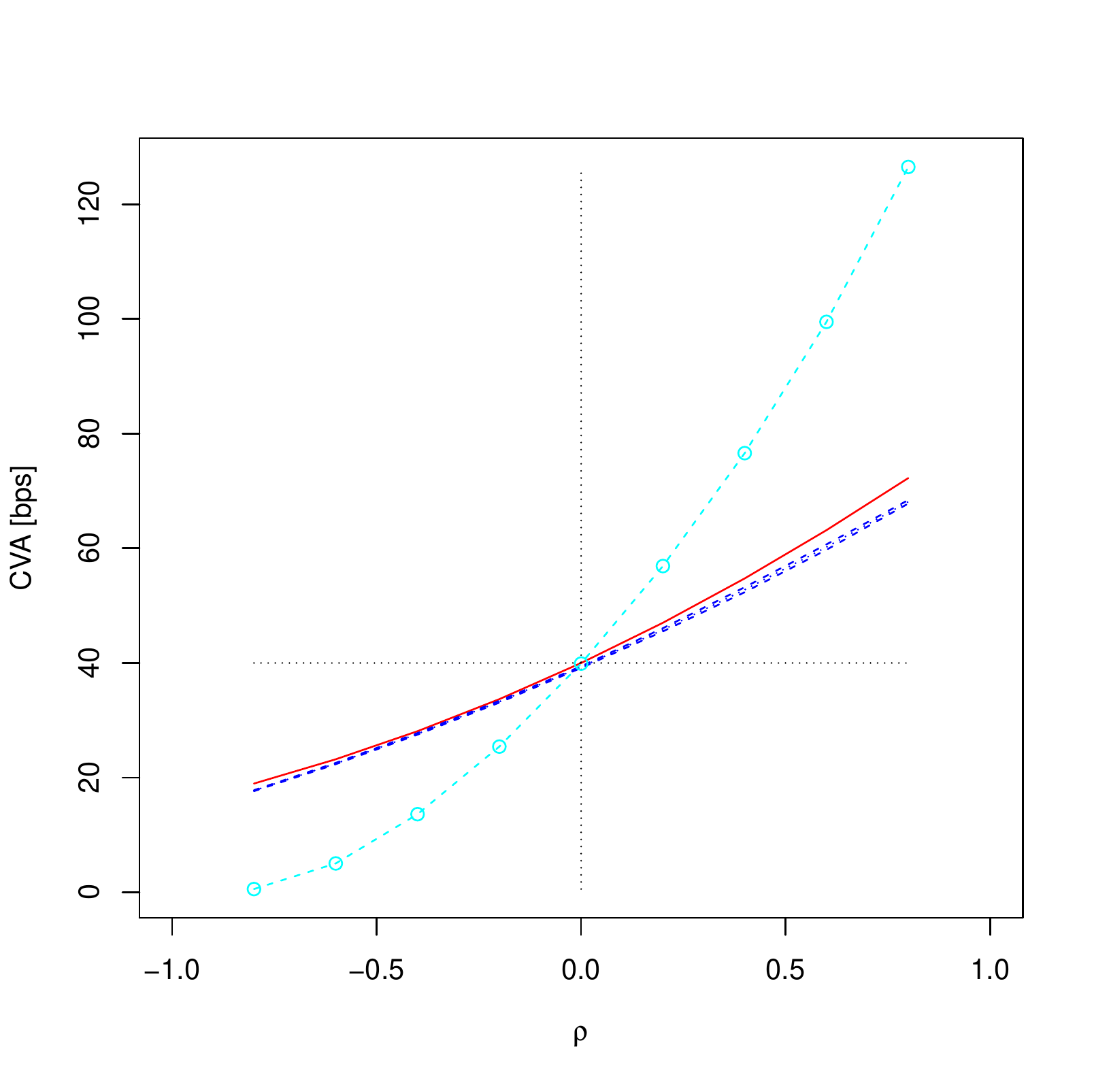}}\hspace{0.02cm}
\subfigure[JCIR]{\includegraphics[width=0.48\columnwidth]{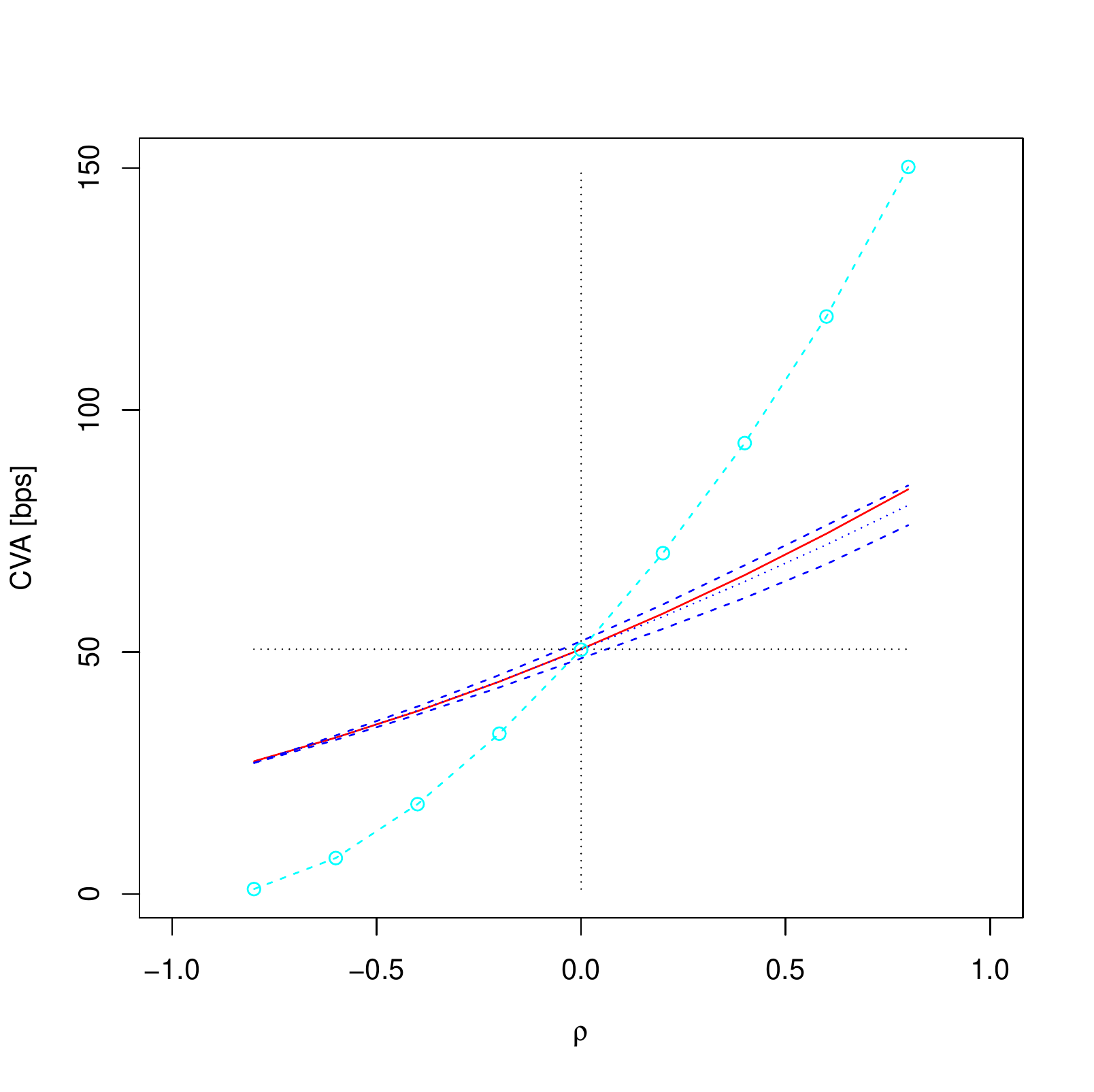}}\\
\subfigure[CIR]{\includegraphics[width=0.48\columnwidth]{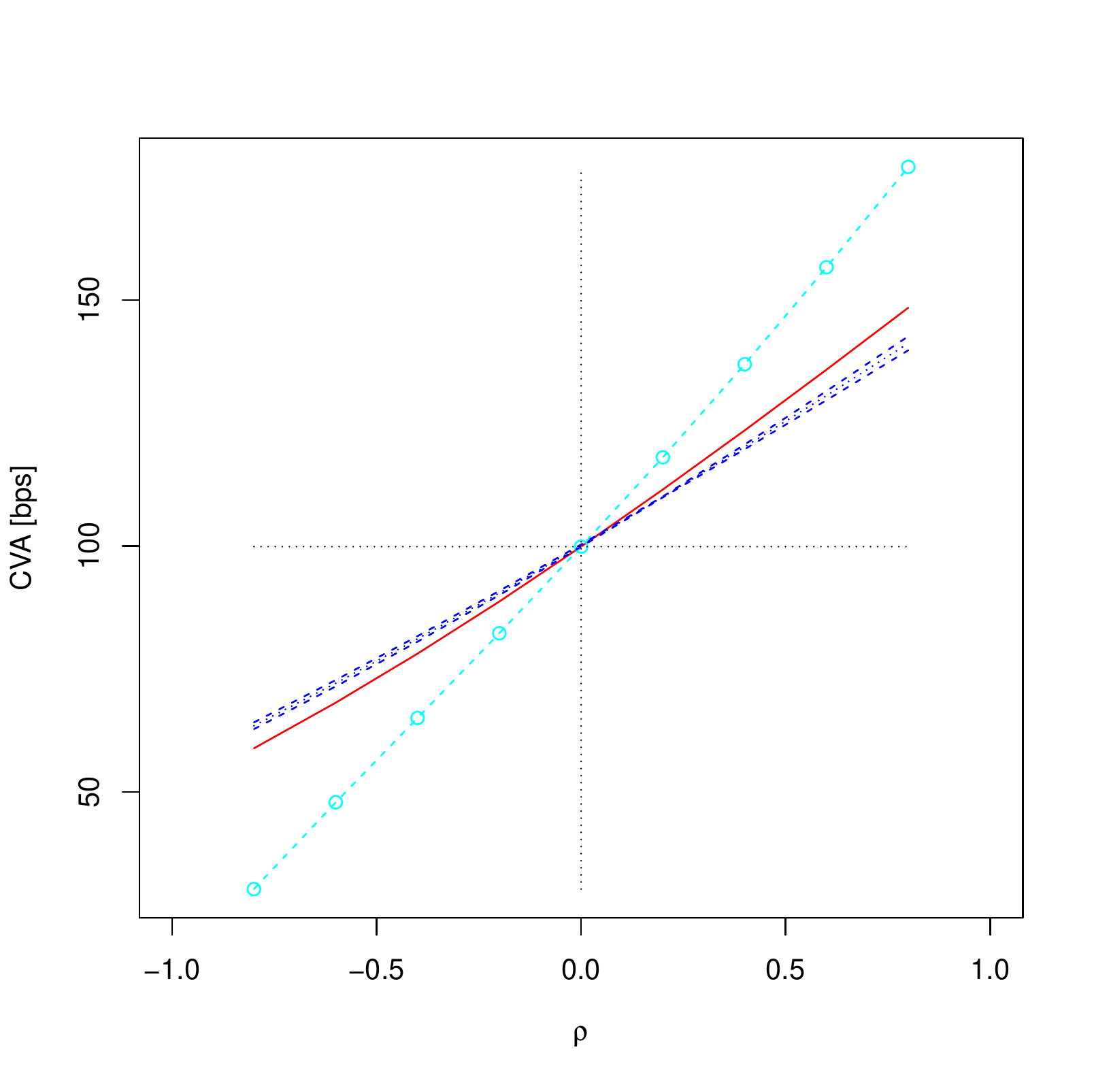}}\hspace{0.02cm}
\subfigure[JCIR]{\includegraphics[width=0.48\columnwidth]{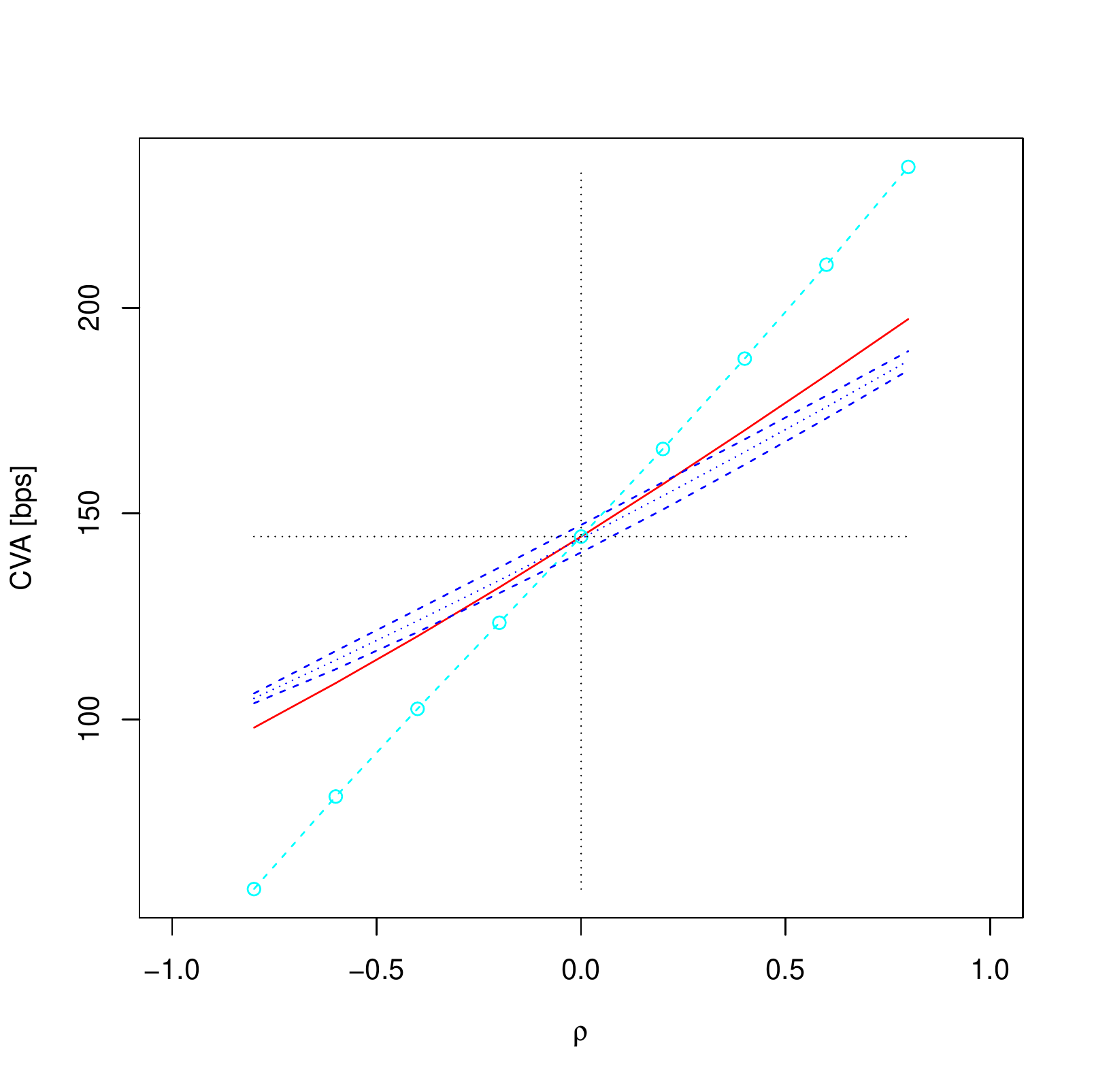}}\\
\caption{CVA Figures for Gaussian copula (dotted cyan), deterministic drift adjustment (red) and Monte Carlo methods (blue, average $\pm$ 2 standard deviations on $10\times 10k$ paths)  (right). Profiles: 3Y Gaussian exposures with $\nu=8\%$ and CIR parameters given by Set 2 (top) and 15Y swap-type exposures with $\nu=2.2\%$ and CIR parameters given by Set 3 (bottom). In both cases, JCIR arrival rate and mean size of jumps are given by $\alpha=\gamma=10\%$.}\label{Fig.CVA.JCIR.DB}
\end{figure}

%
Figure~\ref{Fig.CVA.JCIR.DB} provides the CVA as a function of $\rho$ for CIR and JCIR. 

For the sake of comparison, we also provide the results implied by the Gaussian Copula (static resampling) approach. The idea behind the resmapling method is to assume that $V_t$ and $\tau$ are linked via a given copula for any $t$. The Gaussian copula is specifically handy when the exposure is normally distributed at any point in time, $V_t\sim\cN(\mu(t),\sigma(t))$. To see this, notice first that $V_t$ has the same distribution as a Uniform random variable $U$ mapped through the quantile function $F^{-1}_{V_t}$ of $V_t$:
\beq
V_t\sim F^{-1}_{V_t}(U)\;.\nonumber
\eeq

As $G(\tau)\sim U$ one can parametrize $U$ as a function of $\tau$ using a Gaussian coupling scheme: $U(\tau):=\Phi(\rho \Phi^{-1}(G(\tau))+\sqrt{1-\rho^2}Z)\sim U$; where $Z$ is a standard Normal random variable independent from $\tau$; this amounts to say that $V_t$ and $\tau$ are linked via a Gaussian copula with constant correlation $\rho$. Hence, one can draw samples of $V_t$ \textit{conditionally upon} $\tau=t$ by evaluating $F^{-1}_{V_t}$ at $U(t)$. In the specific case where the exposure is Gaussian, $F^{-1}_{V_t}(x)=\mu(t)+\sigma(t)\Phi^{-1}(x)$ so that finally
\beq
\left. V_t\right|_{\tau=t}\sim F^{-1}_{V_t}(U(t))=\mu(t)+\sigma(t)\rho \Phi^{-1}(G(t))+\sigma(t)\sqrt{1-\rho^2}Z\sim\cN\left(\mu^\rho(t),\sigma^\rho(t)\right)\;,\nonumber
\eeq
where $\mu^\rho(t):=\mu(t)+\rho\sigma(t)\Phi^{-1}(G(t))$ and $\sigma^\rho(t):=\sigma(t)\sqrt{1-\rho^2}$.
%
%

Using~(\ref{eq:AnEPE}), the EPE associated to the Gaussian copula approach takes then the simple analytical form
\beq
\hbox{EPE}(t)=\sigma^\rho(t)\phi\left(\frac{\mu^\rho(t)}{\sigma^\rho(t)}\right)+\mu^\rho(t)\Phi\left(\frac{\mu^\rho(t)}{\sigma^\rho(t)}\right)\;.\nonumber
\eeq

We plot on Fig.\ref{Fig.CVA.JCIR.DB} some CVA figures for CIR, JCIR and the Gaussian copula as a function of the correlation parameter $\rho$. Notice that the Gaussian Copula figures are impacted by the choice of the CIR parameters as they depend on the curve $G(t)$ that is assumed equal to $P^\lambda(0,t)$, which is a function of the parameters driving $\lambda$.
%
%
%
%
\subsubsection{Impact of the discretization scheme and the deterministic approximation}
\label{sec:CVADiscreteScheme}
%
We analyze here the impact of the discretization scheme, the time step $\delta$ as well as the choice of the deterministic approximation $\theta(s,t)$ of $\theta_s^t$,~(\ref{eq:DraftAdjh}) or~(\ref{eq:DraftAdjlb}). One can see from Table~\ref{Tab2} that the impact of the deterministic approximation of $\theta_s^t$ is lower than 1 basis point except when Feller condition is strongly violated due to a very large volatility (Set 4); in that case $h(t)$ and $\bar{\lambda}(t)$ can signficantly differ for large $t$. It is not surprising to observe that the performance of the deterministic approximation deteriorates for large $\rho$ in such volatile cases. Observe that similarly, the impact of the discretization scheme is typically limited to one basis point in all cases except again for Set 4. 

%
\begin{table}
\centering
\begin{tabular}{|c|c|ccc|ccc|ccc|ccc|}
\hline
&$\delta$&\multicolumn{3}{c}{WM(1)}	& \multicolumn{3}{|c}{WM(2)} &	\multicolumn{3}{|c}{MC(1)} &	\multicolumn{3}{|c|}{MC(2)}	\\
\hline
\multirow{2}{*}{Set 1} & 0.01 &	\multirow{2}{*}{20} &	\multirow{2}{*}{36} &	\multirow{2}{*}{57} &	\multirow{2}{*}{21} &	\multirow{2}{*}{36} &	\multirow{2}{*}{57}	& 19 $\pm$ 1	& 35 $\pm$ 2	& 55 $\pm$ 3 & 19 $\pm$ 1 & 36 $\pm$ 3 &55 $\pm$ 1\\
 & 0.001 &	 &	 &	&	&	& & 19 $\pm$ 1	& 36 $\pm$ 1	& 55 $\pm$ 1 & 20 $\pm$ 1 & 36 $\pm$ 1 &55 $\pm$ 1\\
\hline
\multirow{2}{*}{Set 2} & 0.01 &	\multirow{2}{*}{19} &	\multirow{2}{*}{40} &	\multirow{2}{*}{72} &	\multirow{2}{*}{19} &	\multirow{2}{*}{40} &	\multirow{2}{*}{72}	& 18 $\pm$ 0	& 40 $\pm$ 1	& 69 $\pm$ 3 & 18 $\pm$ 1 & 40 $\pm$ 1 & 69 $\pm$ 2\\
 & 0.001 &	 &	 &	&	&	& & 18 $\pm$ 1	& 40 $\pm$ 1	& 69 $\pm$ 2 & 18 $\pm$ 0 & 40 $\pm$ 2 & 69 $\pm$ 2\\
\hline
\multirow{2}{*}{Set 3} & 0.01 &	\multirow{2}{*}{6} &	\multirow{2}{*}{18} &	\multirow{2}{*}{40} &	\multirow{2}{*}{6} &	\multirow{2}{*}{18} &	\multirow{2}{*}{40}	& 7 $\pm$ 1	& 18 $\pm$ 1	& 37 $\pm$ 1 & 7 $\pm$ 0 & 18 $\pm$ 1 & 37 $\pm$ 2\\
 & 0.001 &	 &	 &	&	&	& & 6 $\pm$ 1	& 18 $\pm$ 0	& 37 $\pm$ 1 & 7 $\pm$ 1 & 18 $\pm$ 1 & 36 $\pm$ 2\\
\hline
\multirow{2}{*}{Set 4} & 0.01 &	\multirow{2}{*}{3} &	\multirow{2}{*}{37} &	\multirow{2}{*}{141} &	\multirow{2}{*}{3} &	\multirow{2}{*}{37} &	\multirow{2}{*}{138}	& 6 $\pm$ 1	& 35 $\pm$ 2	& 94 $\pm$ 3 & 14 $\pm$ 1 & 47 $\pm$ 2 & 111 $\pm$ 3\\
 & 0.001 &	 &	 &	&	&	& & 6 $\pm$ 1	& 34 $\pm$ 2	& 93 $\pm$ 5 & 10 $\pm$ 1 & 42 $\pm$ 2 & 104 $\pm$ 5\\
\hline
\end{tabular}
\caption{CVA figures (upfront in bps, rounded) for Gaussian exposure with maturity 3Y and volatility $\nu=8\%$. Methods $WM(1)$ and $WM(2)$ corresponds to the drift-adjustment method with deterministic approximations (\ref{eq:DraftAdjh}) and (\ref{eq:DraftAdjlb}), respectively. Methods $MC(1)$ and $MC(2)$ corresponds to the full Monte Carlo method with discretization scheme (\ref{eq:CIR:Lord}) and (\ref{eq:CIR:Reflex}), respectively. The three quotes per column respectively correspond to upfront CVA in bps for $\rho=-0.8$ (left) $\rho=0$ (middle) and $\rho=0.8$ (right). The confidence intervals have been generated from 10 sets of simulations featuring $10k$ paths each and correspond to global average $\pm$ twice the empirical CVA's standard deviation.}\label{Tab2}
\end{table}
%
\begin{remark} We can use any of the deterministic approximations $\theta(s,t)$ of $\theta_s^t$ as both $h(t)$ and $\bar{\lambda(t)}$ can be easily obtained in the case of the CIR$^{++}$ dynamics. For instance,
$$\bar{\lambda}(s)=\psi(s)+y_0e^{-\kappa s}+\theta(1-e^{-\kappa s})\;,$$
where $\psi$ can be extracted from the market-implied curve $G$. Both deterministic approximations yield very similar results except in extreme scenarii. Therefore, we restrict ourselves to show the results related to the second approximation, replacing $\lambda_s$ by $h(s)$ as in~(\ref{eq:DraftAdjh}).
\end{remark}
%
\section{Conclusion}
\label{sec:conclusion}
%
{Wrong way risk is a well-known key driver of counterparty credit risk. In spite of its primary importance however, it is frequently disregarded. The standard CVA formula provided in the Basel III report for instance does not propose a WWR framework. This is obviously a major shortcoming that may drastically underestimate the figures. Such a simplification is commonly justified by the lack of a better alternative of accounting for wrong way risk in a sound (yet tractable) manner.

In this paper, a new methodology has been proposed to overcome the difficulties of modeling credit risk in a reduced-form setup for tackling WWR when pricing CVA. This method relies on a new equivalent measure called \textit{wrong way} measure. The outcome is that the effect of WWR is embedded in a drift adjustment of the exposure process. This drift adjustment is a stochastic process that generally depends on the stochastic intensity. Consequently, the change-of-measure technique does not lead, strictly speaking, to a dimensionality reduction of the CVA pricing problem. Nevertheless, it is possible to avoid the simulation of the intensity process by approximating the drift adjustment by a deterministic function. In spite of its simplicity, numerical evidence shows that for a broad range of parameter values, the expected positive exposure profiles under WWR are very well approximated when replacing the intensity $\lambda_t$ by the hazard rate $h(t)$ or its expected value $\bar{\lambda}_t$ in the drift adjustment. Therefore, the approximation has a typically limited impact on CVA figures, providing arguably satisfactory estimations given the uncertainty on other key variables like e.g. the recovery rate or the close-out value of the portfolio. Hence, the proposed setup drastically simplifies the management of WWR when pricing CVA.

%
\section*{Appendix}
\label{sec:app}
%
%
\subsection*{Ornstein-Uhlenbeck (OU) formulae}
%
The dynamics of OU (or Vasicek) intensities is given by the SDE
$$dy_t=\kappa(\theta-y_t)dt+\sigma dW^\lambda_t$$
in which case $\lambda$ defined as $\lambda_t=y_t+\psi(t)$ is known as the Hull-White dynamics. This model is very popular for interest-rates modeling. Nevertheless, it is not appropriate for the modeling of stochastic intensities as it is a Gaussian process and hence can take negative values. This inconsistency is revealed by our methodology as in OU, the num\'eraire is not almost surely positive. Hence, the resulting figures can be negative, which is of course impossible. 

Yet, the analytical expressions of the functions $A,B,A_t$ and $B_t$ involved in the drift adjustment are available. Setting $\tau:=t-s$, one finds
\beqn
A^{\hbox{{\tiny OU}}}(s,t)&=&\exp\left\{\left(\theta-\frac{\sigma^2}{2\kappa^2}\right)(B^{\hbox{{\tiny OU}}}(s,t)-\tau)-\frac{\sigma^2}{4\kappa^2}\left(B^{\hbox{{\tiny OU}}}(s,t)\right)^2\right\}\nonumber\\
A^{\hbox{{\tiny OU}}}_t(s,t)&=&A^{\hbox{{\tiny OU}}}(s,t)\left((B^{\hbox{{\tiny OU}}}_t(s,t)-1)\left(\theta-\frac{\sigma^2}{2\kappa^2}\right)-\frac{\sigma^2}{2\kappa}B^{\hbox{{\tiny OU}}}(s,t)B^{\hbox{{\tiny OU}}}_t(s,t)\right)
\nonumber\\
B^{\hbox{{\tiny OU}}}(s,t)&=&\frac{1-e^{-\kappa\tau}}{\kappa}\nonumber\\
B^{\hbox{{\tiny OU}}}_t(s,t)&=&e^{-\kappa\tau}\nonumber
\eeqn
%
%
\subsection*{Cox-Ingersoll-Ross (CIR) formulae}
%
When $y$ is a CIR process, i.e. when
$$dy_t=\kappa(\theta-y_t)dt+\sigma\sqrt{y_t} dW^\lambda_t$$
then $\lambda$ defined as $\lambda_t=y_t+\psi(t)$ is said to be a CIR$^{++}$ process. The process $y$ is always non-negative and there are many circumstances where $\lambda$ remains positive too. One gets 
\beqn
A^{\hbox{{\tiny CIR}}}(s,t)&=&\left(\frac{h\exp(\frac{\kappa+h}{2}\tau)}{e^{h\tau}-1}B^{\hbox{{\tiny CIR}}}(s,t)\right)^{\frac{2\kappa\theta}{\sigma^2}}
\nonumber\\
A^{\hbox{{\tiny CIR}}}_t(s,t)&=&A^{\hbox{{\tiny CIR}}}(s,t)\frac{2\kappa\theta}{\sigma^2}\left(\frac{\frac{\kappa+h}{2}-h e^{h\tau}}{e^{h\tau}-1}+\frac{B^{\hbox{{\tiny CIR}}}_t(s,t)}{B^{\hbox{{\tiny CIR}}}(s,t)}\right)
\nonumber\\
B^{\hbox{{\tiny CIR}}}(s,t)&=&\frac{e^{h\tau}-1}{h+\frac{\kappa+h}{2}\left(e^{h\tau-1}\right)}\nonumber\\
B^{\hbox{{\tiny CIR}}}_t(s,t)&=&e^{h\tau}\left(\frac{B^{\hbox{{\tiny CIR}}}(s,t)h}{e^{h\tau}-1}\right)^2\nonumber
\eeqn
where $h:=\sqrt{\kappa^2+2\sigma^2}$.
%
\subsection*{Cox-Ingersoll-Ross with compound Poisson jumps (JCIR) formulae}
%
Consider jump-diffusion dynamics like JCIR, 
$$dy_t=\kappa(\theta-y_t)dt+\sigma\sqrt{y_t}dW^\lambda_t+dJ_t$$
where $J_t$ is a pure-jump process. A tractable setup is to consider $J_t$ to be a compound Poisson process with exponentially distributed jump sizes with mean $\gamma$ with jump rate $\alpha$. The process $\lambda$ resulting from a deterministic shift $\lambda_t=y_t+\psi(t)$ of this model is called $JCIR^{++}$.

Setting
\beqn
d&:=&\gamma^2-2\kappa\gamma-2\gamma^2\nonumber\\
\nu&:=&\frac{2\alpha\gamma}{d}\nonumber\\
\xi&:=&\frac{h+\kappa+2\gamma}{2}\nonumber\\
\eeqn
one gets
\beqn
A^{\hbox{{\tiny JCIR}}}(s,t)&=&A^{\hbox{{\tiny CIR}}}(s,t)\times\left\{
	\begin{array}{ll}
		\left(\frac{e^{\xi \tau}}{1+\frac{\xi}{h}(e^{h \tau}-1)}\right)^\nu&\hbox{ if }d\neq0\\ 
		\exp\left(-\frac{\alpha\gamma}{\xi}\left(\tau+\frac{e^{-h\tau}-1}{h}\right)\right)&\hbox{ if }d=0
		\end{array}
		\right.\nonumber\\
A^{\hbox{{\tiny JCIR}}}_t(s,t)&=&A^{\hbox{{\tiny JCIR}}}(s,t)\times\left\{
	\begin{array}{ll}
		\left(\frac{A_t^{\hbox{{\tiny CIR}}}(s,t)}{A^{\hbox{{\tiny CIR}}}(s,t)}+\nu\xi\left(1-\frac{e^{h\tau}}{1+\frac{\xi}{h}(e^{h\tau}-1)}\right)\right)&\hbox{ if }d\neq0\\ 
		\frac{\alpha\gamma}{\xi}\left(e^{-h\tau}-1\right)&\hbox{ if }d=0
		\end{array}
		\right.\nonumber\\
B^{\hbox{{\tiny JCIR}}}(s,t)&=&B^{\hbox{{\tiny CIR}}}(s,t)\nonumber\\
B^{\hbox{{\tiny JCIR}}}_t(s,t)&=&B^{\hbox{{\tiny CIR}}}_t(s,t)\nonumber
\eeqn

\ifdefined \MyBib
   \bibliography{\MyBib}
  \bibliographystyle{plain}
\fi

\end{document}